\documentclass{article} 

\usepackage{blindtext}
\usepackage{geometry}
\usepackage[T1]{fontenc}
\usepackage[utf8]{inputenc}
\usepackage[english]{babel}
\usepackage[OT1]{fontenc}
\usepackage[notext]{stix}
\usepackage{hyperref}
\usepackage{tikz}
\usepackage[ruled,vlined,linesnumbered]{algorithm2e}
\usepackage[shortlabels]{enumitem}
\usepackage{booktabs}
\usepackage{multirow}
\usepackage{epigraph}
\usepackage{longtable}
\usepackage{lscape}
\usepackage{caption}
\usepackage{subcaption}
\usepackage{pdfpages}
\usepackage{graphicx}
\usepackage{xspace}
\usepackage{tabularx}
\usepackage{array,booktabs}
\usepackage{amsthm}
\usepackage{mathtools, nccmath}
\usepackage{parskip}
\usepackage{comment}

\usepackage{natbib}
\usepackage[normalem]{ulem}
\let\originalepigraph\epigraph 
\renewcommand\epigraph[2]{\originalepigraph{\textit{#1}}{\textsc{#2}}}

\newcommand{\rojo}[1]           {\textcolor[rgb]{0.98,0.00,0.00}{#1}} 
\newcommand{\azul}[1]           {\textcolor[rgb]{0.00,0.00,1.00}{#1}} 
\newcommand{\OMT}{\textsf{OMT}\xspace}

\newcommand{\mT}{\mathcal{T}}
\newcommand{\mS}{\mathcal{S}}
\newcommand{\mTree}{\mathfrak{T}}

\newcommand{\gapUL}{$g U\hspace{-0.06cm}L\ $}
\newcommand{\gapbUL}{$g\overline{U}\hspace{-0.06cm}L\ $}
\newcommand{\gapbUR}{$g\overline{U}\hspace{-0.06cm}R\ $}
\newcommand{\gapUbL}{$gU\hspace{-0.06cm}\overline{L}\ $}
\newcommand{\meangapUL}{$\overline{g U\hspace{-0.06cm}L}\ $}

\newcommand{\esp}{@{\hspace{0.2cm}}} 

\newcommand{\hhp}{\hspace{-1.8cm}} 
\newcommand{\hp}{\hspace{-0.6cm}\textrm{s.t.:}\hspace{0.25cm}}
 
\newcommand{\Hp}{\hspace{-1.0cm}\hspace{1.23cm}} 

\newtheorem{property}{Property}[section]
\newtheorem{example}{Example}[section]

\newcommand{\msout}[1]{\text{\sout{\ensuremath{#1}}}}
\newcommand{\rv}[1]{\relax\ifmmode\azul{#1}\else\azul{\textbf{#1}}\fi}
\newcommand{\rvv}[2]{\rojo{\ifmmode\msout{#1}\else\sout{#1}\fi} \azul{\ifmmode#2\else\textbf{#2}\fi}}

\title{The Ordered Median Tree Location Problem}
\author{
  Pozo Monta\~no, Miguel A.\footnote{Department of Statistics and Operational Research. University of Seville. E-mail: miguelpozo@us.es,}\\ 
  Puerto Albandoz, Justo\footnote{Department of Statistics and Operational Research. University of Seville. E-mail: puerto@us.es,} \\
  Torrej\'on Valenzuela, Alberto\footnote{Institute of Mathematics of the University of Seville. University of Seville. E-mail: atorrejon@us.es} 
}
\date{\today}

\begin{document}

\maketitle

\begin{abstract}

In this paper, we propose the Ordered Median Tree Location Problem (\OMT). The \OMT is a single-allocation facility location problem where $p$ facilities must be placed on a network connected by a non-directed tree. The objective is to minimize the sum of the ordered weighted averaged allocation costs plus the sum of the costs of connecting the facilities in the tree. We present different MILP formulations for the \OMT based on properties of the minimum spanning tree problem and the ordered median optimization. Given that ordered median location problems are rather difficult to solve we have improved the \OMT solution performance by introducing covering variables in a valid reformulation plus developing two pre-processing phases to reduce the size of this formulations. In addition, we propose a Benders decomposition algorithm to approach the \OMT. We establish an empirical comparison between these new formulations and we also provide enhancements that together with a proper formulation allow to solve medium size instances on general random graphs.


\end{abstract}

\section{Introduction}
\label{sec:Introduction}

Facility location problems are concerned with attaining the optimal placement of facilities in order to minimize costs under certain considerations. In this context, facilities connection is an important feature to include in these problems. \citet{Gupta2001} introduced the Connected Facility Problem (ConnFL), which aims to find the optimal location of a set of facilities that require to be connected by a Steiner tree (see, e.g., \citealp{Ljubic2021}). Several papers and surveys cover different aspects of the ConnFL and new cases of study based on this problem (see, e.g, \citealp{Fortz2015, Ljubic2021}). As detailed in \citet{Gollowitzer2011}, ConnFL is also related to other problems in the literature, for example, rent-or-buy problems (\citealp{Gupta2002}), Steiner tree-star problem (\citealp{Lee1993b, Khuller2002}), general connected facility location problem (\citealp{Bardossy2010}), prize collecting capacitated connected facility location problem (\citealp{Leitner2011}) or the tree of hubs location problem (\citealp{Contreras2010,Pozo2021}).

Although facilities connection can be modeled by using a Steiner tree problem, there are several other possibilities to guarantee such connectivity. Another common approach is exploiting the minimum spanning tree problem (MST) structural properties, (see, e.g., \citealp{Anazawa2001}). The most relevant properties of trees allow that the basic problem of finding a minimum cost spanning tree can be solved efficiently in polynomial time (\citealp{Kruskal1956, Prim1957}) and by means of linear programming tools (\citealp{Miller1960, Edmonds1970, Gavish1983, Martin1991} among several others). Optimization problems related to spanning trees, or simply spanning tree problems, are among the core problems in combinatorial optimization, see, e.g., the former tree of hubs location problem (\citealp{Contreras2010}), the multi-objective spanning tree problem (\citealp{Hamacher1994}) or the Stackelberg minimum spanning tree problem (\citealp{Cardinal2011}). Moreover, spanning trees are found in a wide range of applications (see, e.g., \citealp{Climaco2012}).

To extend and provide a general framework for facility location problems, the Discrete Ordered Median Location Problem (DOMP) was introduced (see \citealp{Nickel2006, Boland2006}), which could be used to model different locations problems, as the $p$-median or the $p$-center problem. DOMP consists in choosing $p$ facility locations and assigning each client to a facility with the smallest allocation cost so to minimize a special objective function, the so-called ordered median function. Given a vector of weights, this function sorts these costs non-decreasingly and then performs the scalar product with the given vector of weights. This adds a sorting feature to the underlying location problem. The objective is to minimize the total allocation cost after applying rank dependent compensation factors. Regarding the rank dependent weights applied to the costs, its application intuitively arises when these weights can be seen as compensation factors that try to diminish unfair situations. DOMP objective function have been successfully applied in the field of location analysis and distribution models (see, e.g., \citealp{Puerto2005, Boland2006, Puerto2008, Marin2009, Kalcsics2010a, Puerto2015, Ljubic2024}). DOMP structure has also been embedded within other well-known problems giving rise to, e.g., ordered weighted average problems (see \citealp{Galand2012, Fernandez2014, Fernandez2017}) or ordered hub location problems (see \citealp{Puerto2011, Puerto2013, Puerto2016, Pozo2021}).

In this paper, we present the Ordered Median Tree Location Problem (\OMT), a single-allocation facility location problem where $p$ facilities must be placed on a network connected by a non-directed tree. The objective is to assign clients to its closest facility in order to minimize the ordered allocation average cost plus the facilities tree connection average cost. In our location problem, every client node can potentially become a facility and these facilities are uncapacitated, i.e., can serve as many clients as needed. Therefore, \OMT has two main modeling aspects to model: the tree structure defined by the network and the rank dependent compensation factors applied to the costs of the system through the ordered median function. Section \ref{sec:ProblemDescription} extends the problem description in more detail. Applications of the \OMT can be given for any general discrete facility location problem that require facilities to be connected by a tree structure network (telecommunications or pipes as an example). The order median component may model the need of compensate those clients worst served (far from their closest facility).

\OMT is a new problem that generalizes both MST and DOMP. \OMT includes as a particular case the MST when $p$ is considered equal to the number of client nodes. If this is the case, clients would self-allocate to themselves giving rise to a MST among all nodes. As the number of facilities $p$ decreases, clients not required to be self-allocated start to appear, as it happens in the ConnFL. Besides, if connectivity costs are set to zero, a ``pure'' DOMP arises. For these reasons, \OMT can be seen as a bi-objective model in which the trade-off between connectivity and allocation costs requires to be balanced. Moreover, by considering the DOMP objective function, we enrich the possibilities in which client allocations are made, distinguishing \OMT from the ConnFL. 
In this paper, we analyze the median, $k$-center and $k$-trimmed mean objectives, but the possibilities are many, e.g., we can model equity criteria such as the range, sum of absolute differences or weighting inversely proportional to the allocation costs, model obnoxious location problems (those in which we want the facilities to be as far apart as possible), model preferences with negative weight vectors, and more.


As aforementioned, solving the \OMT implies solving a generalization of the DOMP and the MST. For this reason, a Benders decomposition arises naturally as a general \OMT solving technique where the previous knowledge from DOMP and MST is jointly exploited (see \citealp{Martins2013} for a successful application of Benders decomposition in a hub location problem with inner tree structure). Benders decomposition (\citealp{Benders1962}) is a method for solving mixed integer programming (MIP) problems that have special structure in the constraint set, i.e., when fixing the complicating variables (integer variables), the mathematical program reduces to an ordinary, easy to solve linear problem. The technique relies on projection and problem separation, followed by solution strategies of dualization, outer linearization and relaxation (\citealp{Lasdon2002, Minoux1986}). In general terms, the complicating variables of the original problem are projected out, resulting into an equivalent model with fewer variables, but many more constraints. To achieve optimality, a large number of these constraints will not be required, suggesting then a strategy of relaxation that ignores all but a few of these constraints. Over the years, different improvements have been proposed in order to improve the performance of the Benders decomposition algorithm, (see \citealp{Geoffrion1972, Magnanti1981, Fischetti2010, Fortz2009, Naoum2013, Conforti2019, Brandenberg2021}). Several successful applications of Benders decomposition to different problems rekindled the interest of the research community for it, motivating the application to the \OMT in the current work. The reader could find a more detailed description about improvements in Benders decomposition in the survey of \citet{Rahmaniani2017}.

The contributions of this paper are the following. 
First, as it is shown in Section 2.2 we present a new problem that was missing in the DOMP and ConnFL/MST state-ot-the-art. Specifically, \OMT generalizes two well-studied, important problems in the sense that both facilities/clients connectivity and clients allocations are jointly included. Therefore, this new problem  arises naturally as a bi-objective model in which we try to balance the trade-off between connectivity and allocation costs. 
Second, we present different MILP formulations for the \OMT that arise naturally from the properties of the DOMP and MST and we compare the number of variables and constraints as a function of the number of ties in the cost matrix. Regarding to the MST subproblem, we provide ad hoc \OMT connectivity cuts and flow formulations. 
Third, we provide theoretical results on the polytopes of the two main \OMT formulations including the number of variables and constraints as a function of the number of ties in the cost matrix. This analysis can be extended to the DOMP and was not previously given in the literature.
Fourth, we present a different alternative formulation that treats the \OMT as a single tree over the complete set of input nodes and we propose a Benders decomposition that arises naturally as a general \OMT solving technique where the previous knowledge from DOMP and MST is jointly exploited. 
Fifth, we derive two new \OMT fast heuristics designed for providing initial solutions in the branching process and we also take advantage of existing DOMP preprocessings for fixing variables. 
Finally, we derive extensive computational results comparing in detail the different formulations, enhancements and solution techniques provided. 

The remainder of the paper is organized as follows. In Section \ref{sec:ProblemDescription}, we formally define the \OMT reviewing a scheme of some well-known related problems in the literature. Section \ref{sec:OMTformulations} presents the catalogue of formulations and algorithms studied for the \OMT. Section \ref{sec:Improvements} shows the improvements made in order to enhance the proposed models, initial solution computation and some preprocessing phases developed for variable fixing. The empirical performance of the resulting \OMT formulations is analyzed in Section \ref{sec:Computationalexperiments}, where we present extensive numerical results and a comparison of these formulations for different particular cases. Finally, some conclusions are summarized in Section \ref{sec:Conclusions}.

\section{Problem description}
\label{sec:ProblemDescription}

\subsection{Notation and definition}
\label{subsec:Notationanddefinition}

In this section we formally introduce the \OMT and fix the notation for the rest of the paper. Let $G = (V,E)$ be a network where $V$ is the set of nodes (assumed to be clients or potential facilities) and $E$ the set of edges connecting nodes. In the \OMT, $p \leq |V|$ facilities must be placed on nodes of $V$ and connected by a non-directed tree. The model assumes single-allocation, i.e., each client node has a unique facility where it is allocated. A weight $c_{ij} \geq 0$ is defined for the cost of allocating client $i$ to facility $j$ or as the design cost of edge $(i,j)$ whether both $i,j \in V$ are selected facilities. According to many \OMT related problems described in the literature, these costs are assumed to be symmetrical between each pair of nodes, $c_{ij} = c_{ji}$ for $i,j \in V$, and nodes can be allocated to themselves with no cost, $c_{ii}=0$ for each $i\in V$, that is the so called free self-service assumption. In addition, the allocation costs of the $|V|$ clients to their corresponding facilities are compensated using scaling factor parameters $\lambda = (\lambda_1, \ldots, \lambda_{|V|})$ (see \citealp{Nickel2006, Boland2006, Kalcsics2010b, Marin2009, Marin2010, Puerto2011, Puerto2013, Puerto2016, Pozo2021} for different ordered median location models). If client $i$ is allocated to facility $j$ at a cost $c_{ij}$ and this cost is non-decreasingly ordered in the $\ell$-th position among this type of costs, then this term would be scaled by $\lambda_\ell$, i.e., the corresponding compensated cost would be $\lambda_\ell c_{ij}$. Compensation of allocation costs is based on the fact that a solution that is good for the system does not have to be acceptable for all single parties if their costs to reach the system are too high in comparison to other parties. We compensate this parties to prevent those sites from not using the system as an act of fairness. For the sake of understandability, we summarize the introduced notation in Table \ref{tab:Notation}.

\begin{table}[h!]
\renewcommand{\esp}{\hspace{0.1cm}}
\begin{center}
\begin{tabular}{@{\hspace{0.1cm}}c@{\hspace{0.1cm}} @{\hspace{0.1cm}}l@{\hspace{0.1cm}}}
    \hline
     $G=(V,E)$        & undirected network where $V$ is the set of nodes and $E$ is the set of edges \\
     $p$              & fixed number of facilities to locate \\
     $c_{ij}$         & design cost of edge $(i,j) \in E$ or allocation cost between client $i$ and a facility \\
                      & placed at node $j$ \\
     $\ell\in V$      & index for the $\ell$-th position of the sorted allocation costs sequence \\
     $\lambda_{\ell}$ & scaling factor for the $\ell$-th allocation cost \\
\hline
\end{tabular}
\caption{Notation introduced for the \OMT.}
\label{tab:Notation}
\end{center}
\end{table}

With the above notation, the \OMT is to find the optimal location of $p$ facilities in a network in such a way that:

\begin{enumerate}
  \item Each client is allocated to exactly one facility.
  \item Facilities are connected together following a tree structure.
  \item Edge costs connecting facilities plus compensated ranked allocation costs are minimized. Since the number of tree edges might be quite different from the number of allocation arcs, the objective function will minimize the average cost of the compensated ranked allocation costs plus the average cost of a tree edge.
\end{enumerate}

Observe that depending on the choices of the $\lambda$-vector we can obtain different criteria to account for the costs in the objective function. For instance, if $\lambda=(0,\ldots,0,1,\stackrel{k}{\ldots},1)$ is considered, the allocation costs component of the objective function would be the sum of the $k$-largest costs ($k$-centrum). This usually provides different solutions or different allocation patterns for problems with different $\lambda$, even though the optimal solution gets the same set of facilities (see \citealp{Puerto2011}). 

\begin{figure}[h!]
\centering
\begin{subfigure}[rm]{.3\textwidth}
\centering
\includegraphics[page=2,width=4cm]{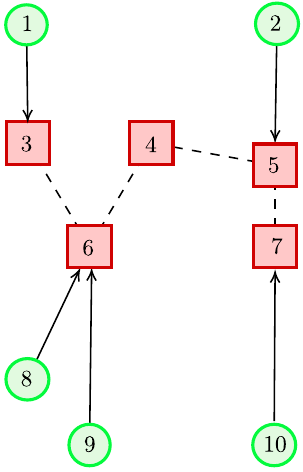}
\caption{}
\end{subfigure}
\begin{subfigure}[rm]{.3\textwidth}
\centering
\includegraphics[page=3,width=4cm]{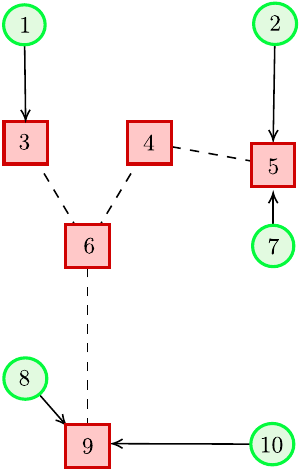}
\caption{}
\end{subfigure}
\begin{subfigure}[rm]{.3\textwidth}
\centering
\includegraphics[page=4,width=4cm]{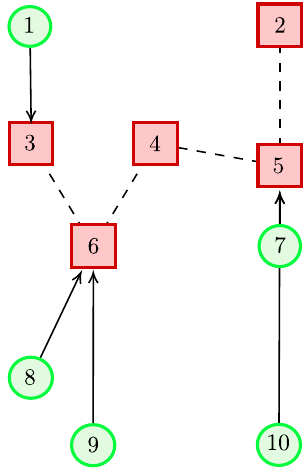}
\caption{}
\end{subfigure}
\begin{minipage}{0.9\textwidth}
\caption{Three different \OMT solutions according to (a) median [objective value: 18.3], (b) $k$-centrum [objective value: 26.0] and (c) $k$-trimmed mean criterion [objective value: 16.0], considering rounded costs proportional to Euclidean distances.}
\label{OMT_example1}
\end{minipage}
\end{figure}

Figure \ref{OMT_example1} depicts different \OMT solutions for a network example of $|V|=10$ clients (\textit{green circles}) and cost matrix (\ref{example_matrix}), where $p=5$ facilities (\textit{red squares}) are located. Considering rounded costs proportional to Euclidean distances, client allocations to facilities (\textit{arrows}) and the tree design connecting facilities (\textit{dashed lines}) are different according to different criteria, namely (a) median $\lambda=(1,\stackrel{10}{\ldots},1)$, (b) $k$-centrum $\lambda=(0,\stackrel{6}{\ldots},0,1,\stackrel{4}{\ldots},1)$ and (c) $k$-trimmed mean criterion $\lambda=(0,0,0,1,1,1,1,0,0,0)$. For the latter, the three most expensive allocations, which correspond to nodes $8$, $9$ and $10$, are ignored for the objective value computation. Due to this reason, for this criterion, those clients could be allocated to any facility, despite in Figure \ref{OMT_example1} appear allocated to its closest facility. 
We recall that \OMT objective function includes the minimization of $p-1$ edges costs plus a variable number of allocation costs that will vary depending on the instance size, the criteria selected, etc. That is, the contribution of the objective function corresponding to the tree design part may change considerably compared to the weight of the compensated ranked allocation costs. 
For that reason, since the number of tree edges might be quite different from the number of allocation arcs, the objective function minimizes the average cost of the compensated ranked allocation costs plus the average cost of a tree edge. 
Averaging the two features in the objective function is needed in order to normalize the trade-off between these two elements. 
As a consequence of this rationale, we have now a higher value in the $k$-centrum optimum because we have the average between the higher costs divided by the number of costs considered in the lambda vector (smaller denominator) instead of the total average as in the median criterion.

A first consideration about the \OMT is that it is an ${\cal NP}$-complete problem, which comes from the hardness results of the different combinatorial problems that give rise to the \OMT (see Section \ref{subsec:Subproblems}). Both the $p$-median location problem and the ordered median problem are known to be ${\cal NP}$-complete problems on its decision version (\citealp{Nickel2006}).

\begin{equation}
\label{example_matrix}
\begin{pmatrix}
\centering
 0 & 25 & 10 & 18 & 28 & 21 & 32 & 35 & 40 & 47 \\
25 &  0 & 27 & 14 & 12 & 28 & 20 & 43 & 45 & 40 \\
10 & 27 &  0 & 15 & 25 & 11 & 27 & 25 & 30 & 39 \\
18 & 14 & 15 &  0 & 10 & 14 & 14 & 29 & 32 & 32 \\
28 & 12 & 25 & 10 &  0 & 21 &  7 & 34 & 34 & 28 \\
21 & 28 & 11 & 14 & 21 &  0 & 20 & 16 & 20 & 28 \\
32 & 20 & 27 & 14 &  7 & 20 &  0 & 29 & 28 & 20 \\
35 & 43 & 25 & 29 & 34 & 16 & 29 &  0 &  7 & 25 \\
40 & 45 & 30 & 32 & 34 & 20 & 28 &  7 &  0 & 20 \\
47 & 40 & 39 & 32 & 28 & 28 & 20 & 25 & 20 &  0 
\end{pmatrix}
\end{equation}

\subsection{Subproblems and related problems}
\label{subsec:Subproblems}

\begin{figure}[h!]
\begin{center}
\includegraphics[page=1,width=10cm]{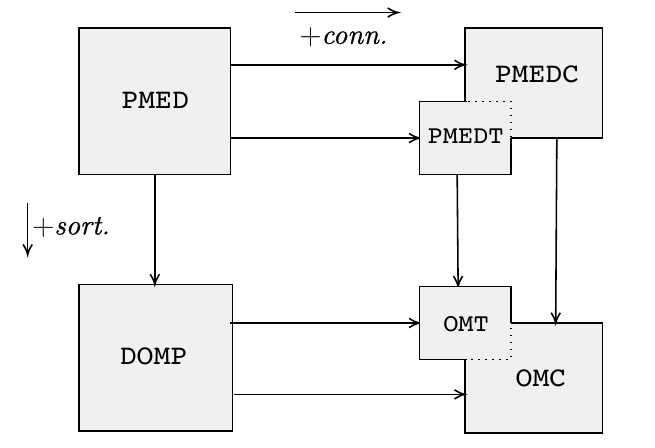}
\caption{Diagram of relationships among the different \OMT subproblems}
\label{Fig:figura1}
\end{center}
\end{figure}

As mentioned, the \OMT has two main modeling aspects: the tree structure defined by the facilities network (tree connectivity) and the rank-dependent compensation factors applied to the operation cost of the system through the ordered median function (sorting). Therefore, the \OMT can be seen as the classical $p$-Median Location Problem (PMED) adding two different features: \textit{tree connectivity} and \textit{sorting}. In this way, the \OMT has as subproblems different well-known problems of the literature. The connectivity imposed to PMED gives rise to the PMED with inner Connected Structure (PMEDC). When the connected structure is a tree, we define the PMED with inner Tree Structure (PMEDT), that is closely related to several known problems in the literature as the tree-star network design problem (\citealp{Nguyen2007}) or the already mentioned connected facility location problem (\citealp{Gollowitzer2011}). The sorting feature can be included in PMED giving rise to the well-known Discrete Ordered Median Location Problem (DOMP) (\citealp{Nickel2006}). The union of both features gives raises to the DOMP with inner Connected Structure (OMC) from which, when the connected structure considered is a tree, we define our case of study, namely the \OMT. The reader is referred to Figure \ref{Fig:figura1} for an overview of the different considered problems and their relationships.

In addition, the \OMT can be understood as a subproblem of the Ordered Median Tree of Hubs Location Problem (OMTHL) (\citealp{Pozo2021}), where flow exchange between origin-destination pairs is considered for an \OMT. The interested reader may also check the tree of hubs location problem (\citealp{Contreras2010}) and the ordered median hub location problem (\citealp{Puerto2011}). Dealing with the \OMT as a subproblem of the OMTHL is valuable in the sense that the model structure is inherited. Nevertheless, the inherent complexity of the OMTHL and the fact that the \OMT holds its own interest, suggests that the \OMT should be studied as a singular problem.

\section{Analyzing the \OMT: formulations, polyhedral descriptions and Benders decomposition}
\label{sec:OMTformulations}

\subsection{Two \OMT formulations}
\label{subsec:Sortingformulations}

In the following we will present a general integer programming formulation for the \OMT. Given the set of edges $E=\{ (i,j) \ | \ i,j\in V, i<j \}$, we define the set of arcs $A = \{(i,j), (j,i)\ | \ (i,j) \in E\} \cup \{(i,i) \ | \ i \in V \}$, where loops connecting a node to itself are allowed. For modeling purposes the following variables are defined:

\begin{itemize}[noitemsep]
    \item $x_{ij}\in \{0,1\}$ for $(i,j) \in A$, equal to 1 if client $i$ is allocated to facility $j$, 0 otherwise.
    \item $z_{ij}\in \{0,1\}$ for $(i,j) \in E$, equal to 1 if an edge $(i,j)$ connects facilities $i,j \in V$, 0 otherwise.
    \item $x_{ij}^{\ell}\in \{0,1\}$ for $(i,j) \in A,\ell \in V$, equal to 1 if client $i$ is allocated to facility $j$ and $c_{ij}$ is ranked in the $\ell$-th position, 0 otherwise.
\end{itemize}

Along the text, we would assume working with complete networks. However, the proposed formulations are valid for general graphs with edge set $E$ and arc set $A$, as long as the connectivity of the network is ensured. The formulation presented below is based on the DOMP three-index formulation first introduced by \citealp{Boland2006}, including the tree connectivity feature of the \OMT. Therefore, the formulation can be defined as follows:

\begingroup
\allowdisplaybreaks
\begin{subequations}
\begin{align}
  & \hhp \hspace{1cm} F1_{x^\ell}^{\mT}: \min \quad \frac{1}{\displaystyle{\sum_{\ell \in V} \lambda_\ell}} \sum_{\ell \in V} \sum_{(i,j) \in A} \lambda_\ell c_{ij} x_{ij}^\ell + \frac{1}{p-1} \sum_{(i,j) \in E} c_{ij} z_{ij}   & & & \label{OMT_f1_c0} &  \\
  & \hp \sum_{i \in V} x_{ii} = p                     &   &                            & \label{OMT_f1_c1}  \\
  & \Hp \sum_{j \in V} x_{ij} = 1                     &   & i \in V                    & \label{OMT_f1_c2}  \\
  & \Hp x_{ij} \leq x_{jj}                            &   & (i,j) \in A: i\neq j       & \label{OMT_f1_c3}  \\
  & \Hp 2 z_{ij} \leq x_{ii} + x_{jj}                 &   & (i,j) \in E                & \label{OMT_f1_c4}  \\
  & \Hp z_{ij} \in \mT                                &   & (i,j) \in E                & \label{OMT_f1_c5}  \\
  & \Hp \sum_{\ell \in V} x_{ij}^\ell = x_{ij}        &   & (i,j) \in A                & \label{OMT_f1_c6}  \\
  & \Hp \sum_{(i,j) \in A} x_{ij}^\ell = 1            &   & \ell \in V                 & \label{OMT_f1_c7}  \\
  & \Hp \sum_{(i,j) \in A} c_{ij}x_{ij}^\ell \leq \sum_{(i,j) \in A} c_{ij}x_{ij}^{\ell+1} &   & \ell \in V: \ell < |V|  & \label{OMT_f1_c8} \\
  & \Hp x_{ij} \in \{0,1\}                            &   & (i,j) \in A                & \label{OMT_f1_c9}  \\
  & \Hp x_{ij}^\ell \in \{0,1\}                       &   & (i,j) \in A,\ell \in V     & \label{OMT_f1_c10} \\
  & \Hp z_{ij} \in \{0,1\}                            &   & (i,j) \in E.               & \label{OMT_f1_c11}
\end{align}
\end{subequations}
\endgroup

The objective function (\ref{OMT_f1_c0}) minimizes the compensated allocation costs weighted by the number of allocations considered (first component) plus the design cost of the tree connecting facilities weighted by the number of edges (second component). In other words, (\ref{OMT_f1_c0}) minimizes the average cost of an allocation plus the average cost of a tree edge. The group of constraints (\ref{OMT_f1_c1}), (\ref{OMT_f1_c2}) and (\ref{OMT_f1_c3}) models the allocation of clients to facilities. Specifically, constraint (\ref{OMT_f1_c1}) fixes the number of facilities in the network to $p$, constraints (\ref{OMT_f1_c2}) ensure that each client is allocated to exactly one facility and constraints (\ref{OMT_f1_c3}) ensure that no client is allocated to a non-facility node. Constraints (\ref{OMT_f1_c4}) relate the allocation variables to the design variables, imposing that each edge of the facilities spanning tree can only be selected if both edge nodes are facilities. Note that these constraints are obtained from the unification of two simpler group of constraints ($z_{ij} \leq x_{ii} \ \& \ z_{ij} \leq x_{jj}$ for $(i,j) \in E$). Constraints (\ref{OMT_f1_c5}) model the connectivity feature of the \OMT describing the facilities spanning tree polytope $\mT$. These constraints are different depending on the tree characterization considered (see Section \ref{subsec:Tdescription}). Finally, constraints (\ref{OMT_f1_c6}), (\ref{OMT_f1_c7}) and (\ref{OMT_f1_c8}) model the sorting feature of the problem. Specifically, constraints (\ref{OMT_f1_c6}) relate the sorting variables to the allocation variables guaranteeing that if the allocation of client $i$ to facility $j$ is selected, then only one sorting position can be assumed by this allocation, constraints (\ref{OMT_f1_c7}) ensures that every sorting position must be occupied and constraints (\ref{OMT_f1_c8}) sort allocation costs in non-decreasing order.

Note that, since $x_{ij} = \sum_{\ell \in V} x_{ij}^\ell$, by constraint (\ref{OMT_f1_c6}) it is possible to relax the integrality constraint (\ref{OMT_f1_c9}), defining $x_{ij} \in [0,1]$ for all $(i,j) \in A$. Furthermore, each $x_{ij}$ can be replaced by a sum $\sum_{\ell \in V} x_{ij}^\ell$ unifying the allocation and sorting set of variables. For the last, despite the number of variables is reasonably reduced, the number of constraints increases and preliminary tests have shown us that such unification is not efficient in computational terms. For this reason disaggregated formulations are presented along the text.

Let us now consider the ordered sequence of unique allocation costs of $c_{ij} \geq 0$ for $i,j \in V$:

$$c_{(0)} := 0 < c_{(1)}<c_{(2)}<\ldots<c_{(|H|)}=\max_{i,j \in V} c_{ij},$$

where $|H|$ is the number of different non-zero elements of the above allocation cost sequence and $H = \{1,\ldots,|H|\}$. This ordering can be used to perform the sorting process of the allocation costs. Let us define the following variables (namely \textit{covering variables} in the following):

\begin{itemize}
    \item $u_{\ell h} \in \{0,1\}$ for $\ell \in V, h \in H$ if the $\ell$-th allocation cost is at least $c_{(h)}$,
\end{itemize}

that is, the $\ell$-th smallest cost is equal to $c_{(h)}$ if and only if $u_{\ell 1} =... =u_{\ell h} = 1$ and $u_{\ell h+1}=...=u_{\ell |H|}=0$.

\begin{example}
Given $c=\begin{pmatrix}
	0 & 2 & 4 \\
	2 & 0 & 0 \\
	4 & 0 & 0
\end{pmatrix}$, $p=1$ and $\lambda=(1,1,1)$ an OMT solution implies $x_{12}^3=x_{22}^1=x_{32}^2=1$ (or equivalently $x_{12}^3=x_{22}^2=x_{32}^1=1$). 
Since $c_{(0)} := 0 < c_{(1)}=2<c_{(2)}=4$, we have 
 $u=\begin{pmatrix}
 	0 & 0 \\
 	0 & 0  \\
 	1 & 0 
 \end{pmatrix}$.
\end{example}

As reviewed in previous works (see \citealp{Puerto2011}, \citealp{Puerto2013} and \citealp{Pozo2021} among others) sorting costs can be rather difficult in large instances when using the $x^{\ell}$ variables previously defined. Besides, formulations using covering variables (see \citealp{Elloumi2004, Nickel2006, Puerto2008, Espejo2009, Marin2009, Marin2010, Puerto2011, Garcia2011}) can be used to downsize the number of variables in use. 

Recalling the general scheme proposed in the previous section, we can rewrite $F1_{x^\ell}^{\mT}$ as follows:

\begingroup
\allowdisplaybreaks
\begin{subequations}
\begin{align}
  & \hhp \hspace{1cm} F1^{\mT}_{u}: \min \quad \frac{1}{\displaystyle{\sum_{\ell \in V}} \lambda_\ell} \sum_{\ell \in V} \sum_{h \in H} \lambda_\ell u_{\ell h} (c_{(h)}-c_{(h-1)}) + \frac{1}{p-1} \sum_{(i,j) \in E} c_{ij} z_{ij}             & & & \label{OMT_f2_c0} \\
  & \hp \sum_{i \in V} x_{ii} = p                 &   &                                 & \label{OMT_f2_c1}  \\
  & \Hp \sum_{j \in V} x_{ij} = 1                 &   & i \in V                         & \label{OMT_f2_c2}  \\
  & \Hp x_{ij} \leq x_{jj}                        &   & (i,j) \in A: i \neq j           & \label{OMT_f2_c3}  \\
  & \Hp 2z_{ij} \leq x_{ii} + x_{jj}              &   & (i,j) \in E                     & \label{OMT_f2_c4}  \\
  & \Hp z_{ij} \in \mT                            &   & (i,j) \in E                     & \label{OMT_f2_c5}  \\
  & \Hp \sum_{\ell \in V} u_{\ell h} = \sum_{i,j \in V: c_{ij} \geq c_{(h)}} x_{ij}     &   & h \in H &   & \label{OMT_f2_c6}  \\
  & \Hp u_{\ell h} \leq u_{\ell+1 h}              &   & h \in H, \ell \in V: \ell < |V| & \label{OMT_f2_c7}  \\
  & \Hp u_{\ell h} \geq u_{\ell h+1}              &   & h \in H, \ell \in V: h < |H|    & \label{OMT_f2_c8}  \\
  & \Hp x_{ij} \in \{0,1\}                        &   & (i,j) \in A                     & \label{OMT_f2_c9}  \\
  & \Hp u_{\ell h} \in \{0,1\}                    &   & h \in H, \ell \in V             & \label{OMT_f2_c11} \\
  & \Hp z_{ij} \in \{0,1\}                        &   & (i,j) \in E.                    & \label{OMT_f2_c10} 
\end{align}
\end{subequations}
\endgroup

Constraints (\ref{OMT_f2_c6}) state that the number of allocations with a cost at least $c_{(h)}$ must be equal to the number of sites that support allocation costs greater than or equal to $c_{(h)}$ and constraints (\ref{OMT_f2_c7}) and (\ref{OMT_f2_c8}) sorts the values of the sorting variables $u$ in non-decreasing order. As indicated in \citealp{Labbe2017} and \citealp{Marin2009}, the extra sorting group of constraints (\ref{OMT_f2_c8}) is redundant, nevertheless, they are binding for some of the results on Section \ref{subsec:Theoreticalresults} and have been proved to strengthen the covering formulation.

Note that $F1^{\mT}_{u}$ is defined over a set $H$ whose cardinality depends on the unique non-zero elements in the cost matrix.
We recall that \OMT assumes free self-service and symmetrical costs, what implies a significant number of repetitions in the cost matrix and a diagonal of zeros. Te be more precise, let
$\hat{\alpha}$ be the number of ties within the part of the cost matrix above the diagonal and by $\hat{\chi_0}$ a binary parameter equal to 1 if there is at least a 0 within the part of the cost matrix, zero otherwise.
Thus,
\begin{itemize}
  \item $0\leq \hat{\alpha}\leq \frac{1}{2}(|V|^2-|V|)-1$.
  \item $\hat{\alpha} = \frac{1}{2}(|V|^2-|V|)-\hat{\chi_0}-|H|$.
\end{itemize}

Note that $\hat{\alpha}$ can also be easily computed as $\hat{\alpha} = \sum_{i=0}^{|H|} (m_i-1)$, where $m_i$ stands for the multiplicity of cost $c_{(h)}$ within the part of the cost matrix above the diagonal.

We provide in Table \ref{tab:theoreticalsizes} theoretical size values in terms of variables ($\#variables$) and constraints ($\#constraints$) of $F1_{x^\ell}^{\mT}$ and $F1^{\mT}_{u}$ \OMT formulations. Note that, the dimensions do not include the redundant set \eqref{OMT_f2_c8} for $F1^{\mT}_{u}$, and both formulations  are specified independently of the $\mT$ used but including \eqref{OMT_f1_c5} and \eqref{OMT_f2_c5} as constraints.

\begin{table}[h!]
\centering
\begin{tabular}{c|l|l}
\hline
                     & $\# variables$                                   & $\#constrains$             \\ \hline
                     &                                                  &                            \\ [-0.1cm] 
$F1_{x^\ell}^{\mT}$: & $|V|^3+\frac{3|V|^2}{2}-\frac{|V|}{2}$           & $3|V|^2 + |V|$             \\ [0.5cm] \hline
                     &                                                  &                            \\ 
$F1^{\mT}_{u}$:      & $|V||H|+\frac{3|V|^2}{2}-\frac{|V|}{2} =$        & $2|V|^2+|H||V|-|V|+1 =$    \\ [0.1cm]
                     &                                                  &                            \\
                     & $\frac{1}{2}|V|^3+|V|^2-(\hat{\chi_0}+\hat{\alpha}+\frac{1}{2})|V|$ & $\frac{1}{2}|V|^3+\frac{3}{2}|V|^2-(\hat{\chi_0}+\hat{\alpha}+1)|V|+1$ \\ 
                     &                                                  &                            \\ \hline
\end{tabular}
\caption{Theoretical model dimensions in the \OMT formulations}
\label{tab:theoreticalsizes}
\end{table}

We observe from Table \ref{tab:theoreticalsizes} the following property:
\begin{property}\label{prop1a}\hspace{0cm}\\
\begin{itemize}
  \item[(a)] $F1^{\mT}_{u}$ has $\frac{1}{2}|V|^3+\frac{1}{2}|V|^2+(\hat{\chi_0}+\hat{\alpha})|V|>0$ less variables than $F1_{x^\ell}^{\mT}$.
  \item[(b)] The difference of constraints between $F1^{\mT}_{u}$ and $F1_{x^\ell}^{\mT}$ is $\frac{1}{2}|V|^3-\frac{3}{2}|V|^2-(\hat{\chi_0}+\hat{\alpha}+2)|V|+1$. Therefore, $F1^{\mT}_{u}$ has less constraints than $F1_{x^\ell}^{\mT}$ if and only if 
  $$\hat{\alpha}\geq\frac{1}{2}|V|^2-\frac{3}{2}|V|-\hat{\chi_0}-1,$$
  that is, $|H| \leq |V|+1$.
\end{itemize}
\end{property}

\begin{proof}
(a) Is straightforward to prove.

(b) In terms of constraints, the difference between $F1_{x^\ell}^{\mT}$ and $F1^{\mT}_{u}$ is the sorting group of constraints, (\ref{OMT_f1_c6})-(\ref{OMT_f1_c8}) and (\ref{OMT_f2_c6})-(\ref{OMT_f2_c7}) respectively. 
For this group, there exists a total of $|V|^2+2|V|-1$ constraints in $F1_{x^\ell}^{\mT}$ and $|H||V|=\frac{1}{2}|V|^3-\frac{1}{2}|V|^2-(\hat{\chi_0}+\hat{\alpha})|V|$ constraints in $F1^{\mT}_{u}$.
Therefore, subtracting both expressions we get that the difference in the number of constraints is $\frac{1}{2}|V|^3-\frac{3}{2}|V|^2-(\hat{\chi_0}+\hat{\alpha}+2)|V|+1$, that is negative only if we have $\hat{\alpha} > \frac{1}{2}|V|^2- \frac{3}{2}|V|-\hat{\chi_0}-2+\frac{1}{|V|}$ (that is $\hat{\alpha}\geq\frac{1}{2}|V|^2-\frac{3}{2}|V|-\hat{\chi_0}-1$) and $|H|<|V|+2-\frac{1}{|V|}$ (that is $|H|\leq|V|+1$).

\end{proof}

Formulation of the ordered median using variables $x^{\ell}$ is the general representation used to describe the problem, while formulation using covering variables $u$ performs best computationally. There exist several more formulations to represent the ordered median problem (see, e.g., \citealp{Labbe2017}). However, the analysis of such other formulations embedded within the \OMT structure is beyond the scope of this paper.

\subsection{Theoretical results on polytopes}
\label{subsec:Theoreticalresults}

In this section, we provide a theoretical comparison of the polytopes regarding our formulations $F1^{\mT}_{x^{\ell}}$ and $F1^{\mT}_{u}$ presented in previous Section \ref{sec:OMTformulations}. Comparing \OMT formulations is equivalent to comparing DOMP formulations because both problems only differ in the connection feature, which is modeled by the same group of constraints for every formulation (see Section \ref{subsec:Tdescription} for more details). Therefore, some of the results we adapt to our \OMT polytopes comparison are presented in \citet{Labbe2017}. Nevertheless, our goal is to state the formal relationships between the linear relaxations of the considered formulations and study the similarities between these polytopes describing in detail the degenerate case of no ties in the cost matrix. Also, note that in this section the \OMT assumptions have been relaxed in an attempt to generalize these results.

\subsubsection{Analyzing polytopes with a general cost matrix}
\label{subsubsec:Theoreticalresultsgeneral}

The results presented in this subsection are valid for a general cost matrix independently of the number of ties. The reader may note that \OMT does not really require free self-service and symmetric cost assumptions. Without these two assumptions our formulation is still valid although it slightly changes the meaning of costs $c_{ii}$ that would turn to be the opening cost of facility $i$ plus de allocation cost of client $i$ to facility $i$ (if a facility is opened $i$, client $i$ has to be allocated to such facility).
Let $\Omega_{x^\ell}^{\mT}$ be the set of points satisfying constraints (\ref{OMT_f1_c1})-(\ref{OMT_f1_c10}) and $\Omega_{u}^{\mT}$ be the set of points satisfying constraints (\ref{OMT_f2_c1})-(\ref{OMT_f2_c11}). Let also $\phi_{x^\ell}(\cdot)$ and $\phi_{u}(\cdot)$ denote the value of the objective functions, (\ref{OMT_f1_c0}) and (\ref{OMT_f2_c0}) respectively, evaluated at a feasible point $(\cdot)$ and $P(\Omega_{\mS}^{\mT})$ for $\mS \in \{x^\ell,u\}$ the polytope defined from the linear relaxation of $\Omega_{\mS}^{\mT}$. For the sake of understandability, we remove the tree notation $\mT$ in this section which is not relevant. A first property to recall is the following.

\begin{property}
\label{prop2}
There exist two mappings $f$ and $g$, 

\begin{align*}
f: \mathbb{R}^{|V|^2} \times \mathbb{R}^{|V|^2} \times \mathbb{R}^{|V|^{3}} & \longmapsto \mathbb{R}^{|V|^{2}} \times \mathbb{R}^{|V|^2} \times\mathbb{R}^{|V||H|}, \\ (x_{ij}, z_{ij}, x_{ij}^\ell) & \longmapsto (x_{ij}, z_{ij}, u_{\ell h}). \\
g: \mathbb{R}^{|V|^{2}} \times \mathbb{R}^{|V|^2} \times\mathbb{R}^{|V||H|} & \longmapsto \mathbb{R}^{|V|^2} \times \mathbb{R}^{|V|^2} \times \mathbb{R}^{|V|^{3}}, \\
(x_{ij}, z_{ij}, u_{\ell h})  & \longmapsto (x_{ij}, z_{ij}, x_{ij}^\ell).
\end{align*}

such that,

\begin{itemize}
  \item For any point $(x_{ij}, z_{ij}, x_{ij}^\ell) \in P(\Omega_{x^\ell})$, then $\phi_{x^\ell}(x_{ij}, z_{ij}, x_{ij}^\ell) = \phi_{u}(f(x_{ij}, z_{ij}, x_{ij}^\ell))$.
  \item For any point $(x_{ij}, z_{ij},  u_{\ell h}) \in P(\Omega_{u})$, then $\phi_{u}(x_{ij}, z_{ij},  u_{\ell h}) = \phi_{x^{\ell}}(g(x_{ij}, z_{ij},  u_{\ell h}))$.
\end{itemize}
\end{property}

These mapping definitions are given in \citet{Labbe2017}. Via these mappings, several polytope comparisons can be attained.

\begin{property} \*
\label{prop3}
\begin{enumerate}[(a)] 
  \item $g(P(\Omega_{u})) \subset P(\Omega_{x^{\ell}})$.
  \item $f(P(\Omega_{x^{\ell}})) \not\subset P(\Omega_{u})$.
\end{enumerate}
\end{property}

\begin{proof}
\begin{enumerate}[(a)] 
  \item See \textit{Theorem 1} in \citet{Labbe2017}.
  \item Consider the network example depicted in Figure \ref{fig:NoExamp3}, where client nodes (\textit{circles}) allocate to a single facility (\textit{square}). 
  
  \begin{figure}[ht]
\centering
\tikzset{every picture/.style={line width=0.75pt}}
\begin{tikzpicture}[x=0.75pt,y=0.75pt,yscale=-1,xscale=1]
\draw  [color={rgb, 255:red, 58; green, 246; blue, 25 }  ,draw opacity=1 ][fill={rgb, 255:red, 225; green, 250; blue, 210 }  ,fill opacity=1 ][line width=1.5]  (261.75,164.88) .. controls (261.75,157.63) and (267.63,151.75) .. (274.88,151.75) .. controls (282.12,151.75) and (288,157.63) .. (288,164.88) .. controls (288,172.12) and (282.12,178) .. (274.88,178) .. controls (267.63,178) and (261.75,172.12) .. (261.75,164.88) -- cycle ;
\draw  [color={rgb, 255:red, 59; green, 255; blue, 70 }  ,draw opacity=1 ][fill={rgb, 255:red, 234; green, 254; blue, 213 }  ,fill opacity=1 ][line width=1.5]  (374,165) .. controls (374,158.1) and (379.6,152.5) .. (386.5,152.5) .. controls (393.4,152.5) and (399,158.1) .. (399,165) .. controls (399,171.9) and (393.4,177.5) .. (386.5,177.5) .. controls (379.6,177.5) and (374,171.9) .. (374,165) -- cycle ;
\draw  [color={rgb, 255:red, 208; green, 2; blue, 27 }  ,draw opacity=1 ][fill={rgb, 255:red, 255; green, 225; blue, 215 }  ,fill opacity=1 ][line width=1.5]  (318,58) -- (344,58) -- (344,82) -- (318,82) -- cycle ;
\draw    (278,150) -- (316.99,83.72) ;
\draw [shift={(318,82)}, rotate = 120.47] [color={rgb, 255:red, 0; green, 0; blue, 0 }  ][line width=0.75]    (6.56,-1.97) .. controls (4.17,-0.84) and (1.99,-0.18) .. (0,0) .. controls (1.99,0.18) and (4.17,0.84) .. (6.56,1.97)  ;
\draw    (288,166) -- (372,165.02) ;
\draw [shift={(374,165)}, rotate = 179.33] [color={rgb, 255:red, 0; green, 0; blue, 0 }  ][line width=0.75]    (6.56,-1.97) .. controls (4.17,-0.84) and (1.99,-0.18) .. (0,0) .. controls (1.99,0.18) and (4.17,0.84) .. (6.56,1.97) ;
\draw    (373,160) -- (289,160) ;
\draw [shift={(287,160)}, rotate = 360] [color={rgb, 255:red, 0; green, 0; blue, 0 }  ][line width=0.75]    (6.56,-1.97) .. controls (4.17,-0.84) and (1.99,-0.18) .. (0,0) .. controls (1.99,0.18) and (4.17,0.84) .. (6.56,1.97);
\draw    (324,84) -- (284.03,150.29) ;
\draw [shift={(283,152)}, rotate = 301.09] [color={rgb, 255:red, 0; green, 0; blue, 0 }  ][line width=0.75]    (6.56,-1.97) .. controls (4.17,-0.84) and (1.99,-0.18) .. (0,0) .. controls (1.99,0.18) and (4.17,0.84) .. (6.56,1.97);
\draw    (339,84) -- (378.01,152.26) ;
\draw [shift={(379,154)}, rotate = 240.26] [color={rgb, 255:red, 0; green, 0; blue, 0 }  ][line width=0.75]    (6.56,-1.97) .. controls (4.17,-0.84) and (1.99,-0.18) .. (0,0) .. controls (1.99,0.18) and (4.17,0.84) .. (6.56,1.97);
\draw    (382,149) -- (344.99,83.74) ;
\draw [shift={(344,82)}, rotate = 60.44] [color={rgb, 255:red, 0; green, 0; blue, 0 }  ][line width=0.75]    (6.56,-1.97) .. controls (4.17,-0.84) and (1.99,-0.18) .. (0,0) .. controls (1.99,0.18) and (4.17,0.84) .. (6.56,1.97);
\draw    (344,58) .. controls (346.93,31.67) and (316.58,31.97) .. (317.84,56.1) ;
\draw [shift={(318,58)}, rotate = 263.42] [color={rgb, 255:red, 0; green, 0; blue, 0 }  ][line width=0.75]    (6.56,-1.97) .. controls (4.17,-0.84) and (1.99,-0.18) .. (0,0) .. controls (1.99,0.18) and (4.17,0.84) .. (6.56,1.97);
\draw    (399,165) .. controls (414.6,179.63) and (409.29,188.55) .. (388.16,178.32) ;
\draw [shift={(386.5,177.5)}, rotate = 27.07] [color={rgb, 255:red, 0; green, 0; blue, 0 }  ][line width=0.75]    (6.56,-1.97) .. controls (4.17,-0.84) and (1.99,-0.18) .. (0,0) .. controls (1.99,0.18) and (4.17,0.84) .. (6.56,1.97);
\draw    (261.75,164.88) .. controls (244.69,175.11) and (250.68,188.32) .. (273.12,178.78) ;
\draw [shift={(274.88,178)}, rotate = 155.26] [color={rgb, 255:red, 0; green, 0; blue, 0 }  ][line width=0.75]    (6.56,-1.97) .. controls (4.17,-0.84) and (1.99,-0.18) .. (0,0) .. controls (1.99,0.18) and (4.17,0.84) .. (6.56,1.97);
\draw (326,21) node [anchor=north west][inner sep=0.75pt]   [align=left] {0};
\draw (367,97) node [anchor=north west][inner sep=0.75pt]   [align=left] {9};
\draw (330,142) node [anchor=north west][inner sep=0.75pt]   [align=left] {6};
\draw (330,169) node [anchor=north west][inner sep=0.75pt]   [align=left] {5};
\draw (305,114) node [anchor=north west][inner sep=0.75pt]   [align=left] {4};
\draw (285,99) node [anchor=north west][inner sep=0.75pt]   [align=left] {3};
\draw (407,186) node [anchor=north west][inner sep=0.75pt]   [align=left] {2};
\draw (240,185) node [anchor=north west][inner sep=0.75pt]   [align=left] {1};
\draw (342,114) node [anchor=north west][inner sep=0.75pt]   [align=left] {8};
\draw (326,61) node [anchor=north west][inner sep=0.75pt]   [align=left] {1};
\draw (269,156) node [anchor=north west][inner sep=0.75pt]   [align=left] {2};
\draw (381,157) node [anchor=north west][inner sep=0.75pt]   [align=left] {3};
\end{tikzpicture}
\caption{Network example.}
\label{fig:NoExamp3}
\end{figure}
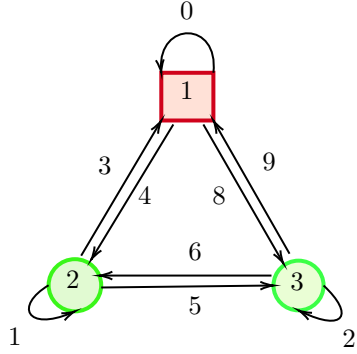

Let the following feasible solution (non-optimal) defined by the values:

\vspace{-0.25cm}

\begin{align*}
  x_{11} & = 1, & x_{12} & = 0, & x_{13} & = 0, \\
  x_{21} & = 1, & x_{22} & = 0, & x_{23} & = 0, \\
  x_{31} & = 1, & x_{32} & = 0, & x_{33} & = 0,
\end{align*}

\vspace{-0.75cm}

\begin{align*}
  z_{12} & = 0, & z_{13} & = 0, & z_{23} & = 0,
\end{align*}

\vspace{-0.75cm}

\begin{align*}
  x^{1}_{11} & = 0.9, & x^{2}_{11} & = 0, & x^{3}_{11} & = 0.1, \\
  x^{1}_{21} & = 0,   & x^{2}_{21} & = 1, & x^{3}_{21} & = 0,   \\
  x^{1}_{31} & = 0.1, & x^{2}_{31} & = 0, & x^{3}_{31} & = 0.9,
\end{align*}

\vspace{-0.75cm}

$$x^{\ell}_{ij} = 0 \ \text{otherwise}.$$

We can verify that the previous solution satisfies the group of constraints defining $P(\Omega_{x^\ell})$ (shown only for sorting constraints):

\begin{itemize}
  \item $\sum_{\ell \in V} x_{ij}^\ell = x_{ij}, \quad (i,j) \in A.$
    \begin{align*}
      x^1_{11} + x^2_{11} + x^3_{11} & = 1 = x_{11}, \\  
      x^1_{21} + x^2_{21} + x^3_{21} & = 1 = x_{21}, \\
      x^1_{31} + x^2_{31} + x^3_{31} & = 1 = x_{31},
    \end{align*}
    $$\text{Otherwise trivially equal to 0.}$$
  \item $\sum_{(i,j) \in A} x_{ij}^\ell = 1, \quad \ell \in V.$
    \begin{align*}
      x^1_{11} + x^1_{21} + x^1_{31} + 0 \ldots & = 1 \\  
      x^2_{11} + x^2_{21} + x^2_{31} + 0 \ldots & = 1 \\  
      x^3_{11} + x^3_{21} + x^3_{31} + 0 \ldots & = 1 \\  
    \end{align*}
  \item $\sum_{(i,j) \in A} c_{ij}x_{ij}^\ell \leq \sum_{(i,j) \in A} c_{ij}x_{ij}^{\ell+1}, \quad \ell \in V: \ell < |V.$
    \begin{align*}
       & c_{11}x^1_{11} + c_{31}x^1_{31} + 0 \ldots = 0.9 \cdot 0 + 0.1 \cdot 9 = 0.9 \leq c_{21}x^2_{21} + 0 \ldots = 1 \cdot 3 = 3 \\  
       & c_{21}x^2_{21} + 0 \ldots = 1 \cdot 3 = 3 \leq c_{11}x^3_{11} + c_{31}x^3_{31} + 0 \ldots = 0.9 \cdot 9 + 0.1 \cdot 0 = 8.1 
    \end{align*}
\end{itemize}

We can give an expression for $u$ variables using $f$ definition, which is given by the system of equations:

\vspace{-0.25cm}

\begin{align}
f: (x_{ij}, z_{ij}, x_{ij}^\ell) & \longmapsto (x_{ij}, z_{ij}, u_{\ell h}), \notag \\ 
u_{\ell h} & = \sum_{i,j \in V: c_{ij} \geq c_{(h)}} x^{\ell}_{ij} \quad \forall h \in H, \ell \in V. \label{fnoties}
\end{align}

\vspace{-0.25cm}

Therefore,

\vspace{-0.25cm}


\begin{align*}
  & u_{11} = 0.1, & & \cdots & &   \cdots    & & \cdots      & & \cdots & & u_{18} = 0.1, \\
  & u_{21} = 1    & & \cdots & & u_{23} = 1, & & u_{24} = 0, & & \cdots & & u_{28} = 0,   \\
  & u_{31} = 0.9, & & \cdots & &   \cdots    & & \cdots      & & \cdots & & u_{38} = 0.9,
\end{align*}

which does not satisfy constraints $u_{\ell h} \leq u_{\ell+1 h}$ for $h \in H, \ell \in V: \ell < |V|$ in the relaxed polytope $P(\Omega_{u})$.
\end{enumerate}

\end{proof}

As a consequence, we have found a feasible solution in $P(\Omega_{x^{\ell}})$ whose image by $f$ is not feasible in $P(\Omega_{u})$. Next, we study the objective function value of the points $(x_{ij}, z_{ij}, x^{\ell}_{ij}) \in P(\Omega_{x^{\ell}})$ such that $f(x_{ij}, z_{ij}, x^{\ell}_{ij}) \notin P(\Omega_{u})$. First, let us consider the following polytope namely $P(\Omega'_{x^{\ell}})$, where constraints (\ref{OMT_f1_c8}) are replaced by constraints

\begin{equation}
  \label{21consequence}
  \sum_{i,j \in V: c_{ij} \geq c_{(h)}} x^{\ell}_{ij} + \sum_{i,j \in V: c_{ij} < c_{(h)}} x^{\ell + 1}_{ij} \leq 1, \quad h \in H, \ell \in V: \ell < |V|,
\end{equation}

which are a consequence of the \textit{staircase inequalities} in \citet{Labbe2017}. Observe that,

\begin{property} \*
\label{prop4}
\begin{enumerate}[(a)] 
  \item $f(P(\Omega'_{x^{\ell}})) = P(\Omega_{u})$.
  \item $g(P(\Omega_{u})) \subset P(\Omega'_{x^{\ell}})$.
\end{enumerate}
\end{property}

\begin{proof}
\begin{enumerate}[(a)] 

 \item Following the results in \citet{Labbe2017}, $g$ satisfies the equality of $f$ by construction, thus $f(g(u)) = u$. This implies $f(P(\Omega'_{x^{\ell}})) \supset P(\Omega_{u})$ and we only need to prove then $f(P(\Omega'_{x^{\ell}})) \subset P(\Omega_{u})$. Given $(x_{ij}, z_{ij}, x^{\ell}_{ij}) \in P(\Omega'_{x^{\ell}})$ we prove that $f(x_{ij}, z_{ij}, x^{\ell}_{ij})$ satisfies constraints (\ref{OMT_f2_c6})-(\ref{OMT_f2_c8}). Constraints (\ref{OMT_f2_c8}) are satisfied by means of $f$ definition, 

\begin{equation} \notag
  u_{\ell h} = \sum_{i,j \in V: c_{ij} \geq c_{(h)}} x_{ij}^{\ell} \geq  \sum_{i,j \in V: c_{ij} \geq c_{(h+1)}} x_{ij}^{\ell} = u_{\ell h+1},
\end{equation}

constraints (\ref{OMT_f2_c6}) are satisfied because of the following 
 
\begin{equation} \notag
  \sum_{\ell \in V} u_{\ell h} = \sum_{\ell \in V} \sum_{i,j \in V: c_{ij} \geq c_{(h)}} x_{ij}^{\ell} =  \sum_{i,j \in V: c_{ij} \geq c_{(h)}} \sum_{\ell \in V} x_{ij}^{\ell} \stackrel{(\ref{OMT_f1_c6})}{=} \sum_{i,j \in V: c_{ij} \geq c_{(h)}} x_{ij},
\end{equation}

and constraints (\ref{OMT_f2_c7}) because

\vspace{-0.25cm}

\begin{equation} \notag
\begin{aligned}
  u_{\ell h} 
  & = \sum_{i,j \in V: c_{ij} \geq c_{(h)}} x_{ij}^{\ell} \stackrel{(\ref{21consequence})}{\leq} 1 - \sum_{i,j \in V: c_{ij} < c_{(h)}} x_{ij}^{\ell+1} \stackrel{(\ref{OMT_f1_c7})}{=} \sum_{(i,j) \in A} x_{ij}^{\ell+1} - \sum_{i,j \in V: c_{ij} < c_{(h)}} x_{ij}^{\ell+1} = \\ 
  & = \sum_{i,j \in V: c_{ij} < c_{(h)}} x_{ij}^{\ell+1} + \sum_{i,j \in V: c_{ij} \geq c_{(h)}} x_{ij}^{\ell+1} - \sum_{i,j \in V: c_{ij} < c_{(h)}} x_{ij}^{\ell+1} = \sum_{i,j \in V: c_{ij} \geq c_{(h)}} x_{ij}^{\ell+1} = u_{\ell+1, h}.
\end{aligned}
\end{equation}

  \item By \textit{Property \ref{prop3}}, we know $g(P(\Omega_{u})) \subset P(\Omega_{x^{\ell}})$. Hence, we just need to prove that constraints (\ref{21consequence}) are satisfied. By constraints (\ref{OMT_f2_c7}) we get
  
\vspace{-0.25cm}

\begin{equation} \notag
\begin{aligned}
  u_{\ell h} \leq u_{\ell+1 h}
  & \implies \sum_{i,j \in V: c_{ij} \geq c_{(h)}} x_{ij}^{\ell} \leq \sum_{i,j \in V: c_{ij} \geq c_{(h)}} x_{ij}^{\ell+1} \implies \sum_{i,j \in V: c_{ij} < c_{(h)}} x_{ij}^{\ell+1} + \sum_{i,j \in V: c_{ij} \geq c_{(h)}} x_{ij}^{\ell} \leq 1.
\end{aligned}
\end{equation}  
  
\end{enumerate}

\end{proof}

Consequently, the following result can be given as a conclusion.

\begin{property}\label{property_phi}
  $\phi_{u} = \phi'_{x^{\ell}} \geq \phi_{x^{\ell}}.$
\end{property}

\begin{proof}
  This property is a consequence of \textit{Property \ref{prop2}} and $P(\Omega_{u}) = f(P(\Omega'_{x^{\ell}})) \subset f(P(\Omega_{x^{\ell}}))$.
\end{proof}
As an observation to Property \ref{property_phi}, our computational experience shows the equality $\phi_{u} = \phi'_{x^{\ell}} = \phi_{x^{\ell}}$ holds in the optimum.

\subsubsection{Analyzing polytopes with no ties in the cost matrix}
\label{subsubsec:Theoreticalresultsnoties}

We analyze next the particular case when no ties exists in the cost matrix. We observe that in such case, the previous definition of $f$ holds while $g$ (see \citealp{Labbe2017}) transforms into a much simpler expression:

\begin{align}
g: \mathbb{R}^{|V|^{2}} \times \mathbb{R}^{|V|^2} \times\mathbb{R}^{|V|^3} & \longmapsto \mathbb{R}^{|V|^2} \times \mathbb{R}^{|V|^2} \times \mathbb{R}^{|V|^{3}},  \notag \\
(x_{ij}, z_{ij}, u_{\ell h})  & \longmapsto (x_{ij}, z_{ij}, x_{ij}^\ell),  \notag \\
\medmath{x_{i j}^{\ell}} & = \medmath{\left\{\begin{array}{ll}
0 & \text { if } \ell<\ell(h) \text { or } x_{ij}=0, \\
\min \left\{x_{ij}-\sum_{k<\ell} x_{ij}^{k}, u_{\ell h}-u_{\ell, h+1}\right\} & \text { if } \ell \geq \ell(h), 
\end{array}\right. } \label{gnoties} 
\end{align}

where $\ell(h)=\min \left\{\ell: u_{\ell h}-u_{\ell, h+1}>0\right\}$ for $h=1, \ldots, H$ and $u_{\ell, H+1}=u_{\ell H}$. We assume that if $u_{\ell h}-u_{\ell, h+1}=0$ for all $\ell \in V$ then $\ell(h)=+\infty$.

Furthermore, when there are no ties in the cost matrix, we can give an expression for $f^{-1}$ using equations in (\ref{fnoties}). Because no ties exist in the cost matrix, we have that each $c_{(h)}$ is uniquely related to a $c_{ij}$, that is, there exists a bijective mapping $\beta$ between the sequence of costs defined as

\begin{align*}
\beta: H        & \longmapsto \{ij \ | \ i,j \in V\}, \\ 
       c_{(h)}  & \longmapsto c_{\beta(h)} = c_{i_h j_h}.
\end{align*}

For every $\ell \in V$, system (\ref{fnoties}) can be expressed in matrix form as follows. Let $\underline{\textbf{u}}_{\ell} = (u_{\ell 1}, \ldots, u_{\ell |V|^2})^{'}$ and $\underline{\textbf{x}}^{\ell} = (x^{\ell}_{11}, \ldots, x^{\ell}_{1|V|}, \ldots, x^{\ell}_{|V|1}, \ldots, x^{\ell}_{|V||V|})^{'}$, we have

\begin{equation}
\label{prop3:eq2}
  \underline{\textbf{u}}_{\ell} = \Sigma_{\ell} \underline{\textbf{x}}^{\ell} \quad \forall \ell \in V,
\end{equation}

where $\Sigma_{\ell}$ is a $|V|^2 \times |V|^2$ matrix. The first row in $\Sigma_{\ell}$ (corresponding to $h=1$) has all coefficients equal to 1 since every cost in the sum is greater than $c_{(1)}$, the second row ($h=2$) has all coefficients equal to 1 except for the coefficient corresponding to some position $i_1 j_1$ which is 0, and so until the last row ($h = |V|^2$), that has all coefficients equal to 0 except for the coefficient corresponding to some position $i_{|V|^2}j_{|V|^2}$ which is 1. Therefore, $\Sigma_{\ell}$ has the following structure

\begin{equation}
\Sigma_{\ell} = \,
\bordermatrix{     
              & \cdots &   & i_1j_1 &   & \cdots &   & i_2j_2 &   & \cdots &   & i_3j_3 &   & \cdots &   & i_{|V|^2}j_{|V|^2} &   & \cdots \cr
    h=1       & \cdots & 1 & 1      & 1 & \cdots & 1 & 1      & 1 & \cdots & 1 & 1      & 1 & \cdots & 1 & 1                  & 1 & \cdots \cr
    h=2       & \cdots & 1 & 0      & 1 & \cdots & 1 & 1      & 1 & \cdots & 1 & 1      & 1 & \cdots & 1 & 1                  & 1 & \cdots \cr
              & \cdots & 1 & 0      & 1 & \cdots & 1 & 0      & 1 & \cdots & 1 & 1      & 1 & \cdots & 1 & 1                  & 1 & \cdots \cr
    \vdots    & \cdots & 1 & 0      & 1 & \cdots & 1 & 0      & 1 & \cdots & 1 & 0      & 1 & \cdots & 1 & 1                  & 1 & \cdots \cr
              &        &   & \vdots &   &        &   & \vdots &   & \ddots &   & \vdots &   &        &   & \vdots             &   & \ddots \cr
    h = |V|^2 & \cdots & 0 & 0      & 0 & \cdots & 0 & 0      & 0 & \cdots & 0 & 0      & 0 & \cdots & 0 & 1                  & 0 & \cdots \cr
}.
\end{equation}

Consequently, $\Sigma_{\ell}$ can be rearranged through elemental matrix transformations to an upper triangular matrix $\Sigma^{upper}_{\ell}$, so we can rewrite system (\ref{prop3:eq2}) as

\begin{equation}
\underline{\textbf{u}}_{\ell} = \Sigma^{upper}_{\ell} 
\begin{pmatrix}
x_{i_1j_1}^{\ell} \\ 
x_{i_2j_2}^{\ell} \\ 
\vdots \\ 
x_{i_{|V|^2}j_{|V|^2}}^{\ell} 
\end{pmatrix}
= \begin{pmatrix}
1 & 1 & 1 & \cdots & 1 & 1 \\ 
0 & 1 & 1 & \cdots & 1 & 1 \\ 
0 & 0 & 1 & \cdots & 1 & 1 \\ 
  &   &   & \vdots &   &   \\ 
0 & 0 & 0 & \cdots & 0 & 1
\end{pmatrix} 
\begin{pmatrix}
x_{i_1j_1}^{\ell} \\ 
x_{i_2j_2}^{\ell} \\ 
\vdots \\ 
x_{i_{|V|^2}j_{|V|^2}}^{\ell} 
\end{pmatrix}.
\end{equation}

Therefore, $\Sigma_{\ell}$ has maximum range which implies that $f$ is an injection and (\ref{prop3:eq2}) defines for every $\ell \in V$ a single unique solution for $x_{ij}^{\ell}$ variables for $i,j \in V$. A general expression for the inverse of $\Sigma^{upper}_{\ell}$ can be given to obtain the solution for $x_{ij}^{\ell}$ variables as follows

\begin{equation}
\underline{\textbf{x}}^{\ell} = \begin{pmatrix}
1 & -1 &  0 & \cdots & 0 &  0 \\ 
0 &  1 & -1 & \cdots & 0 &  0 \\ 
0 &  0 &  1 & \cdots & 0 &  0 \\ 
  &    &    & \vdots &   &    \\ 
0 &  0 &  0 & \cdots & 1 & -1 \\
0 &  0 &  0 & \cdots & 0 &  1
\end{pmatrix} 
\begin{pmatrix}
u_{\ell 1} \\ 
u_{\ell 2} \\ 
\vdots \\ 
u_{\ell |V|^2}
\end{pmatrix},
\end{equation}

being the mapping $f^{-1}$ defined by the following relationship

\begin{equation}
\label{eq:rel_xlu}
x_{i_h j_h}^{\ell} = \left\{\begin{matrix}
u_{\ell h} - u_{\ell h+1} & h = 1,\ldots, |V|^2-1, \\ 
u_{\ell |V|^2}            & h = |V|^2.
\end{matrix}\right.
\end{equation}

Notice that constraints (\ref{OMT_f2_c8}) are needed in order for $f^{-1}$ to be well defined. It can be verified that, in the case of no ties, $f^{-1}$ and $g$ coincide.

\begin{property}
If no ties exists in the cost matrix $(c_{ij})_{i,j \in V}$, then $g=f^{-1}$.
\end{property}

\begin{proof}

If $\ell \geq \ell(h)$, then $x_{ij} - \sum_{k<\ell} x_{ij}^{k} = \sum_{k \in V} x_{ij}^k - \sum_{k<\ell} x_{ij}^{k} = \sum_{k\geq\ell} x_{ij}^{k}  \geq x_{i_h j_h}^{\ell} = u_{\ell h} - u_{\ell h+1} \implies \min \left\{x_{ij}-\sum_{k<\ell} x_{ij}^{k}, u_{\ell h}-u_{\ell, h+1}\right\} = u_{\ell h}-u_{\ell, h+1}$. Otherwise, if $\ell<\ell(h)$ or $x_{ij}=0$, trivially equal to 0.
\end{proof}

As a consequence, we get that $g(P(\Omega'_{u})) = P(\Omega'_{x^{\ell}})$ when no ties exists in the cost matrix.

\subsection{Tree polytope description}
\label{subsec:Tdescription}

This section extends the definition of the tree polytope $\mT$ including some specific considerations that must be reviewed for the application of each tree characterization within the \OMT.

Formulation $F1_{x^\ell}^{\mT}$ and $F1^{\mT}_{u}$ assume constraint $z \in \mT$ to be included in the model. This constraint can be replaced by any representation of the STP polytope, namely \textit{subtour elimination}, \textit{Miller-Tucker-Zemlim} (MTZ), \textit{flow}, etc. (see \citealp{Magnanti1995}, \citealp{Fernandez2017}). Depending on the case, different variables must be considered. The formulation in \citet{Miller1960} uses variables $y_{ij} \in \{0,1\}$ for $(i,j) \in A$, which take value 1 if and only if arc $(i,j)$ belongs to the modeled arborescence and continuous variables $l_i \geq 0$ for $i \in V$, denoting the position that node $i$ occupies in the arborescence respect to the root node. In the \textit{flow based} formulation introduced by \citet{Gavish1983}, continuous flow variables $f_{ij} \geq 0$ for $(i,j) \in A$ are defined on the arcs of the directed network indicating the amount of flow through arc $(i,j)$. The formulation proposed by \citet{Martin1991} models as many arborescences as network nodes using three index decision variables $q_{kij} \in \{0,1\}$ for $(i,j) \in A$, which indicate whether or not arc $(i,j)$ belongs to the arborescence rooted at $k$. In addition, each tree polytope $\mT$ includes constraint $\sum_{(i,j) \in E} z_{ij} = p-1$, which ensures that the tree of facilities has exactly $p-1$ edges.

Some notation must be introduced in order to model $\mT$. Given a subset of nodes $S \subset V$, let us denote by $E(S)$ and $A(S)$ the subsets of edges of $E$ and arcs of $A$ connecting nodes in $S$, i.e., $E(S) = \{(i,j) \in E \ | \ i,j \in S\}$ and $A(S) = \{(i,j) \in A \ | \ i,j \in S\}$. The \textit{cut-set} $\delta(S)$ associated with $S \subset V$ contains all edges with one node in $S$ and the other node outside $S$, i.e. $\delta(S) = \{ (i,j) \in E \ | \ (i \in S,j \in V\setminus S) \ or \ (j \in S,i \in V\setminus S)\}$. Equivalently, in the directed network, the \textit{cut-set directed out} of $S$ is defined as $\delta^+(S) = \{(i,j) \in A \ | \ i \in S,j \in V\setminus S \}$ and the \textit{cut-set directed into} of $S$ is defined as $\delta^-(S) = \{(i,j) \in A \ | \ j \in S,i \in V\setminus S \}$.

In Table \ref{table:formulations_MST} we resume the main properties of the STP formulations that we have considered. The criteria that have guided the selection of the formulations are either their good theoretical properties or some characteristic that seems useful as, for instance, a small number of variables or constraints. It is well known that, as seen in the last column, from the formulations described above $\mT^{sub}$ and $\mT^{km}$ hold the integrality porperty, conversely to $\mT^{mtz}$ and $\mT^{flow}$.

\vspace{0.25cm}

\begin{table}[h!]
\centering
\renewcommand{\arraystretch}{1}
\begin{center}
\begin{tabular}{ c c c  c  c c c }
\hline
Formulation  & notation & main constraints & root & \# vars & \# const. & int \\
\hline
  \small{\begin{tabular}[c]{@{}c@{}} \textbf{Subtour} \\ \citet{Edmonds1970} \end{tabular}} &
  $\mT^{sub}$ &
  \footnotesize $\displaystyle{\sum_{(i,j) \in E(S)} z_{ij} \leq |S|-1, \ S \neq \emptyset, S \subset V}$ &
   & 
  \footnotesize{$O(|E|)$} &
  \footnotesize{$Exp(|V|)$} &
  \small{Yes} \\ \hline
  \small{\begin{tabular}[c]{@{}c@{}} \textbf{Miller-Tucker-Zemlim} \\ \citet{Miller1960} \end{tabular}} &
  $\mT^{mtz}$ &
  \small $l_j \geq l_i + 1 - |V|(1-y_{ij}), \  (i,j) \in A$ &
  \small{$r$} &
  \footnotesize{$O(|E|)$} &
  \footnotesize{$O(|E|)$} &
  \small{No} \\ \hline
  \small{\begin{tabular}[c]{@{}c@{}} \textbf{Flow} \\ \citet{Gavish1983} \end{tabular}} &
  $\mT^{flow}$ &
  \scriptsize{$\displaystyle{\sum_{(i,j) \in \delta^+(i)} f_{ij} - \sum_{(j,i) \in \delta^-(i)} f_{ji}= \left\{\begin{matrix} p-1, & i=r \\ -1,  & i \in V \setminus \{r\} \end{matrix}\right.}$} & 
  \small{$r$} &
  \footnotesize{$O(|E|)$} &
  \footnotesize{$O(|E|)$} &
  \small{No} \\ \hline
  \small{\begin{tabular}[c]{@{}c@{}} \textbf{Kipp Martin}   \\ \citet{Martin1991} \end{tabular}} &
  $\mT^{km}$ &
  \footnotesize{$\displaystyle{\sum_{(i,j)\in \delta^+(i)}q_{kij} \leq \left\{\begin{matrix} 1, & \ k\in V, i\in V:i\neq k \\ 0, & \ k\in V, i = k \end{matrix}\right.}$} & 
  \small{$\forall k$} & 
  \footnotesize{$O(|V||E|)$} & 
  \footnotesize{$O(|V||E|)$} & 
  \small{Yes} \\ \hline
\end{tabular}
\caption{Main properties of the STP formulations considered.}
\label{table:formulations_MST}
\end{center}
\end{table}

As a consequence of the formulations comparison in \citet{Fernandez2017}, we can state the relationship between the different formulations derived from the combination of some sorting representations and tree polyhedra described above. To this end, let $P_{xz}(\Omega_{\mS}^{\mT})$ be the projected polytope onto the space of $x$ and $z$ of the linear relaxation of an \OMT formulation. Then, the following property holds:

\begin{property}
\begin{align}
  & P_{xz}(\Omega_{x^\ell}^{\mT}) =  P_{xz}(\Omega_u^{\mT}). \\
  & P_{xz}(\Omega_{\mS}^{sub})    =  P_{xz}(\Omega_{\mS}^{km}) \subseteq \left\{\begin{matrix} P_{xz}(\Omega_{\mS}^{flow}) \\ \neq \\ P_{xz}(\Omega_{\mS}^{mtz}). \end{matrix}\right.
\end{align}
\end{property}

\begin{proof}
Is derived from the observations in \citet{Fernandez2017}.
\end{proof}

Regarding to $\mT^{sub}$, note that the cardinality of the group of subtour elimination constraints is exponential in the number of nodes. Since many of these constraints are not needed to build the solution polytope, they can be separated in polynomial time by solving a series of minimum cut-problems. An effective branch-and-cut algorithm can be implemented to include dynamically a certain number of these constraints, preventing from using the entire set of subtour elimination constraints and reducing computational effort. Furthermore, a cut-set based group of constraints can be introduced replacing these constrains. Let $S \neq \emptyset, S \subset V$ be a connected component of a solution in the previous algorithm, the following connection cut can be defined:

\begin{align}
 & \sum_{(i,j) \in \delta^+(S)} x_{ij} + \sum_{(i,j) \in \delta^-(S)} x_{ij} + \sum_{(i,j) \in \delta(S)} z_{ij} \geq  1, & \forall S \neq \emptyset, S \subset V.
\end{align}

This cut-set based connection cut implies that at least one allocation or one edge from the tree of facilities must connect $S$ to a node of $V \setminus S$.

Regarding to $\mT^{mtz}$, an arborescence is built rooted at an arbitrary root node in which arcs follow the direction from root to leaves. 
Note that if the given root node $r \in V$ is chosen as a facility, the arborescence will be rooted at $r$. Otherwise, $r$ will be isolated and the root node will be arbitrarily chosen by the model.
$\mT^{flow}$ also relies on a source node $r\in V$ which distributes the flow, but for the \OMT, contrary to $\mT^{mtz}$, the choice of the root node is very influential as care must be taken when spreading the flow. This selection can be done in two ways: (1) adding a set of variables $r_{i} \in \{0,1\}$ for $i \in V$, that are equal to 1 if the facility node $i \in V$ is selected as the source node for the tree of facilities, or otherwise (2) arbitrarily selecting the source node and ensuring flow units are only distributed between facilities, distinguishing whether the source node is a facility or not. If the source node is not a facility, then there must be no flow at the source node and the flow is distributed from the facility to which this source node is allocated.

Therefore, the flow formulation using additional variables includes the following group of constraints:

\begingroup
\allowdisplaybreaks
\begin{subequations}
\begin{align}
  & \hspace{-1cm} \mT^{flow1}: \sum_{i\in V} r_i = 1      &   &              &   & \label{OMT_f11_c1}  & \\
  & \Hp r_i \leq x_{ii}                                   &   &  i \in V     &   & \label{OMT_f11_c2}  & \\
  & \Hp \sum_{(i,j) \in \delta^+(i)}f_{ij} - \sum_{(j,i) \in \delta^-(i)}f_{ji} = (p-1)r_i-(x_{ii}-r_{i}) &   & i \in V &    & \label{OMT_f11_c3}  & \\
  & \Hp f_{ij} \leq (p-1)z_{ij}                           &   & (i,j) \in E  &   & \label{OMT_f11_c4}  & \\
  & \Hp f_{ji} \leq (p-1)z_{ij}                           &   & (i,j) \in E. &   & \label{OMT_f11_c5}  & 
\end{align}
\end{subequations}
\endgroup

On the other hand, the flow formulation without using additional variables includes:

\begingroup
\allowdisplaybreaks 
\begin{subequations}
\begin{align}
  & \hspace{-1.5cm} \mT^{flow2}: \sum_{(i,j) \in \delta^+(i)}f_{ij} - \sum_{(j,i) \in \delta^-(i)}f_{ji} = p x_{ri} - x_{ii} &   & i \in V & \label{OMT_f12_c1} & \\
  & \Hp f_{ij} \leq (p-1)z_{ij}                             &   & (i,j) \in E  &   & \label{OMT_f12_c2}  & \\
  & \Hp f_{ji} \leq (p-1)z_{ij}                             &   & (i,j) \in E. &   & \label{OMT_f12_c3}  &
\end{align}
\end{subequations}
\endgroup

\subsection{Alternative formulation}
\label{Alternative formulation}

Previous $F1_{x^\ell}^{\mT}$ and $F1^{\mT}_{u}$ formulations use the set of design variables $z$ to model a spanning tree connecting only facilities. However, since the \OMT assumes single client-facility allocation, the tree structure of any \OMT solution, uniquely determined by the allocation ($x$) and design ($z$) variables, is globally preserved. This is, considering both the facilities tree and the allocations of clients to their corresponding facilities, we also obtain a larger tree where clients are at the end nodes. This rationale allow us to provide an alternative formulation that models an \OMT solution designing a spanning tree over the complete set of nodes $V$. Holding the definition of variables $x$ and $x^{\ell}$ as before, let now $z_{ij} \in \{0,1\}$ for $(i,j) \in E$, be equal to 1 if $(i,j)$ is an edge connecting either facilities or client-facility pairs, zero otherwise. With this new definition of the $z$ variables, we can reformulate $F1_{x^\ell}^{\mT}$ as follows:

\begingroup
\allowdisplaybreaks
\begin{subequations}
\begin{align}
  & \hhp \hspace{1cm} F2_{x^\ell}^{\mT}: \min \quad \frac{1}{\displaystyle{\sum_{\ell \in V} \lambda_\ell}} \sum_{\ell \in V} \sum_{(i,j) \in A} \lambda_\ell c_{ij} x_{ij}^\ell + \frac{1}{p-1} \sum_{(i,j) \in E} c_{ij} (z_{ij}-x_{ij} - x_{ji})  & & & \label{OMT_f3_c0} & \\
  & \hp \sum_{i \in V} x_{ii} = p                       &   &                           & \label{OMT_f3_c1}  & \\
  & \Hp \sum_{j \in V} x_{ij} = 1                       &   & i \in V                   & \label{OMT_f3_c2}  & \\
  & \Hp 2x_{ij} \leq 1 - x_{ii} + x_{jj}                &   & (i,j) \in A: i\neq j      & \label{OMT_f3_c3}  & \\
  & \Hp x_{ij} + x_{ji} \leq z_{ij}                     &   & (i,j) \in E               & \label{OMT_f3_c4}  & \\
  & \Hp 2z_{ij} \leq x_{ii} + x_{jj} + x_{ij} + x_{ji}  &   & (i,j) \in E               & \label{OMT_f3_c5}  & \\
  & \Hp z_{ij} \in \mT                                  &   & (i,j) \in E               & \label{OMT_f3_c6}  & \\
  & \Hp x_{ij} = \sum_{\ell \in V} x_{ij}^\ell          &   & (i,j) \in A               & \label{OMT_f3_c7}  & \\
  & \Hp \sum_{(i,j) \in A} x_{ij}^\ell = 1              &   & \ell \in V                & \label{OMT_f3_c8}  & \\
  & \Hp \sum_{(i,j) \in A} c_{ij}x_{ij}^\ell \leq \sum_{(i,j) \in A} c_{ij}x_{ij}^{\ell+1}  &   & \ell \in V: \ell < |V|  & \label{OMT_f3_c9} & \\
  & \Hp x_{ij} \in \{0,1\}                              &   & (i,j) \in A \\
  & \Hp x_{ij}^\ell \in \{0,1\}                         &   & (i,j) \in A,\ell \in V \\
  & \Hp z_{ij} \in \{0,1\}                              &   & (i,j) \in E.
\end{align}
\end{subequations}
\endgroup

The objective function (\ref{OMT_f3_c0}) is similar to $F1_{x^\ell}^{\mT}$, but needs to be modified in order to consider only the design cost of the part of the tree connecting facilities, this is, subtracting the associated cost to the edges allocating clients to facilities to the total tree design cost. Now, Constraints (\ref{OMT_f3_c3}) guarantee that client-facility allocations are only possible if $i$ is a client and $j$ is a facility. Note that, these constraints result from unification of two simpler group of constraints $x_{ij} \leq 1 - x_{ii}$ and $x_{ij} \leq x_{jj}$ for $(i,j) \in A: i\neq j$. Constraints (\ref{OMT_f3_c4}) ensure that only one allocation, client $i$ to facility $j$ or vicerversa, can be considered if there exists an edge connecting $i,j \in V$. In this sense, constraints (\ref{OMT_f3_c5}) impose that the only possibilities in which a tree edge can be selected are if both $i$ or $j$ are facilities or if $i$ is a client and $j$ its facility, or viceversa. The remaining constraints hold unvarying.

For this approach, similar $\mT$ descriptions as in Section \ref{subsec:Tdescription} can be considered, but the adaptation is straightforward since the main characteristic to model in this formulation is the tree. Now, the cut-set based group of constraints are simpler, $\sum_{(i,j) \in \delta(S)} z_{ij} \geq 1$, and for the flow formulation it is not necessary to differentiate whether the arbitrary source node selection is a facility or not, reducing formulations complexity. Formulation $F2_u^{\mT}$ for covering variables using this alternative approach is left to the reader.

\subsection{Benders decomposition}
\label{Bendersdecomposition}

The \OMT can be solved by using a Benders decomposition framework (see \citealp{Benders1962}) that we briefly describe in this section. In the classical Benders decomposition algorithm, the original mixed integer problem is divided into two problems, a master problem (MP) and a subproblem (SP), that are solved iteratively. The two problems are related, so the outcome of one directly modifies the outcome of the other. First, the MP is solved to obtain the values of certain fixed variables; with these, we can then solve the SP. Once the SP has been solved, either new feasibility or new optimality cuts are introduced within the MP until the lower and upper bounds coincide. For the \OMT, it arises to use DOMP as MP and MST as SP. For instance, we can illustrate the MP via $F1_{x^\ell}^{\mT}$ as follows:

\begingroup
\allowdisplaybreaks
\begin{subequations}
\begin{align}
  & \hhp \hspace{1cm} F^{MP}: \min \quad \frac{1}{\displaystyle{\sum_{\ell \in V} \lambda_\ell}} \sum_{\ell \in V} \sum_{(i,j) \in A} \lambda_\ell c_{ij} x_{ij}^\ell + \mu & & & \label{OMT_f1b_c0} & \\
  & \hp (\ref{OMT_f1_c1})-(\ref{OMT_f1_c3}), (\ref{OMT_f1_c6})-(\ref{OMT_f1_c10}) \\ 
  & \Hp \mu \geq 0.  
\end{align}
\end{subequations}
\endgroup

Similar descriptions for $F^{MP}$ can be introduced using other formulations in Section \ref{subsec:Sortingformulations}.

In a classical Benders decomposition framework, the MP is solved to optimality keeping both the objective function, which can be split into the sum of the DOMP objective plus the $\mu$ term, and the variables representing the facilities selected, $\bar{x}= \{x_{ii} \ | \ x_{ii}=1, \forall i \in V \}$. If possible, the lower bound is updated using the MP objective. Thereafter, the SP sets as facilities the previously $\bar{x}$ identified in the MP solution and, if possible, the upper bound using the (weighted) sum of the DOMP plus the SP objectives is updated. Finally, in any case, a optimality cut ($Opt.cut$) is added to the MP. This procedure is repeated until both upper and lower bounds coincide.

For the \OMT, the SP is solved by means of a Kipp Martin ($km$) MST dual formulation (see \citet{Labbe2021} for details on the implementation). This $km$ MST formulation and its dual form are adapted to the \OMT structure as follows:

\begingroup
\allowdisplaybreaks
\begin{subequations}
\begin{align}
  & \hhp \hspace{1cm} F^{km}: \min \sum_{(i,j) \in E} c_{ij} z_{ij} & & & \label{OMT_f2b_c0} \\
  & \Hp z_{ij} \leq \bar{x}_{ii}                        &   & (i,j) \in E                   & \label{OMT_f2b_c11} \\
  & \Hp z_{ij} \leq \bar{x}_{jj}                        &   & (i,j) \in E                   & \label{OMT_f2b_c12} \\
  & \Hp \sum_{(i,j) \in E} z_{ij} = p-1                 &   &                               & \label{OMT_f2b_c2} \\
  & \Hp \sum_{\substack{(i',j) \in E: \\ (i'=k \wedge j=i) \\ \lor \\ (i'=i \wedge j=k)}} z_{i'j} + \sum_{(i,j) \in A:j \neq k} q_{kij} \leq 1 & & k,i \in V: i \neq k & \label{OMT_f2b_c3} \\
  & \Hp q_{kij} + q_{kji} = z_{ij}                      &   &  k \in V, (i,j) \in E: i,j \neq k  & \label{OMT_f2b_c4} \\
  & \Hp z_{ij}  \geq 0                                  &   &  (i,j) \in E \\
  & \Hp q_{kij} \geq 0                                  &   &  k \in V, (i,j) \in E.
\end{align}
\end{subequations}
\endgroup

\begingroup
\allowdisplaybreaks
\begin{subequations}
\begin{align}
  & \hhp \hspace{1cm} F^{SP}: \max \alpha (p-1) - \sum_{k,i \in V: i \neq k} \beta_{ki} - \sum_{(i,j) \in E} (\bar{x}_{ii} \tau_{ij} + \bar{x}_{jj} \eta_{ij}) & & & \label{OMT_f3b_c0} \\
  & \Hp \alpha - \beta_{ij} - \beta_{ji} - \sum_{k \in V: k \neq i,j} \gamma^{k}_{ij} - \tau_{ij} - \eta_{ij} \leq c_{ij} &   & (i,j) \in E & \label{OMT_f3b_c1} \\
  & \Hp - \beta_{ki} +  \sum_{\substack{(i',j') \in E: \\ (i'=i \wedge j'=j) \\ \lor \\ (i'=j \wedge j'=i)}} \gamma^{k}_{i'j'} \leq 0 &   & k \in V, (i,j) \in E: i,j \neq k & \label{OMT_f3b_c2} \\
  & \Hp \alpha \in \mathbb{R}        &   &  \\
  & \Hp \beta_{ki} \geq 0            &   & k,i \in V: i \neq k \\
  & \Hp \gamma^k_{ij} \in \mathbb{R} &   & k \in V, (i,j) \in E: i,j \neq k  \\
  & \Hp \tau_{ij}    \geq 0          &   & (i,j) \in E \\
  & \Hp \eta_{ij}    \geq 0          &   & (i,j) \in E.
\end{align}
\end{subequations}
\endgroup

Observe that, constraints (\ref{OMT_f1_c4}) are presented as disaggregated constraints (\ref{OMT_f2b_c11})-(\ref{OMT_f2b_c12}) in the SP because, although it has been computationally proven to report worse results, dealing with continuous $z$ variables force to use the disaggregated formulation. In this way, the $Opt.cut$ introduced have the following expression:

\begin{equation}
  Opt.cut:\ \frac{1}{p-1} \Big[\bar{\alpha}(p-1) - \sum_{k,i \in V: i \neq k} \bar{\beta}_{ku} - \sum_{(i,j) \in E} (x_{ii} \bar{\tau}_{ij} + x_{jj} \bar{\eta}_{ij}) \Big] \leq \mu.
\end{equation}

Note that in every iteration, only one $Opt.cut$ is added. For the general Benders framework, some authors (see, e.g., \citealp{Fischetti2010}) force the addition of feasibility cuts even when the MP is feasible to boost the algorithm. In our case, since our SP is always feasible, it is not necessary to introduce feasibility cuts.

The classical Benders decomposition can be introduced into a branch-and-cut framework (or modern Benders approach), also known as branch-and-Benders-cut, for a more efficient performance as illustrated in Algorithm \ref{alg:Benders_modern}. In this algorithm we initialize the tree $\mTree$, considering solving the MP at the root node, and an empty pool of cuts $\mathcal{P}$. Once solved the MP at the root node, we start the branching procedure. If at a particular node of the process the solution found is fractional, keep branching. Otherwise, when an integer solution is found, the lower bound is updated using the MP objective value of the current node $o' \in \mTree$, the SP is solved fixing the facilities $\bar{x}$ obtained in the MP solution, the upper bound is updated if possible and the optimality constraint $Opt.cut$ is added to a pool $\mathcal{P}$ of cuts. The cuts from pool $\mathcal{P}$ are then included for the MP solution in other nodes of $\mTree$.

\begin{algorithm}
\caption{\OMT branch-and-Benders-cut} \label{alg:Benders_modern}
\SetKwInOut{Input}{input}\SetKwInOut{Output}{output}
\LinesNumbered
Set tree $\mTree=\{o\}$, where $o=F^{MP}$ has no branching constraints \\
Initialize a pool of cuts $\mathcal{P} = \emptyset$ \\
\While{$\mTree$ is nonempty}{
    Select a node $o'\in\mTree$\\
    $\mTree := \mTree \,\backslash\{o'\}$\\
    Solve $o'$ considering $\mathcal{P}$ $\rightarrow \bar{x}$ \\
    \If {$\bar{x}$ is fractional}{
        Branch, resulting in nodes $o''$ and $o'''$\\
        $\mTree := \mTree \cup\{o'',o'''\}$\\
    \Else{
        Solve $F^{SD} \rightarrow (\bar{\alpha}, \bar{\beta}, \bar{\tau}, \bar{\eta}) $\\
        Add optimality constraint $Opt.cut$ to $\mathcal{P}$\\
        $\mTree := \mTree \cup \{o'\}$\\
    }}
}
\end{algorithm} 

This allows to use different strategies for producing the optimality cuts. For instance, these cuts can be generated in every feasible node, only in those nodes that have improved the lower bound beyond a predefined threshold or only when incumbent solutions are found along the search process. If caution is not taken, too many unnecessary cuts can be produced slowing down the attainment of a MP solution instead of speeding it up. Also, in the classical Benders decomposition, every $Opt.cut$ added implies solving to optimality a MP, which is computationally costly. In the modern approach, the pool of cuts $\mathcal{P}$ considered in a specific node of the branch-and-cut procedure may contain cuts that have already been identified, slowing down the efficiency since it would reintroduce repeated cuts when branching at different levels of the search tree. For these reasons, it could be helpful to consider a strategy to initialize the pool of cuts $\mathcal{P}$ prior to starting the procedures using a so called \textit{warm-start phase} (see, e.g., \citealp{Martins2013}). The interest behind this warm phase is to be able to introduce a certain number of cuts at low cost before start branching. In order to introduce this cuts, the solution in the warm-start phase does not need to be computed to optimality, is enough to obtain a feasible solution. Furthermore, a procedure to build cuts from fractional values of the variables can be implemented. 
More details can be found on Appendix A.

The presented branch-and-Benders approach has DOMP as a subproblem which is more difficult than MSTP. Therefore, this branch-and-Benders implies solving a DOMP adding optimality Benders cuts (costly to obtain) what seems to be less efficient than solving a DOMP with some MSTP additional constraints. Thus, this general technique for solving the \OMT would require an alternative way for solving DOMP, may be applying a nested Benders decomposition. In any case,  this analysis is out of the scope of this paper.

\section{Improvements}
\label{sec:Improvements}

\subsection{Initial solution}
\label{subsec:Initialsolution}

The branching processes used as an underlying framework to solve the presented formulations usually take a considerable amount of time in finding an initial feasible solution (upper bound). For this reason, in this section we introduce two \OMT heuristic algorithms to compute an initial solution.

\begin{itemize}

  \item \textbf{DOMP + MST heuristic}. This approach performs first the DOMP and, once the facilities have been assigned and compensated, they are fixed for searching a MST connecting those facilities in polynomial time using Kruskal MST algorithm. 

  
  \item \textbf{PMEDT + DOMP heuristic}. This approach finds $p$ facilities and the tree structure computing the PMEDT by means of a MST competitive formulation, for example the MTZ formulation. Both, facilities and the tree structure, are left fixed and the reassignment of clients to facilities is then performed using the DOMP. This algorithm has been proven to be more efficient providing an initial solution.
  
    
\end{itemize}

\subsection{Variable fixing}
\label{subsec:Variablefixing}

This section addresses the description of some preprocessing steps to reduce the size of the covering formulation $F1^{\mT}_{u}$ in order to improve its efficiency. These variable fixing procedures are based on adaptations of some arguments already used in \citet{Puerto2011, Puerto2013} and \citet{Pozo2021} where fixing a certain number of variables improved the efficiency of the formulations.

Given the ordered sequence of unique costs and the \textit{sorting} factor of the orders $\ell$, the $u_{\ell h}$-variables matrix is expected to have a specific step structure. It might be expected that for small values of $h$, the number of allocations at cost smaller than $c_{(h)}$ should be small and therefore $u_{\ell h}=1$ in the left-bottom part of the $u$-matrix from a certain $\ell$ to the end. Similarly, it might also be expected that many $u_{\ell h}$-variables in the right-upper part of the $u$-matrix will take value 0 in the optimal solution. Indeed, $u_{\ell h}=0$ means that the $\ell$-th sorted allocation cost is less than to $c_{(h)}$, which is very likely to be true if $h$ is sufficiently large and $\ell$ is small enough. For example, since $c_{jj}=0$ for $j \in V$, we have that $u_{\ell h} = 0$ for $\ell \in \{ 1,\ldots,p \}, h \in H$. These preprocessings aim to fix as many of this $u_{\ell h}$-variables as possible.

\subsubsection{Preprocessings for fixing variables to 1}

Given a $h \in H$, in order to fix $u_{\ell h}$-variables to 1 we deal with an auxiliary problem that maximizes the number of allocations satisfying $c_{ij} \leq c_{(h-1)}$, which is equivalent to the number of variables that can assume a 0 value, or what is the same, this will provide the number of allocations satisfying the opposite ($c_{ij} > c_{(h-1)} \equiv c_{ij} \geq c_{(h)}$) and, therefore, the minimum number of $u_{\ell h}$-variables that can assume value of 1. Using the variables previously defined, the formulation of this problem is:

\begin{subequations}
\allowdisplaybreaks
\begin{align}
     & \hhp \hspace{1cm} F^{pre}_{1}: \max H_h^1 := \sum_{i\in V} \sum_{j\in V} x_{ij} &  &  & \label{preproc_1_a} \\
     & \hp \sum_{i\in V} x_{ii} = p                                         &  &                           & \label{preproc_1_b} \\[-0.1cm]
     & \Hp \sum_{j\in V} x_{ij} \leq 1                                      &  & \forall i\in V            & \label{preproc_1_c} \\[-0.1cm]
     & \Hp x_{ij}\leq x_{jj}                                                &  & \forall i,j\in V: i\neq j & \label{preproc_1_d} \\
     & \Hp c_{ij}x_{ij} \leq c_{(h-1)}                                      &  & \forall i,j\in V          & \label{preproc_1_e} \\[-0.05cm]
     & \Hp x_{ij}\in \{0,1\}                                                &  & \forall i,j\in V.         & \label{preproc_1_f}
\end{align}
\end{subequations}

$F^{pre}_{1}$ gives the maximum number (in the worst case) of allocations realized at a cost less than or equal to $c_{(h-1)}$. If $H_h^1$ is the optimal value of problem above, since there are $|V| = N$ total possible allocations, the number of allocations satisfying $c_{ij} \geq c_{(h)}$ must be necessarily greater than or equal to $N-H_h^1$, or equivalently, in any feasible solution of a covering formulation:

$$u^{\ell}_{h}=1, \quad \forall \ell \in \{ H_h^1 + 1, \ldots, N \}.$$

\subsubsection{Preprocessings for fixing variables to 0}

Also, for a given $h \in H$, to fix the maximum number of $u_{\ell h}$-variables possible to 0 we deal with the following auxiliary problem: maximize the number of allocations satisfying $c_{ij} \geq c_{(h)}$, which will provide the minimum number of 0 that the $h$-th column of the $u$-matrix must have. Using the variables defined previously, the formulation of this problem is:

\begin{subequations}
\allowdisplaybreaks
\begin{align}
     & \hhp \hspace{1cm} F^{pre}_{0}: \max H_h^0 := \sum_{i\in V} \sum_{j\in V} x_{ij} &  &  & \label{preproc_0_a} &  \\
     & \hp \sum_{i\in V} x_{ii} = p                                         &  &                           & \label{preproc_0_b} &  \\[-0.1cm]
     & \Hp \sum_{j\in V} x_{ij} \leq 1                                      &  & \forall i\in V            & \label{preproc_0_c} &  \\[-0.1cm]
     & \Hp x_{ij} \leq x_{jj}                                               &  & \forall i,j\in V: i\neq j & \label{preproc_0_d} &  \\
     & \Hp c_{ij} \geq c_{(h)} x_{ij}                                       &  & \forall i,j\in V: i\neq j & \label{preproc_0_e} &  \\[-0.05cm]
     & \Hp x_{ij}\in \{0,1\}                                                &  & \forall i,j\in V.         & \label{preproc_0_f} &
\end{align}
\end{subequations}

In this problem, for constraints (\ref{preproc_0_e}) we could not consider allocations $x_{ii}$ for $i \in V$, as considered in \citealp{Puerto2011}, \citealp{Puerto2013} and \citealp{Pozo2021} among others, because in such case our problem is not feasible. Infeasibility is due to having to consider constraints (\ref{preproc_0_b}) together with constraints (\ref{preproc_0_e}). If we consider allocations such $x_{ii}$ for $i \in V$ we have $c_{ii}=0$ and since $c_{(h)}>0$ for $h \in H$, constraints (\ref{preproc_0_e}) set all $x_{ii}$ to 0, which is not compatible with constraints (\ref{preproc_0_b}). For this reason we must impose $i,j\in V: i\neq j$ in (\ref{preproc_0_e}).

For a given $h \in H$, $F^{pre}_{0}$ is the maximum number of allocations done at a cost greater than or equal to $c_{(h)}$. Therefore, if $H^0_h$ is the optimal value of problem above, the $h$-th column of the $u$-matrix must have at least $N-H^0_h$ zeros. As we know that exactly the $p$ first rows of the $u$-matrix can be fixed to 0, which corresponds to the allocations of the facilities to itselves at a cost 0 not considered in the constraints (\ref{preproc_0_e}), in any feasible solution of the covering formulation satisfies:

$$u^{\ell}_{h}=0, \quad \forall \ell \in \{ 1,\ldots, N-H^0_h+p \}.$$

For the ease of understanding, an example can be found in Appendix B.

\section{Computational experiments}
\label{sec:Computationalexperiments}

This section reports on the results of some computational experiments performed in order to empirically compare the proposed \OMT formulations. Table \ref{tab:OMT_formulations_sum} catalogs all different formulations considering every combination for the tree polytope $\mT \in \{sub,mtz,flow,km\}$, sorting type $\mS \in \{x^{\ell}, u\}$ and whether $\mT$ is built between facilities or in $V$. The interested reader can see Appendix C for the whole catalogue of formulations described in detail.

\begin{table}[ht]
\centering
\begin{tabular}{@{}ccll@{}}
\toprule
$\mT$                   & $\mS$       & (1): $\mT$ in facilities       & (2):  $\mT$ in $V$             \\ \midrule
\multirow{2}{*}{$sub$}  & $x^{\ell}$  & $F1_{x^{\ell}}^{sub1}$, $F1_{x^{\ell}}^{sub2}$ & $F2_{x^{\ell}}^{sub1}$, $F2_{x^{\ell}}^{sub2}$  \\ \cmidrule(l){2-4} 
                        & $u$         & $F1_u^{sub1}$, $F1_u^{sub2}$   & $F2_u^{sub1}$, $F2_u^{sub2}$   \\ \midrule
\multirow{2}{*}{$mtz$}  & $x^{\ell}$  & $F1_{x^{\ell}}^{mtz}$          & $F2_{x^{\ell}}^{mtz}$          \\ \cmidrule(l){2-4} 
                        & $u$         & $F1_u^{mtz}$                   & $F2_u^{mtz}$                   \\ \midrule
\multirow{2}{*}{$flow$} & $x^{\ell}$  & $F1_{x^{\ell}}^{flow1}$, $F1_{x^{\ell}}^{flow2}$ & $F2_{x^{\ell}}^{flow1}$, $F2_{x^{\ell}}^{flow2}$ \\ \cmidrule(l){2-4} 
                        & $u$         & $F1_u^{flow1}$, $F1_u^{flow2}$ & $F2_u^{flow1}$, $F2_u^{flow2}$ \\ \bottomrule
\multirow{2}{*}{$km$}   & $x^{\ell}$  & $F1_{x^{\ell}}^{km}$           & $F2_{x^{\ell}}^{km}$           \\ \cmidrule(l){2-4} 
                        & $u$         & $F1_u^{km}$                    & $F2_u^{km}$                    \\ \bottomrule
\end{tabular}
\caption{OMT formulations summary.}
\label{tab:OMT_formulations_sum}
\end{table}

For the computational study, instances $G=(V,E)$ are generated as complete networks of sizes $|V| \in \{20,30,40,50,60,70,80,90,100\}$ with random integer costs $c_{ij}$ following a uniform distribution in $[1,1e5]$. In order to follow the structure of previous published works, where other ordered median problems were considered, the number of facilities for each group of instances is chosen as $p \in \{\lfloor \frac{N}{4} \rfloor, \lfloor \frac{N}{3} \rfloor, \lfloor \frac{N}{2} \rfloor\}$. Also, we test three commonly studied criteria for the $\lambda$ scaling factor values:

\begin{itemize}
  \item \textbf{Median criterion}: $\lambda=\underbrace{(1,1,....,1)}_{|V|}$.
  \item \textbf{$k$-centrum criterion}: $\lambda=(\underbrace{0,...,0}_{\lfloor\frac{2}{3}|V|\rfloor},\underbrace{1,...,1}_{|V|-\lfloor\frac{2}{3}|V|\rfloor})$.
  \item \textbf{$k$-trimmed mean criterion}: $\lambda=(\underbrace{0,...,0}_{\lfloor\frac{1}{3}|V|\rfloor},\underbrace{1,...,1}_{|V|-\lfloor\frac{2}{3}|V|\rfloor},\underbrace{0,...,0}_{\lfloor\frac{1}{3}|V|\rfloor})$.
\end{itemize}

In all tables, results correspond to groups of 5 instances with same $(|V|,p)$ pair. This gives a benchmark of 135 instances in total. We present average results for each group of instances. All instances were solved with Gurobi 9.1.1 optimizer, under a Windows 10 environment in an Intel(R) Core(TM) i7-2600 CPU 3.40 GHz processor and 16 GB RAM. The CPU time limit fixed for solving each instance is 3600 seconds.

Tables are grouped in blocks. The first block contains the values of the instances parameters, namely $|V|$ and $p$. Then, we give a block of 6 columns for each formulation with the following information:

\begin{itemize}
  \item Columns $|\#|$ indicate the number of instances in the group that could be solved to optimality within the CPU time limit.
  \item Columns $cpu$ report the average computing time in seconds over the groups of instances.
  \item Columns \gapbUR give the percentage relative root gap, computed as $100\frac{obj_{\overline{U}}\ - \ obj_{R}}{obj_{\overline{U}}}$, where $obj_{\overline{U}}$ denotes the best known upper bound obtained in all our experiments and $obj_{R}$ denotes the optimal value of the linear relaxation at the root node.
  \item Columns \gapbUL give the percentage relative lower bound gap, computed as $100\frac{obj_{\overline{U}}\ - \ obj_{L}}{obj_{\overline{U}}}$, where $obj_{L}$ denotes the lower bound at termination.
  \item Columns \gapUbL give the percentage relative upper bound gap, computed as $100\frac{obj_{U}-obj_{\overline{L}}}{obj_{U}}$, where $obj_{U}$ denotes the upper bound at termination and $obj_{\overline{L}}$ denotes the best known lower bound obtained in all our experiments.
  \item Columns \gapUL give the percentage relative gap at termination, computed as $100\frac{obj_{U}-obj_{L}}{obj_{U}}$.
  \item Columns $nod$ indicate the average number of nodes explored in the branch-and-bound ($B\&B$) search tree.
\end{itemize}

Note that, while \gapbUR and \gapbUL provide quality measures of the lower bounds (at the root node and at termination, respectively), \gapUbL  provides a quality measure of the upper bounds. In addition, \gapUL provides a measure of both upper and lower bounds for the average performance. If the reported value of \gapUL is 0 then all instances were solved to optimality. Along all experiments we give as initial solutions those coming from the heuristics developed in Section \ref{sec:Improvements}, so in all cases it is possible to find at least a feasible solution.

\begin{figure}[ht]
\centering
\includegraphics[width=18cm]{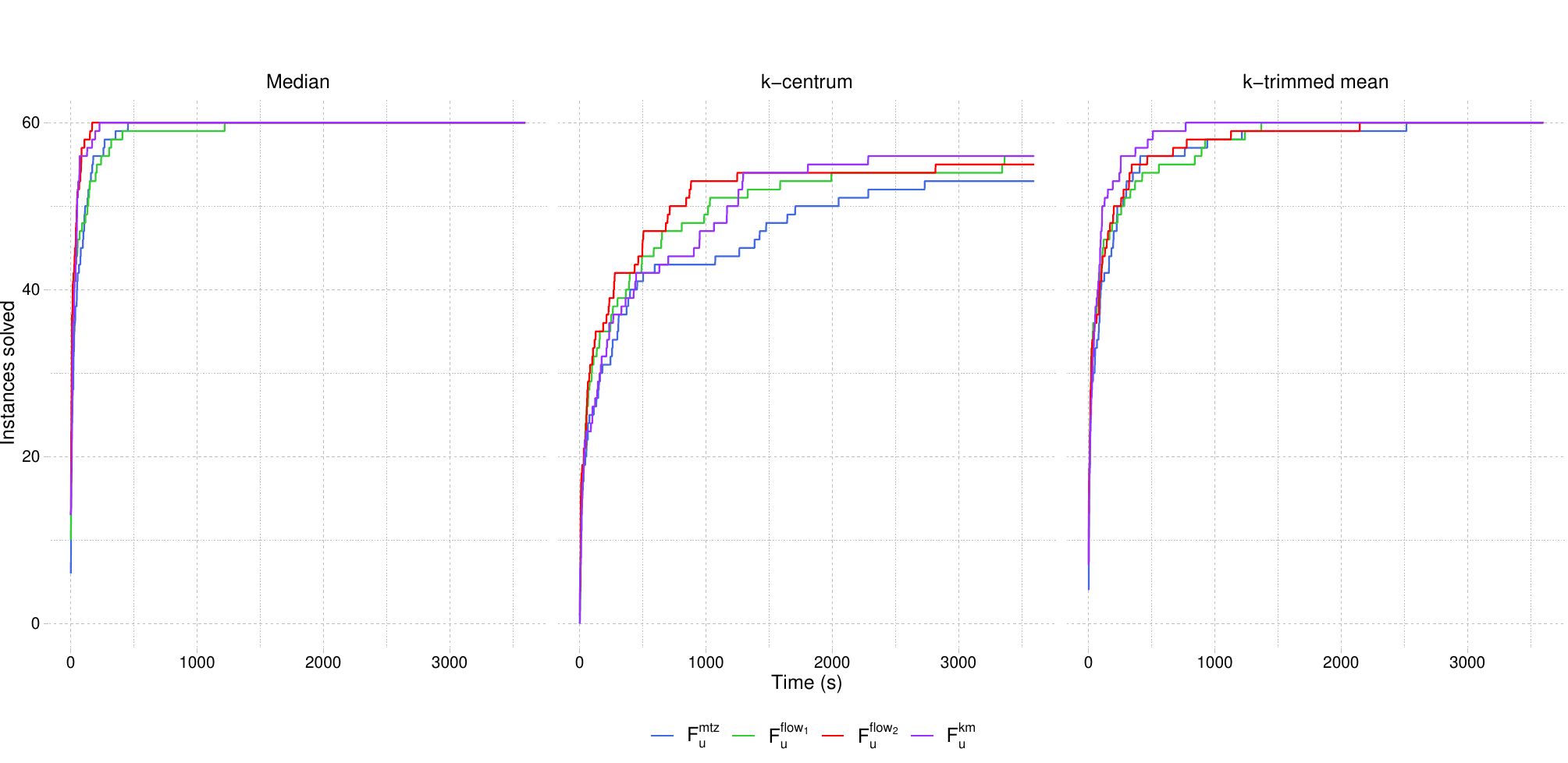}
\caption{$F1_{u}^{(.)}$ formulations comparison for the three different criteria}
\label{OMT_comparison_u_R2}
\end{figure}

The caption just above each block gives the formulation the block refers to. Due to the large number of formulations (see Table \ref{tab:OMT_formulations_sum}), we decided to report just four final formulations. This choice was based on preliminary computational result based on different set of instances. 
We report results among $F1^{(\mT)}_{(\mS)}$ with $\mS \in \{x^{\ell}, u\}$ and $\mT \in \{mtz,flow1,flow2,km\}$ since other approaches (subtour elimination, Benders decomposition and $F2^{(\cdot)}_{(\cdot)}$) were clearly outperformed in our preliminary tests, especially for large instance sizes. All preliminary results are available in Appendix D. Then, the criteria that have guided us to present as final formulations $F1_{x^{\ell}}^{mtz}$, $F1_{x^{\ell}}^{flow1}$, $F1_{u}^{mtz}$ and $F1_{u}^{km}$ are the following:

\begin{enumerate}
  \item Over a total of four formulations we have chosen two types of $F1_{x^{\ell}}^{(\cdot)}$ formulations and two types of $F1_{u}^{(\cdot)}$ formulations. We recall that $F1_{u}^{(\cdot)}$ takes advantage of ties in the cost matrix (see Section 3.1). Besides, choosing formulations among $F1_{x^{\ell}}^{(\cdot)}$ requires an analysis of the gap left (many instances could not be solved in the time limit) and choosing formulations among $F1_{u}^{(\cdot)}$ requires an analysis of the running time (mostly all instances were solved within the time limit).
  \item According to Table \ref{Table_xl_gap}, $F1_{x^{\ell}}^{flow1}$ slightly outperforms to its variant $F1_{x^{\ell}}^{flow2}$. Therefore, given the similarity of these formulations, we discard $F1_{x^{\ell}}^{flow2}$.
  \item According to Table \ref{Table_xl_gap}, $F1_{x^{\ell}}^{mtz}$ slightly outperforms  $F1_{x^{\ell}}^{km}$ (in terms of gaps for $k$-trimmed mean and $k$-centrum criteria).
  \item According to Table \ref{Table_xl_gap}, $F1_{x^{\ell}}^{flow1}$ slightly outperforms  $F1_{x^{\ell}}^{km}$ (again in terms of gaps for $k$-trimmed mean and $k$-centrum criteria).
  \item According to performance profiles in Figure \ref{OMT_comparison_u_R2}, $F1_{u}^{km}$ has shown slightly better performance among all $F1_{u}^{(\cdot)}$ formulations with respect to the $k$-trimmed mean criterion and $k$-centrum criterion (for the most difficult instances). 
  \item Finally, $F1_{u}^{mtz}$ has been chosen so that the reader could have a comparison of the same tree polytope description among both sorting descriptions, that is, by means of $F1_{x^{\ell}}^{mtz}$ and $F1_{u}^{mtz}$. Beyond that, we are interested in reporting $F1_{(\cdot)}^{mtz}$ since $mtz$ has been shown as a promising tree polytope description for being combined with sorting constraints (see \citealp{Fernandez2014}).
\end{enumerate}
 
\begin{table}[ht]
\centering
\begin{tabular}{c|rr|rr|rr}
    \hline
    \multirow{1}{*}{$\mathcal{T}$} & \multicolumn{2}{c|}{Median} & \multicolumn{2}{c|}{$k$-centrum} & \multicolumn{2}{c}{$k$-trimmed mean}  \\
             & $|\#|$ & \meangapUL & $|\#|$ & \meangapUL & $|\#|$ & \meangapUL \\ \hline
    $mtz$    & 57     & 1.26  &  13  & 27.14  &   30  &   9.59    \\
    $flow1$  & 57     & 0.53  &  14  & 25.12  &   32  &   8.85    \\
    $flow2$  & 57     & 0.65  &  13  & 25.90  &   32  &   9.58    \\
    $km$     & 60     &  -    &  13  & 28.47  &   33  &  10.19    \\
    \hline
\end{tabular}
\caption{$F1_{x^{\ell}}^{\mT}$ formulations comparison for the three different criteria, where \meangapUL represents the mean \gapUL value across all runs.}
\label{Table_xl_gap}
\end{table}

A general overview of the \OMT computational experience brings some common observations for all criteria. 
The reader may note that instance sizes have been chosen in such a way that many of the instances cannot be solved to optimality within the time limit established (3600 seconds). 
This is specially remarkable for the $k$-centrum and $k$-trimmed mean criteria that are those that reflect the computational complexity of the sorting problem. 
For that reason we believe cpu time is pointless in this particular study while the reported gaps provide a nicer information of the behavior of each formulation for each criterion and $(|V|,p)$. 
In spite of that, we report them for the interested reader.
A first evidence is that given a fixed tree polytope description, as a consequence of Property \ref{prop2}, values for \gapbUR are equal for every $(|V|,p)$ pair among the two sorting descriptions. 
This can be observed in the tables of results, where for $F1^{mtz}_{x^{\ell}}$ and $F1^{mtz}_{u}$ the values of \gapbUR coincide. 
However, this does not imply that for a fixed sorting description values for \gapbUR should match, e.g., values for $F1^{km}_{u}$ do not coincide with $F1^{mtz}_{u}$ or other tree description.
Reported values of \gapbUL, \gapUbL and \gapUL are smaller for $F1^{(\cdot)}_{u}$ formulations, concluding that $F1^{(\cdot)}_{x^{\ell}}$ formulations are outperformed in these terms. Moreover, $F1^{(\cdot)}_{u}$ formulations are able to certify optimality for most of the instances (median and $k$-trimmed mean criteria) and more often than $F1^{(\cdot)}_{x^{\ell}}$ formulations. Concerning \gapUL values, instances with lower $p$ are rather more difficult to solve in general terms than those with bigger $p$ values. Besides, $nod$ parameter is influenced by how well the problem is described (formulation used, number of fixed variables) and the number of variables and constraints introduced. Starting from the smaller size instances, formulations tend to increase $nod$ values as long as they increase their size. However, at some point $nod$ values start decreasing when variables and constraints considerably increase and computations at each node are more time-consuming. 
In general, the results differences between the $F1^{(\cdot)}_{(\mS)}$ formulations are small within the given sorting description and highly depend on the criterion.

We now comment on some specific considerations regarding each criterion. 
In Table \ref{re:OMTcmedian}, where results for median criterion are reported, all instances reach optimality for $F1^{(\cdot)}_{x^{\ell}}$ formulations when $|V| \leq 40$  and for $F1^{(\cdot)}_{u}$ formulations when $|V| \leq 60$. 
For the median criterion, values for \gapbUR range between $28\%-42\%$.
Concerning $nod$ values, $F1^{(\cdot)}_{x^{\ell}}$ formulations tend to explore more nodes of the $B\&B$ tree for small size instances than $F1^{(\cdot)}_{u}$ formulations. This exploration is gradually increased until $|V|=70$, where computations at each node require more time. On the other hand, $F1^{(\cdot)}_{u}$ formulations need to explore less nodes in order to certify optimality. As shown in previous DOMP problems, median criterion does not reflect the computational complexity of the sorting problem since all lambdas are equal to 1. In Tables \ref{re:OMTckcentrum} and \ref{re:OMTcktrimmean}, where results for $k$-centrum criterion and $k$-trimmed mean criterion are respectively reported, problems to prove optimality start to appear for smaller size instances, being specially difficult for the $k$-centrum criterion. For this criteria, \gapbUR, \gapbUL, \gapUbL and \gapUL values are greater than those reported for median criterion. Moreover, for  $k$-centrum and $k$-trimmed mean, formulations tend to need to explore more $B\&B$ nodes for small size instances than for median criterion and are not able to explore many nodes for large size instances. 
Therefore, we can conclude that, as expected, our formulations perform better for median criterion than for $k$-trimmed mean, and specially $k$-center criterion reports the worst results.

\begin{figure}[ht]
\centering
\includegraphics[width=17cm]{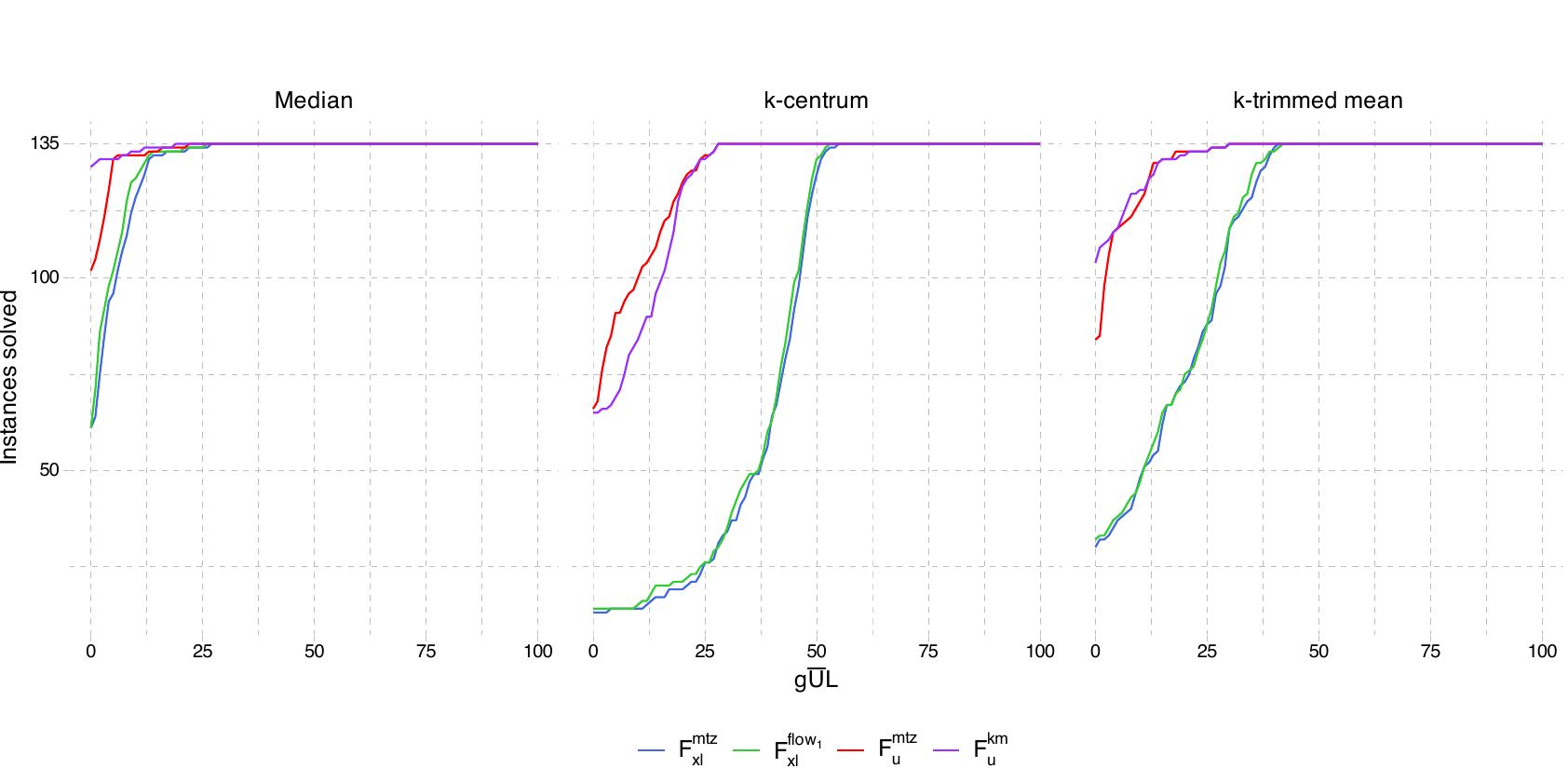}
\caption{Cumulative number of instances solved within a \gapbUL value.}
\label{Fig:gapbULvalue}
\end{figure}

\begin{figure}[ht]
\centering
\includegraphics[width=17cm]{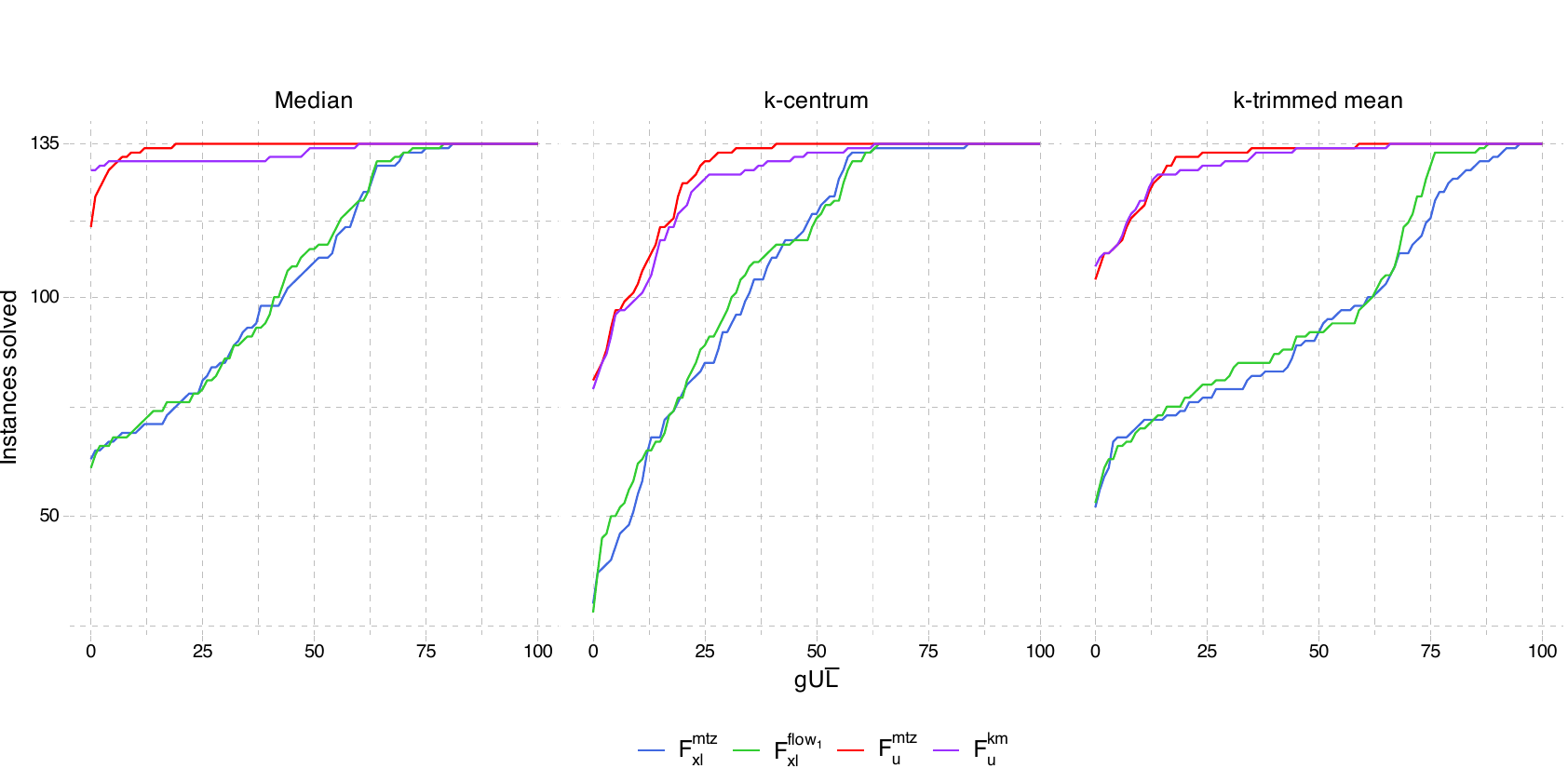}
\caption{Cumulative number of instances solved within a \gapUbL value.}
\label{Fig:gapUbLvalue}
\end{figure}

Besides, the reader may note that particular behavior of gaps \gapbUL and \gapUbL in Tables \ref{re:OMTcmedian}, \ref{re:OMTckcentrum} and \ref{re:OMTcktrimmean}. 
We recall that \gapbUL provide a quality measure of the lower bound at termination, \gapUbL provides a quality measure of the upper bound at termination.
For example, for median criterion \gapUbL is significantly bigger than \gapbUL for $F 1_{x^\ell}^{m t z}$ and $F 1_{x^\ell}^{flow1}$ (it seems difficult for the solver to find good feasible solutions). 
However, for median criterion \gapbUL is bigger than \gapUbL for $F 1_{u}^{m t z}$ (it seems more difficult to increase the lower bound for certifying optimality). 
For the $k$-centrum  criterion \gapbUL is usually bigger than or equal to \gapUbL for mostly all cases. 
However, this comparison depends on the formulation and $(|V|,p)$ for the $k$-trimmed mean criterion. 
These specific comments and previous general ones can also be seen in the cumulative number of instances solved within a \gapbUL and \gapUbL values (Figure \ref{Fig:gapbULvalue} and Figure \ref{Fig:gapUbLvalue} respectively).
In these graphics, $F1^{mtz}_{u}$ shows the best performance in many cases. However, nothing in general, can be concluded for all cases.

\begin{figure}[ht]
\centering
\includegraphics[width=18cm]{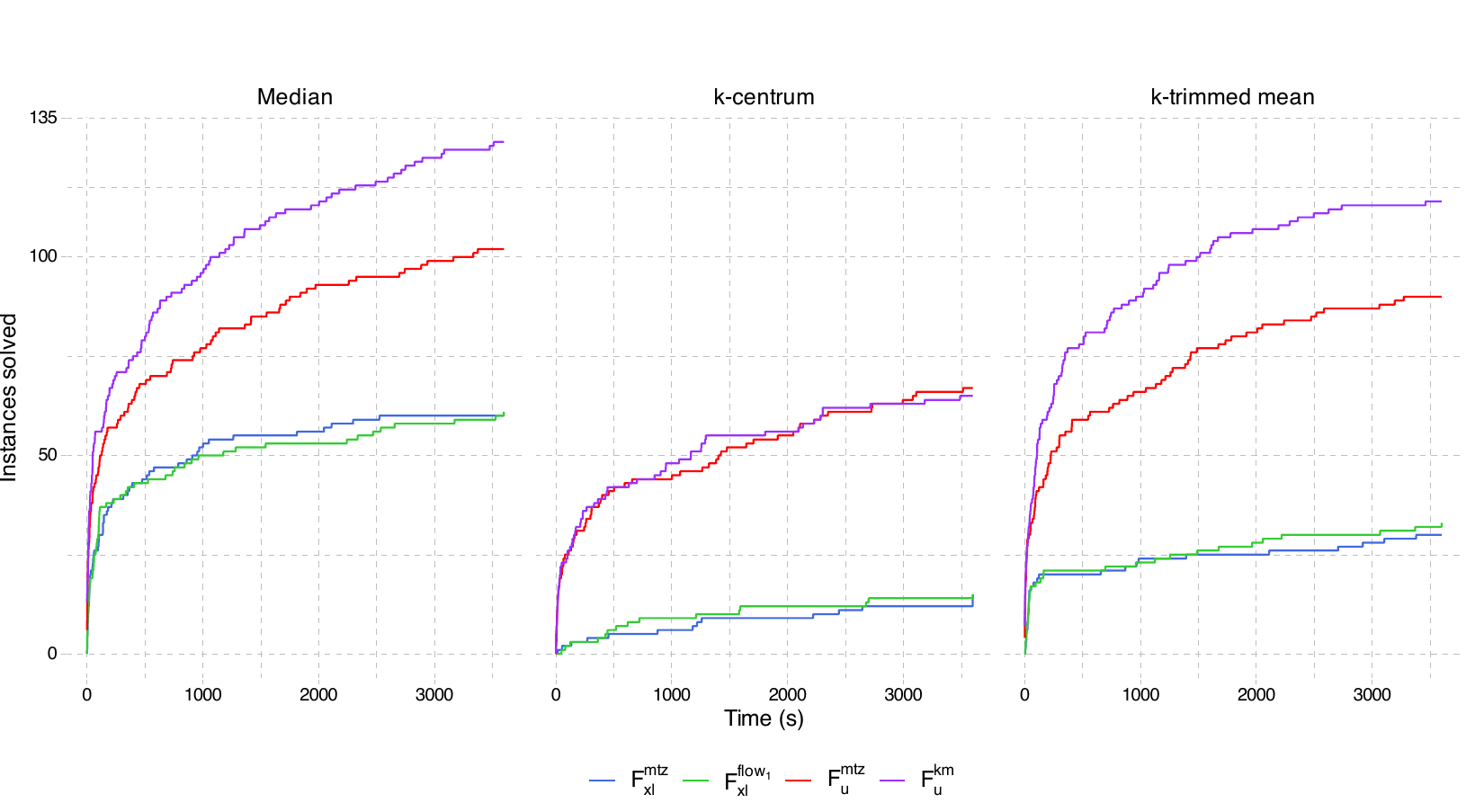}
\caption{Instances solved to optimality within a given time comparison for the final results}
\label{OMT_comparison_rev_4}
\end{figure}

Table \ref{sum_table_final} resumes the final results for mean \gapUL (\meangapUL) values and number of instances solved to optimality across all final runs, and Figure \ref{OMT_comparison_rev_4} depicts the number of instances solved to optimality within a given time. 
On the one hand, these results show the similar performance of the $F1^{(.)}_{x^\ell}$ formulations. 
With respect to the differences between the $F1^{(.)}_{u}$ formulations, in terms of the final number of instances solved and the time required (Figure \ref{OMT_comparison_rev_4}), $F1^{km}_{u}$ seems to perform better than $F1^{mtz}_{u}$. 
However, the poor performance of $F1^{km}_{u}$ regarding small $p$ values makes similar average gaps for $F1^{mtz}_{u}$ and $F1^{km}_{u}$, what makes it hard to draw significant conclusions in general.

\begin{table}[ht]
\centering
\begin{tabular}{c|rr|rr|rr}
    \hline
    \multirow{1}{*}{$\mathcal{T}$} & \multicolumn{2}{c|}{Median} & \multicolumn{2}{c|}{$k$-centrum} & \multicolumn{2}{c}{$k$-trimmed mean}  \\
                                   & $|\#|$ & \meangapUL & $|\#|$ & \meangapUL & $|\#|$ & \meangapUL \\ \hline
    $F1_{x^{\ell}}^{mtz}$          &    60  & 23.13      & 13     &  45.10     &   30   &     37.29    \\
    $F1_{x^{\ell}}^{flow1}$        &    60  & 22.40      & 14     &  43.74     &   32   &     36.21    \\
    $F1_{u}^{mtz}$                 &   102  &  1.25      & 67     &   6.23     &   90   &      2.10    \\
    $F1_{u}^{km}$                  &   129  &  1.49      & 65     &   9.25     &  114   &      1.84    \\
    \hline
\end{tabular}
\caption{Final results summary for mean \gapUL and total instances solved to optimality. }
\label{sum_table_final}
\end{table}

\section{Conclusions}
\label{sec:Conclusions}

In this paper we consider the \OMT, that is a single-allocation location problem where $p$ facilities must be placed on a network and connected by a non-directed tree. The \OMT is a complex network design problem that involves a number of components (network connectivity and sorting of allocation costs) each of which defines by itself a hard combinatorial optimization problem. The in/exclusion of these components give rise to different subproblems of the \OMT that are well-known problems of the literature. We have presented several formulations based on the minimum spanning tree and ordered median properties, as the use of covering variables. In order to improve formulations performance we introduce a series of improvements via providing an initial solution and the development of two preprocessings that reduce the size of covering formulations. We have also developed a Benders decomposition algorithm, which arises naturally in our problem, although it did not improve results of the exact formulations. Finally, we derive extensive computational results comparing in detail the different formulations, enhancements and solution techniques provided.

\section*{\small Acknowledgements}

\small This research was supported by the project "\textit{Nuevos resultados sobre los problemas de diseño y optimización en redes complejas: Aplicaciones al diseño de ciudades inteligentes}" (Proyecto I+D+i FEDER Andalucía, reference US-1256951), "\textit{Retos de la optimización combinatoria en los nuevos modelos de redes complejas y ciencia de datos}" (Proyecto I+D+i Junta de Andalucía, reference P18-FR-1422), "\textit{Optimization on data science and network design problems: Large scale network models meet optimization and data science tools}" (Plan Estatal 2017-2020 Generación Conocimiento - Proyectos I+D+i, reference PID2020-114594GB-C21) and "\textit{Complex networks meet data science}" (Proyectos Fundación "BBVA"). This support is gratefully acknowledged.

\begin{landscape}
\begin{table}[]
\centering
\scalebox{0.7}{
\begin{tabular}{cc|rrrrrrr|rrrrrrr|rrrrrrr|rrrrrrr}
\toprule
\multicolumn{2}{l|}{}  & 
\multicolumn{7}{c|}{$F1_{x^{\ell}}^{mtz}$} & 
\multicolumn{7}{c|}{$F1_{x^{\ell}}^{flow1}$} & 
\multicolumn{7}{c|}{$F1_{u}^{mtz}$} & 
\multicolumn{7}{c}{$F1_{u}^{km}$} \\[0.2cm] \hline
 $|V|$  &  $p$  & 
 $|\#|$ & $cpu$ & \gapbUR & \gapbUL & \gapUbL & \gapUL & $nod$ &
 $|\#|$ & $cpu$ & \gapbUR & \gapbUL & \gapUbL & \gapUL & $nod$ &
 $|\#|$ & $cpu$ & \gapbUR & \gapbUL & \gapUbL & \gapUL & $nod$ &
 $|\#|$ & $cpu$ & \gapbUR & \gapbUL & \gapUbL & \gapUL & $nod$ \\[0.2cm] \hline 
 20 &  5 & 5 & 11.8   & 37.0 &  0.0 &  0.0 &  0.0 &  951 & 5 & 6.3    & 37.0  &  0.0 &  0.0 &  0.0 & 284    & 5 & 1.5    & 37.0 & 0.0 & 0.0 & 0.0 & 1     & 5 & 0.9    & 37.0 & 0.0 & 0.0  & 0.0  & 1  \\ 
 20 &  6 & 5 & 12.2   & 35.0 &  0.0 &  0.0 &  0.0 &  762 & 5 & 10.2   & 35.0  &  0.0 &  0.0 &  0.0 & 741    & 5 & 1.3    & 35.0 & 0.0 & 0.0 & 0.0 & 1     & 5 & 0.7    & 35.0 & 0.0 & 0.0  & 0.0  & 0  \\ 
 20 & 10 & 5 & 6.4    & 28.9 &  0.0 &  0.0 &  0.0 &  516 & 5 & 5.7    & 28.9  &  0.0 &  0.0 &  0.0 & 404    & 5 & 1.2    & 28.9 & 0.0 & 0.0 & 0.0 & 34    & 5 & 0.4    & 28.5 & 0.0 & 0.0  & 0.0  & 0  \\ 
 30 &  7 & 5 & 127.5  & 39.6 &  0.0 &  0.0 &  0.0 & 1005 & 5 & 106.2  & 39.6  &  0.0 &  0.0 &  0.0 & 949    & 5 & 16.2   & 39.6 & 0.0 & 0.0 & 0.0 & 477   & 5 & 11.9   & 39.6 & 0.0 & 0.0  & 0.0  & 1  \\ 
 30 & 10 & 5 & 59.0   & 33.1 &  0.0 &  0.0 &  0.0 & 1110 & 5 & 62.4   & 33.1  &  0.0 &  0.0 &  0.0 & 1531   & 5 & 8.4    & 33.1 & 0.0 & 0.0 & 0.0 & 237   & 5 & 8.2    & 33.1 & 0.0 & 0.0  & 0.0  & 1  \\ 
 30 & 15 & 5 & 47.3   & 32.9 &  0.0 &  0.0 &  0.0 & 2090 & 5 & 44.2   & 33.0  &  0.0 &  0.0 &  0.0 & 710    & 5 & 16.4   & 32.9 & 0.0 & 0.0 & 0.0 & 2213  & 5 & 8.4    & 32.6 & 0.0 & 0.0  & 0.0  & 1  \\ 
 40 & 10 & 5 & 380.3  & 33.3 &  0.0 &  0.0 &  0.0 & 1669 & 5 & 502.2  & 33.3  &  0.0 &  0.0 &  0.0 & 2353   & 5 & 62.7   & 33.3 & 0.0 & 0.0 & 0.0 & 715   & 5 & 41.5   & 33.3 & 0.0 & 0.0  & 0.0  & 462 \\ 
 40 & 13 & 5 & 448.3  & 35.4 &  0.0 &  0.0 &  0.0 & 3080 & 5 & 701.1  & 35.4  &  0.0 &  0.0 &  0.0 & 6339   & 5 & 66.4   & 35.4 & 0.0 & 0.0 & 0.0 & 2400  & 5 & 30.1   & 35.4 & 0.0 & 0.0  & 0.0  & 1  \\ 
 40 & 20 & 5 & 141.2  & 28.6 &  0.0 &  0.0 &  0.0 & 1293 & 5 & 88.9   & 28.6  &  0.0 &  0.0 &  0.0 & 1611   & 5 & 40.9   & 28.6 & 0.0 & 0.0 & 0.0 & 2148  & 5 & 22.3   & 28.6 & 0.0 & 0.0  & 0.0  & 1  \\ 
 50 & 12 & 2 & 2944.8 & 35.8 &  1.4 & 14.0 & 15.1 & 2707 & 4 & 2913.2 & 35.8  &  0.5 &  4.5 &  4.9 & 2874   & 5 & 276.8  & 35.8 & 0.0 & 0.0 & 0.0 & 1165  & 5 & 153.7  & 35.8 & 0.0 & 0.0  & 0.0  & 116 \\ 
 50 & 16 & 5 & 1342.5 & 35.7 &  0.0 &  0.0 &  0.0 & 1908 & 4 & 1563.5 & 35.7  &  0.1 &  0.9 &  1.0 & 2769   & 5 & 145.1  & 35.7 & 0.0 & 0.0 & 0.0 & 1615  & 5 & 51.5   & 35.7 & 0.0 & 0.0  & 0.0  & 1  \\ 
 50 & 25 & 5 & 473.8  & 28.6 &  0.0 &  0.0 &  0.0 & 3616 & 4 & 927.2  & 28.6  &  0.5 &  0.0 &  0.5 & 12136  & 5 & 83.9   & 28.6 & 0.0 & 0.0 & 0.0 & 2028  & 5 & 44.6   & 28.5 & 0.0 & 0.0  & 0.0  & 1 \\ 
 60 & 15 & 0 & 3601.8 & 40.9 &  6.0 & 40.6 & 44.1 & 60   & 0 & 3601.5 & 40.9  &  4.3 & 31.4 & 34.4 &   39   & 5 & 827.7  & 40.9 & 0.0 & 0.0 & 0.0 & 1571  & 5 & 618.9  & 40.9 & 0.0 & 0.0  & 0.0  & 16 \\ 
 60 & 20 & 0 & 3600.9 & 35.5 &  4.5 & 24.9 & 28.2 & 592  & 0 & 3600.4 & 35.5  &  3.1 & 22.4 & 24.6 &  126   & 5 & 594.0  & 35.5 & 0.0 & 0.0 & 0.0 & 4050  & 5 & 375.4  & 35.5 & 0.0 & 0.0  & 0.0  & 20 \\ 
 60 & 30 & 1 & 3090.7 & 33.1 &  1.8 &  0.7 &  2.6 & 5686 & 1 & 3347.5 & 33.1  &  1.4 &  1.4 &  2.7 & 4322   & 5 & 820.2  & 33.1 & 0.0 & 0.0 & 0.0 & 6466  & 5 & 150.8  & 32.8 & 0.0 & 0.0  & 0.0  & 1 \\ 
 70 & 17 & 0 & 3603.1 & 37.5 &  8.7 & 58.1 & 61.8 & 14   & 0 & 3602.2 & 37.5  &  8.6 & 62.5 & 65.7 &    2   & 4 & 2571.9 & 37.5 & 0.6 & 0.0 & 0.6 & 2403  & 5 & 1468.3 & 37.5 & 0.0 & 0.0  & 0.0  & 25 \\ 
 70 & 23 & 0 & 3601.9 & 37.1 &  3.7 & 31.2 & 33.6 & 46   & 1 & 3375.2 & 37.1  &  3.7 & 24.7 & 27.2 &  125   & 4 & 2175.1 & 37.1 & 0.3 & 0.0 & 0.3 & 3503  & 5 & 371.2  & 37.1 & 0.0 & 0.0  & 0.0  & 1 \\ 
 70 & 35 & 2 & 3029.1 & 30.8 &  0.7 &  4.7 &  5.4 & 1224 & 0 & 3601.6 & 30.8  &  1.6 & 14.5 & 15.8 &  477   & 3 & 1933.5 & 30.8 & 1.2 & 0.0 & 1.2 & 9547  & 5 & 449.0  & 30.7 & 0.0 & 0.0  & 0.0  & 1 \\ 
 80 & 20 & 0 & 3600.7 & 39.3 & 11.6 & 61.1 & 65.6 & 3    & 0 & 3600.1 & 39.3  & 10.4 & 59.6 & 63.9 &    1   & 1 & 3365.4 & 39.3 & 2.6 & 2.2 & 4.7 & 3284  & 5 & 2076.7 & 39.3 & 0.0 & 0.0  & 0.0  & 607 \\ 
 80 & 26 & 0 & 3606.4 & 37.0 &  6.0 & 43.5 & 46.9 & 13   & 0 & 3602.0 & 37.0  &  4.6 & 39.3 & 42.0 &    4   & 3 & 2432.1 & 37.0 & 1.7 & 0.3 & 2.0 & 2327  & 5 & 1248.0 & 37.0 & 0.0 & 0.0  & 0.0  & 1 \\ 
 80 & 40 & 0 & 3601.9 & 31.2 &  3.6 & 19.0 & 21.8 & 740  & 1 & 3517.3 & 31.2  &  1.9 & 20.6 & 21.9 &  511   & 3 & 2016.6 & 31.2 & 1.0 & 0.0 & 1.0 & 12871 & 5 & 754.3  & 31.0 & 0.0 & 0.0  & 0.0  & 1 \\ 
 90 & 22 & 0 & 3600.3 & 41.7 & 14.5 & 59.9 & 64.3 &    1 & 0 & 3600.3 & 41.7  & 12.9 & 63.1 & 66.6 &    1   & 2 & 3097.7 & 41.7 & 5.9 & 5.4 & 7.5 & 1930  & 4 & 2873.8 & 41.7 & 3.7 & 9.4  & 9.4  & 456 \\ 
 90 & 30 & 0 & 3602.0 & 38.4 &  6.4 & 47.3 & 50.9 &    1 & 0 & 3600.8 & 38.4  &  4.7 & 47.6 & 50.1 &    1   & 2 & 3163.1 & 38.4 & 1.6 & 0.4 & 2.0 & 4826  & 5 & 1423.9 & 38.4 & 0.0 & 0.0  & 0.0  & 1 \\ 
 90 & 45 & 0 & 3601.2 & 29.3 &  2.6 & 32.4 & 34.2 &    2 & 0 & 3600.6 & 29.3  &  1.1 & 28.8 & 29.5 &   60   & 3 & 2433.2 & 29.3 & 0.4 & 0.2 & 0.6 & 18184 & 5 & 1140.6 & 29.3 & 0.0 & 0.0  & 0.0  & 1 \\ 
100 & 25 & 0 & 3601.5 & 39.4 & 12.9 & 66.5 & 69.2 &    1 & 0 & 3600.3 & 39.4  & 11.6 & 57.7 & 60.7 &    1   & 0 & 3614.1 & 39.4 & 7.8 & 5.9 & 8.5 & 1241  & 1 & 3432.3 & 39.4 & 5.7 & 30.0 & 30.2 & 1 \\ 
100 & 33 & 0 & 3601.2 & 35.4 &  6.6 & 58.5 & 61.2 &    1 & 0 & 3600.3 & 35.4  &  5.2 & 50.5 & 52.9 &    1   & 0 & 3606.4 & 35.4 & 2.7 & 0.7 & 3.2 & 861   & 4 & 2339.8 & 35.4 & 0.2 & 0.7  & 0.7  & 1 \\ 
100 & 50 & 0 & 3600.3 & 29.7 &  1.9 & 17.9 & 19.5 &    1 & 0 & 3603.1 & 29.7  &  1.4 & 39.4 & 40.3 &    1   & 2 & 2959.8 & 29.7 & 2.2 & 0.0 & 2.2 & 10014 & 5 & 2095.2 & 29.7 & 0.0 & 0.0  & 0.0  & 1 \\ 
\bottomrule
\end{tabular}}
\caption{\OMT summary results for median criterion}
\label{re:OMTcmedian}
\end{table}
\end{landscape}

\begin{landscape}
\begin{table}[]
\centering
\scalebox{0.7}{
\begin{tabular}{cc|rrrrrrr|rrrrrrr|rrrrrrr|rrrrrrr}
\toprule
\multicolumn{2}{l|}{}  &
\multicolumn{7}{c|}{$F1_{x^{\ell}}^{mtz}$} &
\multicolumn{7}{c|}{$F1_{x^{\ell}}^{flow1}$} &
\multicolumn{7}{c|}{$F1_{u}^{mtz}$} &
\multicolumn{7}{c}{$F1_{u}^{km}$} \\[0.2cm] \hline
  $|V|$  &  $p$  &
  $|\#|$ & $cpu$ & \gapbUR & \gapbUL & \gapUbL & \gapUL & $nod$ &
  $|\#|$ & $cpu$ & \gapbUR & \gapbUL & \gapUbL & \gapUL & $nod$ &
  $|\#|$ & $cpu$ & \gapbUR & \gapbUL & \gapUbL & \gapUL & $nod$ &
  $|\#|$ & $cpu$ & \gapbUR & \gapbUL & \gapUbL & \gapUL & $nod$ \\[0.2cm] \hline
  20 &   5 & 5 & 504.4  & 62.7 & 0.0  & 0.0  & 0.0  & 64813  & 5 & 307.3  & 62.7 & 0.0  & 0.0  & 0.0  & 47977  & 5 & 8.7    & 62.7 &  0.0 &  0.0 &  0.0 & 2824  & 5 & 10.7   & 62.7 & 0.0  & 0.0  & 0.0  & 715 \\
  20 &   6 & 2 & 2739.8 & 65.0 & 6.0  & 0.0  & 6.0  & 319609 & 4 & 1981.0 & 65.0 & 2.5  & 0.0  & 2.5  & 204699 & 5 & 15.2   & 65.0 &  0.0 &  0.0 &  0.0 & 3192  & 5 & 13.0   & 65.0 & 0.0  & 0.0  & 0.0  & 983 \\
  20 &  10 & 5 & 1472.8 & 56.8 & 0.0  & 0.0  & 0.0  & 260421 & 5 & 1056.2 & 56.8 & 0.0  & 0.0  & 0.0  & 185366 & 5 & 7.9    & 56.8 &  0.0 &  0.0 &  0.0 & 2312  & 5 & 2.6    & 56.5 & 0.0  & 0.0  & 0.0  & 1 \\
  30 &   7 & 0 & 3600.4 & 65.4 & 30.5 & 3.2  & 32.7 & 31958  & 0 & 3600.2 & 65.4 & 28.9 & 1.0  & 29.6 & 33589  & 5 & 188.9  & 65.4 &  0.0 &  0.0 &  0.0 & 7802  & 5 & 136.4  & 65.4 & 0.0  & 0.0  & 0.0  & 2938 \\
  30 &  10 & 0 & 3600.3 & 61.2 & 25.5 & 0.3  & 25.7 & 36798  & 0 & 3600.4 & 61.2 & 24.0 & 0.5  & 24.4 & 39749  & 5 & 248.8  & 61.2 &  0.0 &  0.0 &  0.0 & 11008 & 5 & 95.4   & 61.2 & 0.0  & 0.0  & 0.0  & 2013 \\
  30 &  15 & 1 & 3600.2 & 59.9 & 13.4 & 0.0  & 13.4 & 235432 & 0 & 3600.2 & 59.9 & 11.6 & 0.0  & 11.6 & 165266 & 5 & 66.4   & 59.9 &  0.0 &  0.0 &  0.0 & 7673  & 5 & 28.6   & 59.7 & 0.0  & 0.0  & 0.0  & 623 \\
  40 &  10 & 0 & 3600.3 & 63.8 & 37.0 & 3.8  & 39.3 & 16941  & 0 & 3600.2 & 63.8 & 37.1 & 5.3  & 40.2 & 12991  & 5 & 1237.8 & 63.8 &  0.0 &  0.0 &  0.0 & 45176 & 5 & 621.2  & 63.8 & 0.0  & 0.0  & 0.0  & 13081 \\
  40 &  13 & 0 & 3600.2 & 61.0 & 31.9 & 7.0  & 36.8 & 15750  & 0 & 3600.3 & 61.0 & 31.4 & 3.2  & 33.6 & 15149  & 5 & 529.7  & 61.0 &  0.0 &  0.0 &  0.0 & 17580 & 5 & 285.2  & 61.0 & 0.0  & 0.0  & 0.0  & 4850 \\
  40 &  20 & 0 & 3600.3 & 59.0 & 27.0 & 0.3  & 27.2 & 41430  & 0 & 3600.6 & 59.0 & 26.0 & 0.6  & 26.4 & 40789  & 4 & 908.6  & 59.0 &  1.0 &  0.0 &  1.0 & 26517 & 5 & 334.3  & 59.0 & 0.0  & 0.0  & 0.0  & 3828 \\
  50 &  12 & 0 & 3600.7 & 63.6 & 44.0 & 22.8 & 55.4 & 2328   & 0 & 3601.0 & 63.6 & 42.8 & 22.3 & 54.4 & 2502   & 1 & 3426.7 & 63.6 &  4.7 &  2.9 &  4.8 & 28627 & 1 & 3337.1 & 63.6 & 3.5  & 3.1  & 3.9  & 14862 \\
  50 &  16 & 0 & 3600.5 & 63.5 & 40.5 & 19.3 & 52.1 & 2830   & 0 & 3600.7 & 63.5 & 39.8 & 8.8  & 45.1 & 1844   & 4 & 1955.3 & 63.5 &  0.4 &  0.0 &  0.4 & 23990 & 5 & 1311.3 & 63.5 & 0.0  & 0.0  & 0.0  & 5234 \\
  50 &  25 & 0 & 3600.2 & 56.7 & 32.3 & 7.2  & 37.2 & 5241   & 0 & 3600.2 & 56.7 & 32.0 & 2.4  & 33.6 & 3435   & 4 & 1034.1 & 56.7 &  0.2 &  0.0 &  0.2 & 32517 & 5 & 1000.7 & 56.6 & 0.0  & 0.0  & 0.0  & 1781 \\
  60 &  15 & 0 & 3601.0 & 67.4 & 47.9 & 31.9 & 62.2 & 114    & 0 & 3601.2 & 67.4 & 47.0 & 32.4 & 61.7 & 97     & 3 & 3080.3 & 67.4 &  2.1 &  6.3 &  2.2 & 24493 & 0 & 3601.4 & 67.4 & 13.3 &  6.2 & 13.4 & 5031 \\ 
  60 &  20 & 0 & 3600.6 & 61.7 & 43.4 & 17.9 & 52.9 & 2079   & 0 & 3600.6 & 61.7 & 42.0 & 18.9 & 52.5 & 1745   & 3 & 3012.2 & 61.7 &  1.2 &  1.1 &  1.2 & 28959 & 2 & 3026.0 & 61.7 &  6.4 &  1.1 &  6.4 & 4690 \\ 
  60 &  30 & 0 & 3600.5 & 60.9 & 37.1 &  6.3 & 40.8 & 5730   & 0 & 3600.5 & 60.9 & 37.1 &  5.3 & 40.2 & 4108   & 3 & 2053.1 & 60.9 &  0.7 &  0.3 &  0.7 & 36946 & 5 & 2061.6 & 60.9 &  0.0 &  0.3 &  0.0 & 5250 \\ 
  70 &  17 & 0 & 3603.9 & 63.9 & 46.6 & 49.0 & 69.2 & 24     & 0 & 3602.0 & 63.9 & 47.1 & 51.6 & 71.0 & 15     & 0 & 3602.9 & 63.9 & 12.3 & 12.2 & 12.6 & 7941  & 0 & 3601.8 & 63.9 & 18.2 & 11.8 & 18.1 & 1629 \\ 
  70 &  23 & 0 & 3602.5 & 62.7 & 42.7 & 22.7 & 55.2 & 221    & 0 & 3602.8 & 62.7 & 41.8 & 24.8 & 55.9 & 22     & 2 & 3246.9 & 62.7 &  4.8 &  1.5 &  5.2 & 7571  & 0 & 3602.5 & 62.7 & 11.6 &  1.7 & 12.1 & 1859 \\ 
  70 &  35 & 0 & 3600.9 & 59.9 & 42.3 &  8.8 & 47.3 & 1149   & 0 & 3600.8 & 59.9 & 41.7 &  7.9 & 46.3 & 285    & 1 & 3145.3 & 59.9 &  2.9 &  0.0 &  2.9 & 24419 & 2 & 3496.1 & 59.9 &  4.5 &  0.0 &  4.5 & 3371 \\ 
  80 &  20 & 0 & 3600.3 & 65.7 & 49.8 & 55.2 & 72.2 & 10     & 0 & 3600.8 & 65.7 & 49.2 & 55.3 & 71.9 & 2      & 0 & 3603.7 & 65.7 & 22.3 & 27.2 & 29.9 & 4682  & 0 & 3604.0 & 65.7 & 21.2 & 19.7 & 21.8 & 44 \\ 
  80 &  26 & 0 & 3601.8 & 64.8 & 47.1 & 35.0 & 62.6 & 6      & 0 & 3601.8 & 64.8 & 46.3 & 32.9 & 60.7 & 3      & 0 & 3603.3 & 64.8 & 14.6 &  8.8 & 15.2 & 3180  & 0 & 3602.2 & 64.8 & 17.0 & 14.0 & 22.3 & 14 \\ 
  80 &  40 & 0 & 3602.5 & 60.3 & 42.0 & 16.8 & 51.5 & 44     & 0 & 3601.0 & 60.3 & 42.1 & 18.7 & 52.9 & 16     & 1 & 3426.6 & 60.3 &  2.7 &  0.0 &  2.7 & 13274 & 0 & 3600.8 & 60.3 &  6.5 &  0.0 &  6.5 & 2172 \\ 
  90 &  22 & 0 & 3602.1 & 66.1 & 50.2 & 51.4 & 69.9 & 3      & 0 & 3600.6 & 66.1 & 49.0 & 52.4 & 69.9 & 1      & 0 & 3600.3 & 66.1 & 19.7 & 20.0 & 20.4 & 3377  & 0 & 3605.0 & 66.1 & 21.2 & 22.0 & 23.7 & 445 \\ 
  90 &  30 & 0 & 3603.1 & 64.4 & 44.4 & 31.8 & 59.1 & 15     & 0 & 3601.9 & 64.4 & 44.2 & 28.7 & 57.0 & 5      & 0 & 3600.6 & 64.4 & 10.3 &  7.4 & 10.5 & 4525  & 0 & 3600.8 & 64.4 & 14.6 &  8.1 & 15.4 & 21 \\ 
  90 &  45 & 0 & 3600.8 & 58.3 & 41.7 & 19.4 & 53.1 & 62     & 0 & 3602.0 & 58.3 & 39.9 & 17.5 & 50.4 & 6      & 1 & 3263.6 & 58.3 &  1.9 &  0.0 &  1.9 & 14687 & 0 & 3600.3 & 58.3 &  9.4 &  0.3 &  9.7 & 502 \\ 
 100 &  25 & 0 & 3601.0 & 65.4 & 50.2 & 53.8 & 70.3 & 1      & 0 & 3600.2 & 65.4 & 49.2 & 54.4 & 70.1 & 1      & 0 & 3602.1 & 65.4 & 24.4 & 27.4 & 29.2 & 2126  & 0 & 3601.4 & 65.4 & 24.0 & 46.1 & 47.1 & 3 \\ 
 100 &  33 & 0 & 3603.2 & 63.0 & 46.1 & 47.6 & 65.7 & 1      & 0 & 3601.4 & 63.0 & 45.6 & 44.5 & 63.2 & 1      & 0 & 3601.4 & 63.0 & 18.1 & 17.9 & 18.3 & 2024  & 0 & 3602.5 & 63.0 & 17.9 & 22.7 & 22.9 & 1 \\ 
 100 &  50 & 0 & 3602.8 & 59.4 & 44.0 & 34.7 & 59.9 & 1      & 0 & 3600.8 & 59.4 & 43.2 & 29.3 & 55.8 & 1      & 0 & 3600.9 & 59.4 &  8.8 &  8.8 &  8.8 & 2973  & 0 & 3604.6 & 59.4 & 14.8 & 16.7 & 22.0 & 1 \\ 
\bottomrule
\end{tabular}}
\caption{\OMT results summary for $k$-centrum criterion}
\label{re:OMTckcentrum}
\end{table}
\end{landscape}

\begin{landscape}
\begin{table}[]
\centering
\scalebox{0.7}{
\begin{tabular}{cc|rrrrrrr|rrrrrrr|rrrrrrr|rrrrrrr}
\toprule
\multicolumn{2}{l|}{}  &
\multicolumn{7}{c|}{$F1_{x^{\ell}}^{mtz}$} &
\multicolumn{7}{c|}{$F1_{x^{\ell}}^{flow1}$} &
\multicolumn{7}{c|}{$F1_{u}^{mtz}$} &
\multicolumn{7}{c}{$F1_{u}^{km}$} \\[0.2cm] \hline
  $|V|$  &  $p$  &
  $|\#|$ & $cpu$ & \gapbUR & \gapbUL & \gapUbL & \gapUL & $nod$ &
  $|\#|$ & $cpu$ & \gapbUR & \gapbUL & \gapUbL & \gapUL & $nod$ &
  $|\#|$ & $cpu$ & \gapbUR & \gapbUL & \gapUbL & \gapUL & $nod$ &
  $|\#|$ & $cpu$ & \gapbUR & \gapbUL & \gapUbL & \gapUL & $nod$ \\[0.2cm] \hline
  20 &  5 & 5 & 36.6   & 50.7 & 0.0  & 0.0  & 0.0  & 2762  & 5 & 26.2   & 50.7 & 0.0  & 0.0  & 0.0  & 3047  & 5 & 3.0    & 50.7 &  0.0 &  0.0 &  0.0 & 125  & 5 & 3.8    & 50.7 &  0.0 &  0.0 &  0.0 & 56 \\
  20 &  6 & 5 & 25.9   & 55.2 & 0.0  & 0.0  & 0.0  & 2779  & 5 & 30.1   & 55.2 & 0.0  & 0.0  & 0.0  & 1641  & 5 & 1.3    & 55.2 &  0.0 &  0.0 &  0.0 & 1    & 5 & 0.9    & 55.2 &  0.0 &  0.0 &  0.0 & 0 \\
  20 & 10 & 5 & 13.2   & 30.2 & 0.0  & 0.0  & 0.0  & 2517  & 5 & 13.7   & 30.2 & 0.0  & 0.0  & 0.0  & 5279  & 5 & 2.1    & 30.2 &  0.0 &  0.0 &  0.0 & 1    & 5 & 1.2    & 29.9 &  0.0 &  0.0 &  0.0 & 0 \\
  30 &  7 & 3 & 3320.5 & 63.8 & 7.1  & 0.0  & 7.1  & 73408 & 2 & 3245.6 & 63.8 & 6.4  & 0.0  & 6.4  & 88561 & 5 & 36.1   & 63.8 &  0.0 &  0.0 &  0.0 & 1529 & 5 & 30.6   & 63.8 &  0.0 &  0.0 &  0.0 & 413 \\
  30 & 10 & 3 & 2334.0 & 52.7 & 4.1  & 0.0  & 4.1  & 87074 & 4 & 2021.7 & 52.7 & 2.1  & 0.0  & 2.1  & 89506 & 5 & 10.4   & 52.7 &  0.0 &  0.0 &  0.0 & 16   & 5 & 6.9    & 52.6 &  0.0 &  0.0 &  0.0 & 1 \\
  30 & 15 & 5 & 183.7  & 36.5 & 0.0  & 0.0  & 0.0  & 12498 & 5 & 95.0   & 36.5 & 0.0  & 0.0  & 0.0  & 4367  & 5 & 19.0   & 36.5 &  0.0 &  0.0 &  0.0 & 878  & 5 & 15.4   & 36.5 &  0.0 &  0.0 &  0.0 & 1 \\
  40 & 10 & 0 & 3600.2 & 62.2 & 22.8 & 0.1  & 22.9 & 28542 & 1 & 3323.9 & 62.2 & 18.4 & 0.0  & 18.4 & 22141 & 5 & 174.4  & 62.2 &  0.0 &  0.0 &  0.0 & 2423 & 5 & 94.6   & 62.2 &  0.0 &  0.0 &  0.0 & 975 \\
  40 & 13 & 0 & 3600.1 & 53.9 & 19.9 & 1.8  & 21.2 & 24033 & 0 & 3600.3 & 53.9 & 19.2 & 1.2  & 20.1 & 26373 & 5 & 102.4  & 53.9 &  0.0 &  0.0 &  0.0 & 919  & 5 & 107.7  & 53.9 &  0.0 &  0.0 &  0.0 & 14 \\
  40 & 20 & 3 & 2202.7 & 32.1 & 1.5  & 0.0  & 1.5  & 34646 & 5 & 1176.4 & 32.1 & 0.0  & 0.0  & 0.0  & 20527 & 5 & 89.3   & 32.1 &  0.0 &  0.0 &  0.0 & 833  & 5 & 89.9   & 32.1 &  0.0 &  0.0 &  0.0 & 1 \\
  50 & 12 & 0 & 3600.2 & 58.0 & 25.6 & 12.5 & 34.8 & 1488  & 0 & 3601.0 & 58.0 & 24.9 & 9.9  & 32.3 & 2793  & 5 & 925.2  & 58.0 &  0.0 &  0.0 &  0.0 & 5295 & 5 & 474.6  & 58.0 &  0.0 &  0.0 &  0.0 & 18481 \\
  50 & 16 & 0 & 3600.1 & 55.2 & 18.0 & 1.8  & 19.5 & 5109  & 0 & 3600.1 & 55.2 & 19.2 & 2.7  & 21.3 & 4825  & 5 & 426.9  & 55.2 &  0.0 &  0.0 &  0.0 & 696  & 5 & 84.7   & 55.0 &  0.0 &  0.0 &  0.0 & 1 \\
  50 & 25 & 1 & 3054.2 & 32.2 & 4.0  & 0.0  & 4.0  & 12362 & 0 & 3600.3 & 32.2 & 5.7  & 0.0  & 5.7  & 15313 & 5 & 265.5  & 32.2 &  0.0 &  0.0 &  0.0 & 1503 & 5 & 92.8   & 32.1 &  0.0 &  0.0 &  0.0 & 1 \\
  60 & 15 & 0 & 3600.8 & 65.5 & 33.8 & 48.2 & 65.8 & 3583  & 0 & 3600.7 & 65.5 & 33.0 & 39.5 & 59.6 & 1338  & 5 & 1921.8 & 65.5 &  0.0 &  0.0 &  0.0 & 5514 & 5 & 1105.8 & 65.5 &  0.0 &  0.0 &  0.0 & 3910 \\ 
  60 & 20 & 0 & 3600.5 & 53.2 & 22.0 & 15.9 & 34.4 & 1531  & 0 & 3600.5 & 53.2 & 24.6 & 18.6 & 38.7 & 2728  & 1 & 3057.5 & 53.2 &  1.6 &  0.4 &  2.0 & 486  & 5 & 273.6  & 53.2 &  0.0 &  0.0 &  0.0 & 1 \\ 
  60 & 30 & 0 & 3600.3 & 32.7 &  9.3 &  0.2 &  9.4 & 2178  & 0 & 3600.5 & 32.7 &  9.4 &  0.6 & 10.0 & 2325  & 5 & 1576.9 & 32.7 &  0.0 &  0.0 &  0.0 & 2626 & 5 & 155.8  & 32.7 &  0.0 &  0.0 &  0.0 & 1 \\ 
  70 & 17 & 0 & 3602.0 & 57.5 & 31.0 & 64.4 & 75.3 & 51    & 0 & 3602.9 & 57.5 & 30.2 & 59.9 & 71.8 & 52    & 3 & 2787.4 & 57.5 &  0.9 &  0.0 &  0.9 & 4735 & 2 & 2589.8 & 57.5 &  1.3 &  0.0 &  1.3 & 5319 \\ 
  70 & 23 & 0 & 3601.6 & 51.1 & 26.5 & 44.1 & 58.8 & 387   & 0 & 3601.1 & 51.1 & 25.5 & 31.6 & 49.3 & 136   & 2 & 2550.9 & 51.1 &  2.2 &  0.0 &  2.2 & 535  & 5 & 293.3  & 51.1 &  0.0 &  0.0 &  0.0 & 1 \\ 
  70 & 35 & 0 & 3600.6 & 30.9 &  9.2 &  6.8 & 15.4 & 568   & 0 & 3621.6 & 30.9 & 10.2 &  1.9 & 11.9 & 920   & 5 & 1333.7 & 30.9 &  0.0 &  0.0 &  0.0 & 2683 & 5 & 274.8  & 30.9 &  0.0 &  0.0 &  0.0 & 1 \\ 
  80 & 20 & 0 & 3603.8 & 60.4 & 28.9 & 71.0 & 79.4 & 27    & 0 & 3604.6 & 60.4 & 28.4 & 70.9 & 78.9 & 18    & 0 & 3602.2 & 60.4 &  6.2 & 11.8 &  7.6 & 1854 & 1 & 3406.0 & 60.4 &  7.3 &  9.1 &  2.4 & 2759 \\ 
  80 & 26 & 0 & 3603.0 & 50.3 & 22.5 & 67.3 & 75.7 & 32    & 0 & 3602.3 & 50.3 & 22.1 & 66.7 & 74.4 & 19    & 3 & 2141.3 & 50.3 & 12.9 & 11.9 &  1.9 & 556  & 4 & 1423.3 & 50.3 & 13.5 & 12.1 &  0.3 & 13 \\ 
  80 & 40 & 0 & 3601.1 & 32.8 & 13.3 & 34.1 & 33.1 & 94    & 0 & 3601.6 & 32.8 & 13.7 & 30.2 & 38.3 & 42    & 3 & 2554.8 & 32.8 & 13.8 & 15.6 &  0.3 & 2390 & 5 & 604.6  & 32.8 & 14.2 & 15.6 &  0.0 & 1 \\ 
  90 & 22 & 0 & 3601.7 & 64.4 & 38.2 & 80.0 & 86.4 & 1     & 0 & 3602.9 & 64.4 & 35.9 & 73.9 & 81.7 & 1     & 0 & 3605.1 & 64.4 & 11.0 &  8.4 & 11.4 & 1794 & 0 & 3613.4 & 64.4 &  8.3 & 18.6 & 18.7 & 11 \\ 
  90 & 30 & 0 & 3604.5 & 53.2 & 28.5 & 69.5 & 78.2 & 9     & 0 & 3602.5 & 53.2 & 28.6 & 64.7 & 74.6 & 6     & 2 & 2514.5 & 53.2 &  1.6 &  0.6 &  1.9 & 41   & 4 & 2725.8 & 53.2 &  0.3 &  0.3 &  0.3 & 1 \\ 
  90 & 45 & 0 & 3603.5 & 32.2 & 14.1 & 36.8 & 45.8 & 24    & 0 & 3601.4 & 32.2 & 14.2 & 51.6 & 58.4 & 2     & 0 & 3604.1 & 32.2 &  1.9 &  0.0 &  1.9 & 1981 & 5 & 1628.7 & 32.2 &  0.0 &  0.0 &  0.0 & 1 \\ 
 100 & 25 & 0 & 3602.0 & 62.8 & 37.3 & 75.8 & 83.2 & 1     & 0 & 3601.7 & 62.8 & 37.0 & 71.0 & 79.7 & 1     & 0 & 3602.2 & 62.8 & 12.2 & 20.6 & 22.5 & 38   & 0 & 3603.4 & 62.8 & 10.6 & 24.6 & 25.2 & 4 \\ 
 100 & 33 & 0 & 3601.4 & 53.8 & 28.7 & 78.5 & 84.5 & 1     & 0 & 3602.1 & 53.8 & 27.8 & 71.3 & 79.1 & 1     & 0 & 3675.0 & 53.8 &  3.2 &  1.4 &  3.2 & 77   & 4 & 2456.0 & 53.8 &  1.4 &  1.4 &  1.4 & 5 \\ 
 100 & 50 & 0 & 3601.6 & 30.1 & 14.7 & 60.3 & 66.2 & 2     & 0 & 3601.4 & 30.1 & 13.8 & 59.3 & 65.0 & 1     & 1 & 3246.0 & 30.1 &  1.1 &  0.0 &  1.1 & 658  & 4 & 2055.2 & 30.1 &  0.1 &  0.2 &  0.2 & 261 \\ 
\bottomrule
\end{tabular}}
\caption{\OMT results summary for $k$-trimmed mean criterion}
\label{re:OMTcktrimmean}
\end{table}
\end{landscape}

\newpage

\appendix

\renewcommand{\thesection}{} 
\renewcommand{\thesubsection}{\Alph{section}.\arabic{subsection}}

\newpage

\section{Appendix A: Extended information about the \OMT Benders decomposition}
\label{AppendixA}

Our mathematical formulation can be solved using a Benders decomposition framework (see \citealp{Benders1962}) that we briefly described in this section.

\subsection{Classical Benders decomposition}
\label{subsec:ClassicalBendersdecomposition}

In the classical Benders decomposition algorithm, the original mixed integer problem is divided into two problems, a master problem (MP) and a subproblem (SP), that are solved iteratively. The two problems are related, so the outcome of one directly modifies the outcome of the other. First, the MP is solved to obtain the values of certain fixed variables; with these, we can then solve the SP. Once the SP has been solved, either new feasibility or new optimality cuts are introduced within the MP until the lower and upper bounds coincide. For the \OMT, it arises to use DOMP as MP and MST as SP. For instance, we can illustrate the MP via $F1_{x^\ell}^{\mT}$ as follows:

\begingroup
\allowdisplaybreaks
\begin{subequations}
\begin{align}
  & \hhp \hspace{1cm} F^{MP}: \min \quad \frac{1}{\displaystyle{\sum_{\ell \in V} \lambda_\ell}} \sum_{\ell \in V} \sum_{(i,j) \in A} \lambda_\ell c_{ij} x_{ij}^\ell + \mu & & & \label{OMT_f1b_c0} & \\
  & \hp \sum_{i \in V} x_{ii} = p                     &   &                            & \label{OMT_f1b_c1}  & \\
  & \Hp \sum_{j \in V} x_{ij} = 1                     &   & i \in V                    & \label{OMT_f1b_c2}  & \\
  & \Hp x_{ij} \leq x_{jj}                            &   & (i,j) \in A: i\neq j       & \label{OMT_f1b_c3}  & \\
  & \Hp x_{ij} = \sum_{\ell \in V} x_{ij}^\ell        &   & (i,j) \in A                & \label{OMT_f1b_c6}  & \\
  & \Hp \sum_{(i,j) \in A} x_{ij}^\ell = 1            &   & \ell \in V                 & \label{OMT_f1b_c7}  & \\
  & \Hp \sum_{(i,j) \in A} c_{ij}x_{ij}^\ell \leq \sum_{(i,j) \in A} c_{ij}x_{ij}^{\ell+1} &   & \ell \in V: \ell < |V|  & \label{OMT_f1b_c8} & \\
  & \Hp x_{ij} \in \{0,1\}                            &   & (i,j) \in A \\
  & \Hp x_{ij}^\ell \in \{0,1\}                       &   & (i,j) \in A,\ell \in V \\
  & \Hp \mu \geq 0.  
\end{align}
\end{subequations}
\endgroup

Similar formulations of the MP can be introduced using other formulations reviewed in Section \ref{sec:OMTformulations}. 

A classical Benders decomposition framework is then illustrated in Algorithm \ref{alg:Benders_classical}. For this algorithm we keep the best upper and lower bounds found up to each iteration, $UB$ and $LB$. The procedure is then as follows. The MP is solved to optimality and both the objective function $obj^{MP}$, which can be split into the sum of the OM objective ($obj^{OM}$) plus the $\mu$ term, and the variables representing the facilities selected $\bar{x}= \{x_{ii} \ | \ x_{ii}=1, \forall i \in V \}$ are stored. At this moment, if possible, update the lower bound using the MP objective. Thereafter, solve the MST subproblem setting as facilities the previously $\bar{x}$ identified in the MP solution and, if possible, update the upper bound using the (weighted) sum of the OM plus the SP objectives, $obj^{OM}+obj^{SP}$. Finally, in any case, add the optimality constraint $Opt.cut$ to the MP. This procedure is repeated until the bounds are equalized.

\begin{algorithm}
\caption{\OMT classical Benders decomposition} \label{alg:Benders_classical}
\SetKwInOut{Input}{input}\SetKwInOut{Output}{output}
\LinesNumbered
$UB := \infty$ \\
$LB := 0$ \\
\vspace{0.25cm}
\While{$UB \leq LB$}{
Solve to optimality $F^{MP} \rightarrow (obj^{MP},\bar{x})$, where $obj^{MP} = obj^{OM} + \mu$ \\
\If{$obj^{MP}>LB$}{update LB}
Solve to optimality $F^{SP} \rightarrow obj^{SP}$\\
\If{$obj^{OM} + obj^{SP} < UB$}{update UB}
Add optimality constraint $Opt.cut$ to $F^{MP}$}
\end{algorithm}

As aforementioned, the subproblem to deal with is the MST between the facilities computed in the MP solution, which can be solved in polynomial time using Kruskal algorithm. Once solved, the objective value of the subproblem $obj^{SP}$ is used to build the following optimality constraint to add to the MP:

\begin{equation}
    Opt.cut: \ \frac{obj^{SP}}{p-1}\Bigg[\sum_{i\in V} \bar{x}_{ii} - (p-1) \Bigg] \leq \mu.
\end{equation}

For this $Opt.cut$, if the same $p$ facilities that have been selected as inputs for the SP are selected in the MP, then the $\mu$ variable is increased in the amount of the cost of the tree obtained as output in the SP. Otherwise, if at least one different facility is selected, then $\mu$ does not change. In other words, $\mu$ is null until the same $p$ facilities are selected again, where $\mu$ turns strictly positive and the algorithm ends.

Note that in every iteration, only one $Opt.cut$ is added. For the general Benders framework, some authors (e.g. see \citealp{Fischetti2010}) force the addition of feasibility cuts even when the MP is feasible to boost the algorithm. In our case, since our subproblem is always feasible, it is not necessary to introduce feasibility cuts, but we can force a series of suboptimal cuts to be introduced in each iteration after the optimal cut is included.

Directly using $Opt.cut$ avoids having to go through the dual formulation of our SP, as in most of the Benders decomposition frameworks. However, there exists an implementation of the Kipp Martin ($km$) MST dual formulation in \citet{Labbe2021} which can be use for this purpose. The $km$ MST formulation and its dual form are adapted to the \OMT structure as follows:

\begingroup
\allowdisplaybreaks
\begin{subequations}
\begin{align}
  & \hhp \hspace{1cm} F^{km}: \min \sum_{(i,j) \in E} c_{ij} z_{ij} & & & \label{OMT_f2b_c0} \\
  & \Hp z_{ij} \leq \bar{x}_{ii}                        &   & (i,j) \in E                   & \label{OMT_f2b_c11} \\
  & \Hp z_{ij} \leq \bar{x}_{jj}                        &   & (i,j) \in E                   & \label{OMT_f2b_c12} \\
  & \Hp \sum_{(i,j) \in E} z_{ij} = p-1                 &   &                               & \label{OMT_f2b_c2} \\
  & \Hp \sum_{\substack{(i',j) \in E: \\ (i'=k \wedge j=i) \\ \lor \\ (i'=i \wedge j=k)}} x_{i'j} + \sum_{(i,j) \in A:j \neq k} q_{kij} \leq 1 & & k,i \in V: i \neq k & \label{OMT_f2b_c3} \\
  & \Hp q_{kij} + q_{kji} = z_{ij}                      &   &  k \in V, (i,j) \in E: i,j \neq k  & \label{OMT_f2b_c4} \\
  & \Hp z_{ij}  \geq 0                                  &   &  (i,j) \in E \\
  & \Hp q_{kij} \geq 0                                  &   &  k \in V, (i,j) \in E.
\end{align}
\end{subequations}
\endgroup

\begingroup
\allowdisplaybreaks
\begin{subequations}
\begin{align}
  & \hhp \hspace{1cm} F^{SP}: \max \alpha (p-1) - \sum_{k,i \in V: i \neq k} \beta_{ki} - \sum_{(i,j) \in E} (\bar{x}_{ii} \tau_{ij} + \bar{x}_{jj} \eta_{ij}) & & & \label{OMT_f3b_c0} \\
  & \Hp \alpha - \beta_{ij} - \beta_{ji} - \sum_{k \in V: k \neq i,j} \gamma^{k}_{ij} - \tau_{ij} - \eta_{ij} \leq c_{ij} &   & (i,j) \in E & \label{OMT_f3b_c1} \\
  & \Hp - \beta_{ki} +  \sum_{\substack{(i',j') \in E: \\ (i'=i \wedge j'=j) \\ \lor \\ (i'=j \wedge j'=i)}} \gamma^{k}_{i'j'} \leq 0 &   & k \in V, (i,j) \in E: i,j \neq k & \label{OMT_f3b_c2} \\
  & \Hp \alpha \in \mathbb{R}        &   &  \\
  & \Hp \beta_{ki} \geq 0            &   & k,i \in V: i \neq k \\
  & \Hp \gamma^k_{ij} \in \mathbb{R} &   & k \in V, (i,j) \in E: i,j \neq k  \\
  & \Hp \tau_{ij}    \geq 0          &   & (i,j) \in E \\
  & \Hp \eta_{ij}    \geq 0          &   & (i,j) \in E.
\end{align}
\end{subequations}
\endgroup

Observe that, constraints (\ref{OMT_f1_c4}) are presented as disaggregated constraints (\ref{OMT_f2b_c11})-(\ref{OMT_f2b_c12}) in the SP because, although it has been computationally proven to report worse results, dealing with continuous $z$ variables force to use the disaggregated formulation. In this way, a different $Opt.cut$ can be introduced using $F^{SD}$:

\begin{equation}
  Opt.cut:\ \frac{1}{p-1} \Big[\bar{\alpha}(p-1) - \sum_{k,i \in V: i \neq k} \bar{\beta}_{ku} - \sum_{(i,j) \in E} (x_{ii} \bar{\tau}_{ij} + x_{jj} \bar{\eta}_{ij}) \Big] \leq \mu.
\end{equation}

\subsection{Modern Benders decomposition}
\label{subsec:ModernBendersdecomposition}

The classical Benders decomposition can be introduced into a branch-and-cut framework (also known as branch-and-Benders-cut) for a more efficient approach as illustrated in Algorithm \ref{alg:Benders_modern}. In this algorithm we initialize the ramification tree $\mTree$, considering solving the MP at the root node, and an empty pool of cuts $\mathcal{P}$. Once solved the MP at the root node, we start the branching procedure. If at a particular node of the ramification process the solution found is fractional, keep branching. Otherwise, when an integer solution is found, the lower bound is updated using the MP objective value of the current ramification node $o' \in \mTree$, the SP is solved fixing the facilities $\bar{x}$ obtained in the MP solution, the upper bound is updated if possible and the optimality constraint $Opt.cut$ is added to a pool $\mathcal{P}$ of cuts. The cuts from pool $\mathcal{P}$ are then included for the MP solution in other nodes of $\mTree$.

\begin{algorithm}
\caption{\OMT branch-and-Benders-cut} \label{alg:Benders_modern}
\SetKwInOut{Input}{input}\SetKwInOut{Output}{output}
\LinesNumbered
Set tree $\mTree=\{o\}$, where $o=F^{MP}$ has no branching constraints \\
Initialize a pool of cuts $\mathcal{P} = \emptyset$ \\
\While{$\mTree$ is nonempty}{
    Select a node $o'\in\mTree$\\
    $\mTree := \mTree \,\backslash\{o'\}$\\
    Solve $o'$ considering $\mathcal{P}$ $\rightarrow \bar{x}$ \\
    \If {$\bar{x}$ is fractional}{
        Branch, resulting in nodes $o''$ and $o'''$\\
        $\mTree := \mTree \cup\{o'',o'''\}$\\
    \Else{
        Solve $F^{SD} \rightarrow (\bar{\alpha}, \bar{\beta}, \bar{\tau}, \bar{\eta}) $\\
        Add optimality constraint $Opt.cut$ to $\mathcal{P}$\\
        $\mTree := \mTree \cup \{o'\}$\\
    }}
}
\end{algorithm}

Embedding Benders cuts in a branch-and-cut framework allows to use different strategies for producing these cuts. For instance, the cuts can be generated in every feasible node, only in those nodes that have improved the lower bound beyond a predefined threshold or only when incumbent solutions are found along the search process. If caution is not taken, too many unnecessary cuts can be produced slowing down the obtainment of master problem solution instead of speeding it up.

\subsection{Warm-start phase}
\label{subsec:Warmstartphase}

In the classical Benders decomposition, every $Opt.cut$ added implies solving to optimality a MP, which is computationally costly. In the modern approach, the pool of cuts $\mathcal{P}$ considered in a specific node of the branch-and-cut procedure may contain cuts that have already been identified, slowing down the algorithm efficiency since it would reintroduce repeated cuts when branching at different levels of $\mTree$. For these reasons, it could be helpful to consider a strategy to initialize the pool of cuts $\mathcal{P}$ prior to starting the procedures using a so called \textit{warm-start phase} (e.g. see \citealp{Martins2013}).

This warm-start phase can be introduced in both classical and modern approaches. For the modern approach, the idea is to store the cuts generated in an independent warm-start phase in the pool of cuts $\mathcal{P}$ settled at the beginning of Algorithm \ref{alg:Benders_modern}. Given $\textsf{runtime}^{MP}$, elapsed time for a MP iteration, and $\textsf{runtime}$, total elapsed time for the completition of the algorithm, the parameters considered for tuning the warm-start phase are the following:

\begin{itemize}
  \item $\textsf{max.time}^{MP}$. Time limit that every solution computation of the MP in the warm-start phase can take to obtain the best feasible solution possible. Once reached, the respective $Opt.cut$ is added and a new MP is considered for solving.
  \item $\textsf{max.gap}^{MP}$. Gap percentage allowed to remain between the best lower and upper bound found in every MP of the warm-start phase, $LB^{MP}$ and $UB^{MP}$, computed as $100\cdot\frac{UB^{MP}-LB^{MP}}{LB^{MP}}$. Once reached, the respective $Opt.cut$ is added and a new MP is considered for solving.
  \item $\textsf{max.time}$. Total time limit that the entire warm-start phase can take. Once reached, the warm-start phase is finished and the algorithm starts searching for optimal solutions.
  \item $\textsf{max.gap}$. Gap percentage allowed to remain between best lower and upper bound found in the entire warm-start phase, computed as in $\textsf{max.gap}^{MP}$ but using the overall bounds $LB$ and $UB$. Once reached, the warm-start phase is finished and the algorithm starts searching for optimal solutions.
\end{itemize}

The interest behind this warm-star phase is to be able to introduce a certain number of cuts at low cost in the beginning of the procedures avoiding, on many occasions, that these cuts that have already been introduced are considered again, improving the overall performance of the of the Benders decomposition algorithms. In order to introduce this cuts, the solution in the warm-start phase does not need to be computed to optimality, it is enough if it is feasible. Furthermore, a procedure to build cuts from fractional values of the variables can be implemented.

\newpage

\section{Appendix B: \OMT covering fixing preprocessings example}
\label{AppendixB}

\begin{figure}[ht]
\centering
\tikzset{every picture/.style={line width=0.75pt}} 
\begin{tikzpicture}[x=0.75pt,y=0.75pt,yscale=-1,xscale=1]
\draw  [color={rgb, 255:red, 4; green, 250; blue, 61 }  ,draw opacity=1 ][fill={rgb, 255:red, 226; green, 250; blue, 224 }  ,fill opacity=1 ][line width=1.5]  (85,74.5) .. controls (85,66.49) and (91.49,60) .. (99.5,60) .. controls (107.51,60) and (114,66.49) .. (114,74.5) .. controls (114,82.51) and (107.51,89) .. (99.5,89) .. controls (91.49,89) and (85,82.51) .. (85,74.5) -- cycle ;
\draw  [color={rgb, 255:red, 4; green, 250; blue, 61 }  ,draw opacity=1 ][fill={rgb, 255:red, 226; green, 250; blue, 224 }  ,fill opacity=1 ][line width=1.5]  (214,204) .. controls (214,195.72) and (220.72,189) .. (229,189) .. controls (237.28,189) and (244,195.72) .. (244,204) .. controls (244,212.28) and (237.28,219) .. (229,219) .. controls (220.72,219) and (214,212.28) .. (214,204) -- cycle ;
\draw  [color={rgb, 255:red, 4; green, 250; blue, 61 }  ,draw opacity=1 ][fill={rgb, 255:red, 226; green, 250; blue, 224 }  ,fill opacity=1 ][line width=1.5]  (214,74) .. controls (214,65.72) and (220.72,59) .. (229,59) .. controls (237.28,59) and (244,65.72) .. (244,74) .. controls (244,82.28) and (237.28,89) .. (229,89) .. controls (220.72,89) and (214,82.28) .. (214,74) -- cycle ;
\draw  [color={rgb, 255:red, 4; green, 250; blue, 61 }  ,draw opacity=1 ][fill={rgb, 255:red, 226; green, 250; blue, 224 }  ,fill opacity=1 ][line width=1.5]  (84,205) .. controls (84,196.72) and (90.72,190) .. (99,190) .. controls (107.28,190) and (114,196.72) .. (114,205) .. controls (114,213.28) and (107.28,220) .. (99,220) .. controls (90.72,220) and (84,213.28) .. (84,205) -- cycle ;
\draw [line width=0.75]    (114,74.5) -- (214,74) ;
\draw [line width=0.75]    (114,205) -- (214,204.5) ;
\draw [line width=0.75]    (99.5,89) -- (99,190) ;
\draw [line width=0.75]    (229,89) -- (228.5,190) ;
\draw [line width=0.75]    (110,85) -- (216,195) ;
\draw [line width=0.75]    (111,194) -- (219,84) ;
\draw  [color={rgb, 255:red, 4; green, 250; blue, 61 }  ,draw opacity=1 ][fill={rgb, 255:red, 226; green, 250; blue, 224 }  ,fill opacity=1 ][line width=1.5]  (542,74) .. controls (542,65.72) and (548.72,59) .. (557,59) .. controls (565.28,59) and (572,65.72) .. (572,74) .. controls (572,82.28) and (565.28,89) .. (557,89) .. controls (548.72,89) and (542,82.28) .. (542,74) -- cycle ;
\draw  [color={rgb, 255:red, 4; green, 250; blue, 61 }  ,draw opacity=1 ][fill={rgb, 255:red, 226; green, 250; blue, 224 }  ,fill opacity=1 ][line width=1.5]  (412,205) .. controls (412,196.72) and (418.72,190) .. (427,190) .. controls (435.28,190) and (442,196.72) .. (442,205) .. controls (442,213.28) and (435.28,220) .. (427,220) .. controls (418.72,220) and (412,213.28) .. (412,205) -- cycle ;
\draw [line width=0.75]    (444,74.49) -- (542,74) ;
\draw [shift={(442,74.5)}, rotate = 359.71] [color={rgb, 255:red, 0; green, 0; blue, 0 }  ][line width=0.75]    (6.56,-2.94) .. controls (4.17,-1.38) and (1.99,-0.4) .. (0,0) .. controls (1.99,0.4) and (4.17,1.38) .. (6.56,2.94)   ;
\draw [line width=0.75]    (442,205) -- (540,204.51) ;
\draw [shift={(542,204.5)}, rotate = 179.71] [color={rgb, 255:red, 0; green, 0; blue, 0 }  ][line width=0.75]    (6.56,-2.94) .. controls (4.17,-1.38) and (1.99,-0.4) .. (0,0) .. controls (1.99,0.4) and (4.17,1.38) .. (6.56,2.94)   ;
\draw [line width=0.75]  [dash pattern={on 4.5pt off 4.5pt}]  (441,90) -- (541.5,189.06) ;
\draw    (85,74.5) .. controls (58,63) and (91,35) .. (99.5,60) ;
\draw    (411,75) .. controls (378.5,61.21) and (407.11,27.05) .. (425.18,58.52) ;
\draw [shift={(426,60)}, rotate = 242.1] [color={rgb, 255:red, 0; green, 0; blue, 0 }  ][line width=0.75]    (6.56,-2.94) .. controls (4.17,-1.38) and (1.99,-0.4) .. (0,0) .. controls (1.99,0.4) and (4.17,1.38) .. (6.56,2.94)   ;
\draw    (229,59) .. controls (236,33) and (272,61) .. (244,74) ;
\draw    (99,220) .. controls (80,245) and (62,209) .. (84,205) ;
\draw    (244,204) .. controls (269,214) and (239,244) .. (229,219) ;
\draw    (557,219) .. controls (568.82,249.54) and (603.93,221.86) .. (573.44,204.77) ;
\draw [shift={(572,204)}, rotate = 27.26] [color={rgb, 255:red, 0; green, 0; blue, 0 }  ][line width=0.75]    (6.56,-2.94) .. controls (4.17,-1.38) and (1.99,-0.4) .. (0,0) .. controls (1.99,0.4) and (4.17,1.38) .. (6.56,2.94)   ;
\draw  [color={rgb, 255:red, 208; green, 2; blue, 27 }  ,draw opacity=1 ][fill={rgb, 255:red, 250; green, 226; blue, 226 }  ,fill opacity=1 ][line width=1.5]  (411.5,60.06) -- (441,60.06) -- (441,90) -- (411.5,90) -- cycle ;
\draw  [color={rgb, 255:red, 208; green, 2; blue, 27 }  ,draw opacity=1 ][fill={rgb, 255:red, 250; green, 226; blue, 226 }  ,fill opacity=1 ][line width=1.5]  (541.5,189.06) -- (571,189.06) -- (571,219) -- (541.5,219) -- cycle ;
\draw (94,68) node [anchor=north west][inner sep=0.75pt]   [align=left] {{\fontfamily{helvet}\selectfont 1}};
\draw (93,200) node [anchor=north west][inner sep=0.75pt]   [align=left] {{\fontfamily{helvet}\selectfont 4}};
\draw (224,69) node [anchor=north west][inner sep=0.75pt]   [align=left] {{\fontfamily{helvet}\selectfont 2}};
\draw (224,199) node [anchor=north west][inner sep=0.75pt]   [align=left] {{\fontfamily{helvet}\selectfont 3}};
\draw (159,55) node [anchor=north west][inner sep=0.75pt]   [align=left] {{\fontfamily{helvet}\selectfont 2}};
\draw (137,95) node [anchor=north west][inner sep=0.75pt]   [align=left] {{\fontfamily{helvet}\selectfont 1}};
\draw (190,117) node [anchor=north west][inner sep=0.75pt]   [align=left] {{\fontfamily{helvet}\selectfont 5}};
\draw (85,131) node [anchor=north west][inner sep=0.75pt]   [align=left] {{\fontfamily{helvet}\selectfont 3}};
\draw (159,206.75) node [anchor=north west][inner sep=0.75pt]   [align=left] {{\fontfamily{helvet}\selectfont 2}};
\draw (232,129) node [anchor=north west][inner sep=0.75pt]   [align=left] {{\fontfamily{helvet}\selectfont 4}};
\draw (421,71) node [anchor=north west][inner sep=0.75pt]   [align=left] {{\fontfamily{helvet}\selectfont 1}};
\draw (421,199) node [anchor=north west][inner sep=0.75pt]   [align=left] {{\fontfamily{helvet}\selectfont 4}};
\draw (552,68) node [anchor=north west][inner sep=0.75pt]   [align=left] {{\fontfamily{helvet}\selectfont 2}};
\draw (551,200.09) node [anchor=north west][inner sep=0.75pt]   [align=left] {{\fontfamily{helvet}\selectfont 3}};
\draw (486,55) node [anchor=north west][inner sep=0.75pt]   [align=left] {{\fontfamily{helvet}\selectfont 2}};
\draw (494,121) node [anchor=north west][inner sep=0.75pt]   [align=left] {{\fontfamily{helvet}\selectfont 1}};
\draw (486,206.75) node [anchor=north west][inner sep=0.75pt]   [align=left] {{\fontfamily{helvet}\selectfont 2}};
\draw (60,52) node [anchor=north west][inner sep=0.75pt]   [align=left] {{\fontfamily{helvet}\selectfont 0}};
\draw (60,207) node [anchor=north west][inner sep=0.75pt]   [align=left] {{\fontfamily{helvet}\selectfont 0}};
\draw (258,53) node [anchor=north west][inner sep=0.75pt]   [align=left] {{\fontfamily{helvet}\selectfont 0}};
\draw (257,209) node [anchor=north west][inner sep=0.75pt]   [align=left] {{\fontfamily{helvet}\selectfont 0}};
\draw (382,53) node [anchor=north west][inner sep=0.75pt]   [align=left] {{\fontfamily{helvet}\selectfont 0}};
\draw (590,210) node [anchor=north west][inner sep=0.75pt]   [align=left] {{\fontfamily{helvet}\selectfont 0}};
\end{tikzpicture}
\caption{\OMT instance example (left) and its optimal solution (right).}
\label{fig:OMT_example_covering_fixing}
\end{figure}
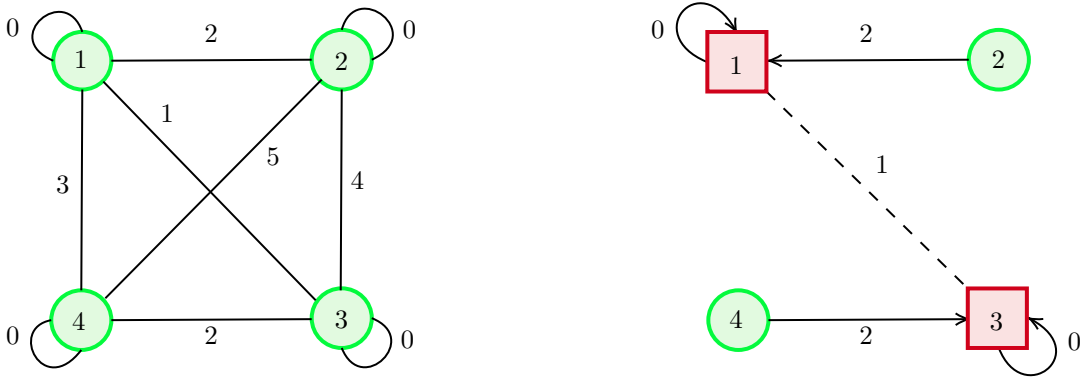

Let consider the complete instance of the \OMT with $|V|=4$ in Figure \ref{fig:OMT_example_covering_fixing} (left) and the allocation costs specified over its edges. We want to locate $p=2$ facilities considering a scaling vector of $\lambda = (1,\overset{N}{\ldots},1)$ (median criterion). Note that we assume that $c_{ii}=0$ for $i \in V$. The optimal allocations (arrows) and facilities (squares) for this instance are depicted in Figure \ref{fig:OMT_example_covering_fixing} (right).

In the optimal solution we get a total allocation cost of 4, given by $c_{21}$ plus $c_{43}$ (also plus $c_{11}$ and $c_{33}$ that are zero). Facilities are connected by means of $(1,3)$, holding itself the tree of facilities structure expected in any solution of the \OMT.

In order to illustrate how the covering formulation preprocessings for fixing a certain number of $u_{\ell h}$-variables to 1 or 0 using the auxiliary problems $F^{pre}_{1}$ and $F^{pre}_{0}$, we must first consider the ordered sequence of unique costs $c_{(0)} := 0 < c_{(1)}<c_{(2)}<\ldots<c_{(|H|)}=\max_{i,j \in V} c_{ij}$ of our problem as in Table \ref{tab:OMT_example_ordered_sequence}.

\begin{table}[ht]
\centering
\begin{tabular}{@{}llllll@{}}
\toprule
$c_{(0)}$ & $c_{(1)}$ & $c_{(2)}$ & $c_{(3)}$ & $c_{(4)}$ & $c_{(5)}$ \\ \midrule
0         & 1         & 2         & 3         & 4         & 5         \\ \bottomrule
\end{tabular}
\caption{Sequence of unique ordered costs for the \OMT example.}
\label{tab:OMT_example_ordered_sequence}
\end{table}

Note that cost 0 is repeated 4 times, since we have 4 possible allocations of facilities to themselves, and cost 2 is also repeated 2 times. For the covering formulation, the $u$-matrix of dimensions $|V| \times |H|$ of the optimal solution is given by the matrix

$$u_{\ell h} = \begin{pmatrix}
0 & 0 & 0 & 0 & 0 \\
0 & 0 & 0 & 0 & 0 \\
1 & 1 & 0 & 0 & 0 \\
1 & 1 & 0 & 0 & 0 
\end{pmatrix},$$

which means that the $u_{\ell h}$ variables activated are $u_{31}, u_{32}, u_{41}, u_{42} = 1$. Note that we are not considering the $c_{(0)}$ column in the $u$-matrix, which in case of being considered must be $u_{\ell 0}=1$ for $\ell \in V$. We can verify over this matrix that the allocation cost objective holds using covering variables:

\begin{align*}
\sum_{\ell \in V} \sum_{h \in H} \lambda_\ell u_{\ell h} (c_{(h)}-c_{(h-1)}) 
& = u_{31} (c_{(1)}-c_{(0)}) + u_{32} (c_{(2)}-c_{(1)}) + u_{41}(c_{(1)}-c_{(0)}) + u_{42}(c_{(2)}-c_{(1)}) = \\ 
& = (1-0) + (2-1) + (1-0) + (2-1) = 4.
\end{align*}

Now, if we want to previously fix a certain number of $u_{\ell h}$-variables to 1, we deal with the auxiliary problem $F^{pre}_{1}$. In this problem we try to maximize the number of allocations satisfying $c_{ij} \leq c_{(h-1)}$ from which, given that there exists a total of $|V|$ possible allocations, we can get the minimum number of allocations satisfying $c_{ij} > c_{(h-1)} \equiv c_{ij} \geq c_{(h)}$, as specified in the covering formulation by constraints \ref{OMT_f2_c6}. For example, if $h=1$, we are looking for allocations satisfying $c_{ij}x_{ij} \leq c_{(0)} = 0$. Only the 4 allocations at a cost 0 of the potential facilities to themselves can be considered in this case, but because constraints (\ref{preproc_1_b}) must be satisfied, only 2 of these allocations can be selected and, therefore, the solution is $H_1^1 = 2$. This means that there are at least $N - H_1^1 = 2$ allocations that satisfy the opposite and, as the ones are placed in the lower part of the $u$-matrix, we have $u_{\ell 1}=1$ for any $\ell \in \{N - H_1^1 + 1, \ldots, N \} = \{3,4\}$. For $h=2$, we consider again 2 of the 4 allocations of a facility to itself and only one more allocation, which is $x_{13}$, and as a consequence we have $H_2^1 = 3$. Hence, only $N - H_2^1 = 1$ variable can be fixed, which corresponds to $u_{42}=1$. Following the same reasoning, we get $H^1_h=4$ for $h \geq 3$, so we are not able to fix any more variables.

On the other hand, for fixing $u_{\ell h}$-variables to 0 we deal with the auxiliary problem $F^{pre}_{0}$. In our example, for $h=5$ we have only one possible allocation that satisfies $c_{ij} \geq c_{(5)} x_{ij} = 5x_{ij}$ plus the 2 allocations coming from constraints (\ref{preproc_0_b}). This means that as much $H^0_5 = 3$ of the $u_{\ell 5}$-variables could be fixed to 1 (there could be less), so the 5-th column of the $u$-matrix must have at least $N-H^0_5 = 4-3 = 1$ zeros. However, as we know that the $p$ first rows of the $u$-matrix can be fixed to 0, we end up having $u_{\ell 5}=0$ for any $\ell \in \{ 1,\ldots, N-H^0_5+p \} = \{ 1,2,3 \}$. For $h \leq 4$ the optimal solution of the auxiliary problem is $H^0_h = 4$ and consequently we are only able to fix the $p$ first variables of each column that we knew in advance.

We can summarize the previous information in the following table:

\begin{table}[ht]
\centering
\begin{tabular}{@{}c|cccc@{}}
\toprule
$h$ &
  $\overset{ H_h^1 }{ \overbrace{\text{\# alloc} | c_{ij} \leq c_{(h-1)}} }$ &
  $N-H_h^1$ &
  $\overset{ H_h^0 }{ \overbrace{\text{\# alloc} | c_{ij} \geq c_{(h-1)}} }$ &
  $N-H_h^0+p$ \\ \midrule
1 & 2 & 2 & 4 & 2 \\
2 & 3 & 1 & 4 & 2 \\
3 & 4 & 0 & 4 & 2 \\
4 & 4 & 0 & 4 & 2 \\
5 & 4 & 0 & 3 & 3 \\ \bottomrule
\end{tabular}
\caption{Fixing solutions for the \OMT example.}
\label{tab:fig:OMT_example_covering_fixing_solution}
\end{table}

Thus, solving both auxiliary problems, we get the following final preprocessing matrix of $u_{\ell h}$-variables fixed, where $NF$ means not fixed by any of the auxiliary problems:

$$preproc(u_{\ell h}) = \begin{pmatrix}
0  & 0  & 0  & 0  & 0 \\
0  & 0  & 0  & 0  & 0 \\
1  & NF & NF & NF & 0 \\
1  & 1  & NF & NF & NF 
\end{pmatrix}.$$

\newpage

\section{Appendix C: \OMT formulations description}
\label{AppendixC}

\subsection{\OMT subtour elimination formulation}
\label{subsec:OMTsubtoureliminationformulation}

\begingroup
\allowdisplaybreaks
\begin{subequations}
\begin{align}
  & \hhp \hspace{1cm} F1_{x}^{sub1}: \min \quad \frac{1}{\sum_{\ell \in V} \lambda_\ell} \sum_{\ell \in V} \sum_{(i,j) \in A} \lambda_\ell c_{ij} x_{ij}^\ell + \frac{1}{p-1} \sum_{(i,j) \in E} c_{ij} z_{ij} & & & \label{OMT_sub_x_1_0} & \\
  & \hp \sum_{i \in V} x_{ii} = p                     &   &                                 & \label{OMT_sub_x_1_1}  & \\
  & \Hp \sum_{j \in V} x_{ij} = 1                     &   & i \in V                         & \label{OMT_sub_x_1_2}  & \\
  & \Hp x_{ij} \leq x_{jj}                            &   & (i,j) \in A: i \neq j           & \label{OMT_sub_x_1_3}  & \\
  & \Hp 2 z_{ij} \leq x_{ii} + x_{jj}                 &   & (i,j) \in E                     & \label{OMT_sub_x_1_4}  & \\
  & \Hp \sum_{(i,j) \in E} z_{ij} = p-1               &   &                                 & \label{OMT_sub_x_1_5}  & \\
  & \Hp \sum_{i,j \in S: i<j} z_{ij} \leq |S|-1       &   & S \neq \emptyset, S \subset V   & \label{OMT_sub_x_1_6}  & \\
  & \Hp \sum_{\ell \in V} x_{ij}^\ell = x_{ij}        &   & (i,j) \in A                     & \label{OMT_sub_x_1_7}  & \\
  & \Hp \sum_{(i,j) \in A} x_{ij}^\ell = 1            &   & \ell \in V                      & \label{OMT_sub_x_1_8}  & \\
  & \Hp \sum_{(i,j) \in A} c_{ij}x_{ij}^\ell \leq \sum_{(i,j) \in A} c_{ij}x_{ij}^{\ell+1}  &   & \ell \in V: \ell < |V|  & \label{OMT_sub_x_1_9} & \\
  & \Hp x_{ij} \in \{0,1\}                            &   & (i,j) \in A \\
  & \Hp x_{ij}^\ell \in \{0,1\}                       &   & (i,j) \in A, \ell \in V \\
  & \Hp z_{ij} \in \{0,1\}                            &   & (i,j) \in E.
\end{align}
\end{subequations}
\endgroup

\begin{itemize}[noitemsep]
  \item[] (\ref{OMT_sub_x_1_0}): Weighted minimization of the compensated allocation cost of the system plus the design cost of the tree connecting facilities (objective).
  \item[] (\ref{OMT_sub_x_1_1}): Exactly $p$ facilities (allocation). 
  \item[] (\ref{OMT_sub_x_1_2}): Each client is allocated to exactly one facility (allocation).
  \item[] (\ref{OMT_sub_x_1_3}): No client is allocated to a non-facility (allocation).
  \item[] (\ref{OMT_sub_x_1_4}): Edges of the tree of facilities can only be selected if both nodes are facilities (allocation).
  \item[] (\ref{OMT_sub_x_1_5}): Exactly $p-1$ edges connecting facilities (connectivity).
  \item[] (\ref{OMT_sub_x_1_6}): No subtours are allowed to be formed between facilities edges (connectivity). 
  \item[] (\ref{OMT_sub_x_1_7}): Relationship between sorting and allocation variables (sorting).
  \item[] (\ref{OMT_sub_x_1_8}): Every sort position must be consider (sorting).
  \item[] (\ref{OMT_sub_x_1_9}): Correct sorting of the costs sequence (sorting).
\end{itemize}

\begingroup
\allowdisplaybreaks
\begin{subequations}
\begin{align}
  & \hhp \hspace{1cm} F1_{x}^{sub2}: \min \quad \frac{1}{\sum_{\ell \in V} \lambda_\ell} \sum_{\ell \in V} \sum_{(i,j) \in A} \lambda_\ell c_{ij} x_{ij}^\ell + \frac{1}{p-1} \sum_{(i,j) \in E} c_{ij} z_{ij} & & & \label{OMT_sub_x_2_0} & \\
  & \hp \sum_{i \in V} x_{ii} = p                     &   &                                  & \label{OMT_sub_x_2_1}  & \\
  & \Hp \sum_{j \in V} x_{ij} = 1                     &   & i \in V                          & \label{OMT_sub_x_2_2}  & \\
  & \Hp x_{ij} \leq x_{jj}                            &   & (i,j) \in A: i \neq j            & \label{OMT_sub_x_2_3}  & \\
  & \Hp 2 z_{ij} \leq x_{ii} + x_{jj}                 &   & (i,j) \in E                      & \label{OMT_sub_x_2_4}  & \\
  & \Hp \sum_{(i,j) \in E} z_{ij} = p-1               &   &                                  & \label{OMT_sub_x_2_5}  & \\
  & \Hp \displaystyle{\sum_{i \in S, j \in V \setminus S}} x_{ij} + \displaystyle{\sum_{i \in S, j \in V\setminus S}} x_{ji} + \displaystyle{\sum_{i \in S, j \in V \setminus S:i<j}} z_{ij} + \displaystyle{\sum_{i \in S, j \in V\setminus S:i>j}} z_{ji} \geq 1  &   & \forall S \neq \emptyset, S \subset V                & \label{OMT_sub_x_2_6}  & \\
  & \Hp \sum_{\ell \in V} x_{ij}^\ell = x_{ij}        &   & (i,j) \in A                      & \label{OMT_sub_x_2_7}  & \\
  & \Hp \sum_{(i,j) \in A} x_{ij}^\ell = 1            &   & \ell \in V                       & \label{OMT_sub_x_2_8}  & \\
  & \Hp \sum_{(i,j) \in A} c_{ij} x_{ij}^\ell \leq \sum_{(i,j) \in A} c_{ij} x_{ij}^{\ell+1} &   & \ell \in V: \ell < |V|  & \label{OMT_sub_x_2_9} & \\
  & \Hp x_{ij} \in \{0,1\}                            &   & (i,j) \in A \\
  & \Hp x_{ij}^\ell \in \{0,1\}                       &   & (i,j) \in A, \ell \in V \\
  & \Hp z_{ij} \in \{0,1\}                            &   & (i,j) \in E. 
\end{align}
\end{subequations}
\endgroup

\begin{itemize}[noitemsep]
  \item[] (\ref{OMT_sub_x_2_0}): Weighted minimization of the compensated allocation cost of the system plus the design cost of the tree connecting facilities (objective).
  \item[] (\ref{OMT_sub_x_2_1}): Exactly $p$ facilities (allocation). 
  \item[] (\ref{OMT_sub_x_2_2}): Each client is allocated to exactly one facility (allocation).
  \item[] (\ref{OMT_sub_x_2_3}): No client is allocated to a non-facility (allocation).
  \item[] (\ref{OMT_sub_x_2_4}): Edges of the tree of facilities can only be selected if both nodes are facilities (allocation).
  \item[] (\ref{OMT_sub_x_2_5}): Exactly $p-1$ edges connecting facilities (connectivity).
  \item[] (\ref{OMT_sub_x_2_6}): At least one allocation or one edge from the tree of facilities must connect $S$ to a node of $V \setminus S$ (connectivity). 
  \item[] (\ref{OMT_sub_x_2_7}): Relationship between sorting and allocation variables (sorting).
  \item[] (\ref{OMT_sub_x_2_8}): Every sort position must be consider (sorting).
  \item[] (\ref{OMT_sub_x_2_9}): Correct sorting of the costs sequence (sorting).
\end{itemize}

\begingroup
\allowdisplaybreaks
\begin{subequations}
\begin{align}
  & \hhp \hspace{1cm} F2_x^{sub1}: \min \quad \frac{1}{\displaystyle{\sum_{\ell \in V} \lambda_\ell}} \sum_{\ell \in V} \sum_{(i,j) \in A} \lambda_\ell c_{ij} x_{ij}^\ell + \frac{1}{p-1} \sum_{(i,j) \in E} c_{ij} (z_{ij}-x_{ij} - x_{ji})  & & & \label{OMT_sub_x_3_0} & \\
  & \hp \sum_{i \in V} x_{ii} = p                       &   &                                       & \label{OMT_sub_x_3_1}  & \\
  & \Hp \sum_{j \in V} x_{ij} = 1                       &   & i \in V                               & \label{OMT_sub_x_3_2}  & \\
  & \Hp 2x_{ij} \leq 1 - x_{ii} + x_{jj}                &   & (i,j) \in A: i \neq j                 & \label{OMT_sub_x_3_3}  & \\
  & \Hp x_{ij} + x_{ji} \leq z_{ij}                     &   & (i,j) \in E                           & \label{OMT_sub_x_3_4}  & \\
  & \Hp 2z_{ij} \leq x_{ii} + x_{jj} + x_{ij} + x_{ji}  &   & (i,j) \in E                           & \label{OMT_sub_x_3_5}  & \\
  & \Hp \sum_{(i,j) \in E} z_{ij} = N-1                 &   &                                       & \label{OMT_sub_x_3_6}  & \\
  & \Hp \sum_{i,j \in S: i<j} z_{ij} \leq |S|-1         &   & S \neq \emptyset, S \subset V         & \label{OMT_sub_x_3_7}  & \\
  & \Hp \sum_{\ell \in V} x_{ij}^\ell = x_{ij}          &   & (i,j) \in A                           & \label{OMT_sub_x_3_8}  & \\
  & \Hp \sum_{(i,j) \in A} x_{ij}^\ell = 1              &   & \ell \in V                            & \label{OMT_sub_x_3_9}  & \\
  & \Hp \sum_{(i,j) \in A} c_{ij}x_{ij}^\ell \leq \sum_{(i,j) \in A} c_{ij}x_{ij}^{\ell+1}      &   & \ell \in V: \ell < |V|  & \label{OMT_sub_x_3_10} & \\
  & \Hp x_{ij} \in \{0,1\}                              &   &  (i,j) \in A \\
  & \Hp x_{ij}^\ell \in \{0,1\}                         &   &  (i,j) \in A, \ell \in V \\
  & \Hp z_{ij} \in \{0,1\}                              &   &  (i,j) \in E.
\end{align}
\end{subequations}
\endgroup

\begin{itemize}[noitemsep]
  \item[] (\ref{OMT_sub_x_3_0}): Weighted minimization of the compensated allocation cost of the system plus the design cost of the tree subtracting in the design cost part the cost associated to the edges that represent the allocation of client to facilities (objective).
  \item[] (\ref{OMT_sub_x_3_1}): Exactly $p$ facilities (allocation). 
  \item[] (\ref{OMT_sub_x_3_2}): Each client is allocated to exactly one facility (allocation).
  \item[] (\ref{OMT_sub_x_3_3}): Client-facility allocations are only possible if $i$ is a client and $j$ is a facility (allocation).
  \item[] (\ref{OMT_sub_x_3_4}): Only one allocation, whether client $i$ to facility $j$ or viceversa, can be considered if there exists an edge connecting $i,j \in V$ (allocation).
  \item[] (\ref{OMT_sub_x_3_5}): Only possibilities in which an edge can be selected is if both $i$ or $j$ are facilities or if either $i$, or $j$, is a client and $j$, or $i$ respectively, its facility (allocation).
  \item[] (\ref{OMT_sub_x_3_6}): Exactly $p-1$ edges connecting facilities (connectivity).
  \item[] (\ref{OMT_sub_x_3_7}): No subtours are allowed to be formed between facilities edges (connectivity). 
  \item[] (\ref{OMT_sub_x_3_8}): Relationship between sorting and allocation variables (sorting).
  \item[] (\ref{OMT_sub_x_3_9}): Every sort position must be consider (sorting).
  \item[] (\ref{OMT_sub_x_3_10}): Correct sorting of the costs sequence (sorting).
\end{itemize}

The cardinality of the set of subtour elimination constraints is exponential in the number of nodes. An effective algorithm can be implemented using a branch-and-cut ramification process to add dynamically in polynomial time a certain number of these constraints. Formulations $F2_{x}^{sub2}$, $F1_{u}^{sub1}$, $F1_{u}^{sub2}$, $F2_{u}^{sub1}$ and $F2_{u}^{sub2}$ are left to the reader.

\subsection{\OMT MTZ formulation}
\label{subsec:OMTMTZformulation}

Let $D=(V,A,r)$ be a rooted weighted directed network where $A$ the set of arcs and $r$ is a root node. An \textit{arborescence} of $D$ is a subgraph $D' = (V,A',r)$ where $A'\subseteq A$ such that each non-root node has exactly one incoming (or outcoming) edge (thus $|A'| = |V|-1$) and $D'$ has no cycles. The \textit{Miller-Tucker-Zemlin} formulation (MTZ) builds an arborescence rooted at a arbitrarily selected root node $r\in V$, in which arcs follow the direction from root to leaves.

\begingroup
\allowdisplaybreaks
\begin{subequations}
\begin{align}
  & \hhp \hspace{1cm} F1_x^{mtz}: \min \quad \frac{1}{\sum_{\ell \in V} \lambda_\ell} \sum_{\ell \in V} \sum_{(i,j) \in A} \lambda_\ell c_{ij} x_{ij}^\ell + \frac{1}{p-1} \sum_{(i,j) \in E} c_{ij} z_{ij} & & & \label{OMT_mtz_x_1_0} & \\
  & \hp \sum_{i \in V} x_{ii} = p                       &   &                               & \label{OMT_mtz_x_1_1}  & \\
  & \Hp \sum_{j \in V} x_{ij} = 1                       &   & i \in V                       & \label{OMT_mtz_x_1_2}  & \\
  & \Hp x_{ij} \leq x_{jj}                              &   & (i,j) \in A: i \neq j         & \label{OMT_mtz_x_1_3}  & \\
  & \Hp 2 z_{ij} \leq x_{ii} + x_{jj}                   &   & (i,j) \in E                   & \label{OMT_mtz_x_1_4}  & \\
  & \Hp \sum_{(i,j) \in E} z_{ij} = p-1                 &   &                               & \label{OMT_mtz_x_1_5}  & \\
  & \Hp \sum_{(j,i) \in A}y_{ji} = 1                    &   & i \in V \setminus \{r\}       & \label{OMT_mtz_x_1_6}  & \\
  & \Hp y_{ij} + y_{ji} = z_{ij}                        &   & (i,j) \in E                   & \label{OMT_mtz_x_1_7}  & \\
  & \Hp l_j \geq l_i + 1 - N(1-y_{ij})                  &   & (i,j) \in A : i \neq j        & \label{OMT_mtz_x_1_8}  & \\
  & \Hp l_r = 1                                         &   &                               & \label{OMT_mtz_x_1_9}  & \\
  & \Hp 2 \leq l_i \leq p                               &   & i \in V \setminus \{r\}       & \label{OMT_mtz_x_1_10}  & \\
  & \Hp \sum_{\ell \in V} x_{ij}^\ell = x_{ij}          &   & (i,j) \in A                   & \label{OMT_mtz_x_1_11}  & \\
  & \Hp \sum_{(i,j) \in A} x_{ij}^\ell = 1              &   & \ell \in V                    & \label{OMT_mtz_x_1_12}  & \\
  & \Hp \sum_{(i,j) \in A} c_{ij}x_{ij}^\ell \leq \sum_{(i,j) \in A} c_{ij}x_{ij}^{\ell+1}  &   & \ell \in V: \ell < |V|  & \label{OMT_mtz_x_1_13} & \\
  & \Hp x_{ij} \in \{0,1\}                              &   & (i,j) \in A \\
  & \Hp x_{ij}^\ell \in \{0,1\}                         &   & (i,j) \in A, \ell \in V \\
  & \Hp z_{ij} \in \{0,1\}                              &   & (i,j) \in E \\
  & \Hp y_{ij} \in \{0,1\}                              &   & (i,j) \in A \\
  & \Hp l_i \geq 0                                      &   & i \in V.
\end{align}
\end{subequations}
\endgroup

\begin{itemize}[noitemsep]
  \item[] (\ref{OMT_mtz_x_1_0}): Weighted minimization of the compensated allocation cost of the system plus the design cost of the tree of facilities (objective).
  \item[] (\ref{OMT_mtz_x_1_1}): Exactly $p$ facilities (allocation).
  \item[] (\ref{OMT_mtz_x_1_2}): Each client is allocated to exactly one facility (allocation).
  \item[] (\ref{OMT_mtz_x_1_3}): No client is allocated to a non-facility (allocation).
  \item[] (\ref{OMT_mtz_x_1_4}): Edges of the tree of facilities can only be selected if both nodes are facilities (allocation).
  \item[] (\ref{OMT_mtz_x_1_5}): Exactly $p-1$ edges connecting facilities (connectivity).
  \item[] (\ref{OMT_mtz_x_1_6}): Exactly one arc goes into a non-root node of the arborescence (connectivity).
  \item[] (\ref{OMT_mtz_x_1_7}): If considering an edge between facilities, the arborescence includes one of the corresponding arcs (connectivity).
  \item[] (\ref{OMT_mtz_x_1_8}): If $(i,j)$ arc is considered in the arborescence, then the position of $j$ must be higher than $i$ (connectivity).
  \item[] (\ref{OMT_mtz_x_1_9}) - (\ref{OMT_mtz_x_1_10}): The root node is in 1st position of the arborescence and the non-root nodes must be distributed between the 2nd and $p$-th positions (connectivity).
  \item[] (\ref{OMT_mtz_x_1_11}): Relationship between sorting and allocation variables (sorting).
  \item[] (\ref{OMT_mtz_x_1_12}): Every sort position must be consider (sorting).
  \item[] (\ref{OMT_mtz_x_1_13}): Correct sorting of the costs sequence (sorting).
\end{itemize}

\begingroup
\allowdisplaybreaks
\begin{subequations}
\begin{align}
  & \hhp \hspace{1cm} F2_x^{mtz}: \min \quad \frac{1}{\displaystyle{\sum_{\ell \in V} \lambda_\ell}} \sum_{\ell \in V} \sum_{(i,j) \in A} \lambda_\ell c_{ij} x_{ij}^\ell + \frac{1}{p-1} \sum_{(i,j) \in E} c_{ij} (z_{ij}-x_{ij}-x_{ji})  & & & \label{OMT_mtz_x_2_0} & \\
  & \hp \sum_{i \in V} x_{ii} = p                       &   &                               & \label{OMT_mtz_x_2_1}   & \\
  & \Hp \sum_{j \in V} x_{ij} = 1                       &   & i \in V                       & \label{OMT_mtz_x_2_2}   & \\
  & \Hp 2x_{ij} \leq 1 - x_{ii} + x_{jj}                &   & (i,j) \in A: i\neq j          & \label{OMT_mtz_x_2_3}   & \\
  & \Hp x_{ij} + x_{ji} \leq z_{ij}                     &   & (i,j) \in E                   & \label{OMT_mtz_x_2_4}   & \\
  & \Hp 2z_{ij} \leq x_{ii} + x_{jj} + x_{ij} + x_{ji}  &   & (i,j) \in E                   & \label{OMT_mtz_x_2_5}   & \\
  & \Hp \sum_{(i,j) \in E} z_{ij} = N-1                 &   &                               & \label{OMT_mtz_x_2_6}   & \\
  & \Hp \sum_{(j,i) \in A}y_{ji} = 1                    &   & i \in V \setminus \{r\}       & \label{OMT_mtz_x_2_7}   & \\
  & \Hp y_{ij} + y_{ji} = z_{ij}                        &   & (i,j) \in E                   & \label{OMT_mtz_x_2_8}   & \\
  & \Hp l_j \geq l_i + 1 - N(1-y_{ij})                  &   & (i,j) \in A: i \neq j         & \label{OMT_mtz_x_2_9}   & \\
  & \Hp l_r = 1                                         &   &                               & \label{OMT_mtz_x_2_10}  & \\
  & \Hp 2 \leq l_i \leq N                               &   & i \in V \setminus \{r\}       & \label{OMT_mtz_x_2_11}  & \\
  & \Hp \sum_{\ell \in V} x_{ij}^\ell = x_{ij}          &   & (i,j) \in A                   & \label{OMT_mtz_x_2_12}  & \\
  & \Hp \sum_{(i,j) \in A} x_{ij}^\ell = 1              &   & \ell \in V                    & \label{OMT_mtz_x_2_13}  & \\
  & \Hp \sum_{(i,j) \in A} c_{ij}x_{ij}^\ell \leq \sum_{(i,j) \in A} c_{ij}x_{ij}^{\ell+1}  &   &  \ell \in V: \ell < |V|  & \label{OMT_mtz_x_2_14} & \\
  & \Hp x_{ij} \in \{0,1\}                              &   & (i,j) \in A \\
  & \Hp x_{ij}^\ell \in \{0,1\}                         &   & (i,j) \in A, \ell \in V \\
  & \Hp z_{ij} \in \{0,1\}                              &   & (i,j) \in E \\
  & \Hp y_{ij} \in \{0,1\}                              &   & (i,j) \in A \\
  & \Hp l_i \geq 0                                      &   & i \in V.
\end{align}
\end{subequations}
\endgroup

\begin{itemize}[noitemsep]
  \item[] (\ref{OMT_mtz_x_2_0}): Weighted minimization of the compensated allocation cost of the system plus the design cost of the tree subtracting in the design cost part the cost associated to the edges that represent the allocation of client to facilities (objective).
  \item[] (\ref{OMT_mtz_x_2_1}): Exactly $p$ facilities (allocation). 
  \item[] (\ref{OMT_mtz_x_2_2}): Each client is allocated to exactly one facility (allocation).
  \item[] (\ref{OMT_mtz_x_2_3}): Client-facility allocations are only possible if $i$ is a client and $j$ is a facility (allocation).
  \item[] (\ref{OMT_mtz_x_2_4}): Only one allocation, whether client $i$ to facility $j$ or viceversa, can be considered if there exists an edge connecting $i,j \in V$ (allocation).
  \item[] (\ref{OMT_mtz_x_2_5}): Only possibilities in which an edge can be selected is if both $i$ or $j$ are facilities or if either $i$, or $j$, is a client and $j$, or $i$ respectively, its facility (allocation).
  \item[] (\ref{OMT_mtz_x_2_6}): Exactly $p-1$ edges connecting facilities (connectivity).
  \item[] (\ref{OMT_mtz_x_2_7}): Exactly one arc goes into a non-root node of the arborescence (connectivity).
  \item[] (\ref{OMT_mtz_x_2_8}): If considering an edge between facilities, the arborescence includes one of the corresponding arcs (connectivity).
  \item[] (\ref{OMT_mtz_x_2_9}): If $(i,j)$ arc is considered in the arborescence, then the position of $j$ must be higher than $i$ (connectivity).
  \item[] (\ref{OMT_mtz_x_2_10}) - (\ref{OMT_mtz_x_2_11}): The root node is in 1st position of the arborescence and the non-root nodes must be distributed between the 2nd and $p$-th positions (connectivity). 
  \item[] (\ref{OMT_mtz_x_2_12}): Relationship between sorting and allocation variables (sorting).
  \item[] (\ref{OMT_mtz_x_2_13}): Every sort position must be consider (sorting).
  \item[] (\ref{OMT_mtz_x_2_14}): Correct sorting of the costs sequence (sorting).
\end{itemize}

Formulations $F1_{u}^{mtz}$ and $F2_{u}^{mtz}$ are left to the reader.

\subsection{\OMT flow based formulations}
\label{subsec:OMTflowbasedformulations}

The flow formulation also relies on a source node $r\in V$ which distributes the flow but for the \OMT, contrary to the MTZ formulation, the choice of the root node is very influential as care must be taken when spreading the flow. This selection can be done in two ways:

\vspace{-0.25cm}

\begin{itemize}[noitemsep]
    \item[$\diamond$] Adding a set of variables $r_{i} \in \{0,1\}$ for $i \in V$, is 1 if the node $i \in V$ is selected as the source node for the tree of facilities.
    \item[$\diamond$] Arbitrarily selecting the source node and distributing the flow along the tree, distinguishing whether the node selected as the source is a facility or not.
\end{itemize}

\vspace{-0.25cm}


\begin{itemize}
  \item \textbf{Formulation using additional variables}
\end{itemize}

\vspace{-0.5cm}

\begingroup
\allowdisplaybreaks
\begin{subequations}
\begin{align}
  & \hhp \hspace{1cm} F1_x^{flow1}: \min \quad \frac{1}{\sum_{\ell \in V} \lambda_\ell} \sum_{\ell \in V} \sum_{(i,j) \in A} \lambda_\ell c_{ij} x_{ij}^\ell + \frac{1}{p-1} \sum_{(i,j) \in E} c_{ij} z_{ij} & & & \label{OMT_flow_x_1_0} & \\
  & \hp \sum_{i \in V} x_{ii} = p                                       &   &                             & \label{OMT_flow_x_1_1}  & \\
  & \Hp \sum_{j \in V} x_{ij} = 1                                       &   & i \in V                     & \label{OMT_flow_x_1_2}  & \\
  & \Hp x_{ij} \leq x_{jj}                                              &   & (i,j) \in A: i \neq j       & \label{OMT_flow_x_1_3}  & \\
  & \Hp 2z_{ij} \leq x_{ii} + x_{jj}                                    &   & (i,j) \in E                 & \label{OMT_flow_x_1_4}  & \\
  & \Hp \sum_{(i,j) \in E} z_{ij} = p-1                                 &   &                             & \label{OMT_flow_x_1_5}  & \\
  & \Hp \sum_{i\in V} r_i = 1                                           &   &                             & \label{OMT_flow_x_1_6}  & \\
  & \Hp r_i \leq x_{ii}                                                 &   & i \in V                     & \label{OMT_flow_x_1_7}  & \\
  & \Hp \sum_{(i,j) \in \delta^+(i)}f_{ij} - \sum_{(j,i) \in \delta^-(i)}f_{ji} = (p-1)r_i-(x_{ii}-r_{i}) & & i \in V &    & \label{OMT_flow_x_1_8}  & \\
  & \Hp f_{ij} \leq (p-1)z_{ij}                                         &   & (i,j) \in E                 & \label{OMT_flow_x_1_9}  & \\
  & \Hp f_{ji} \leq (p-1)z_{ij}                                         &   & (i,j) \in E                 & \label{OMT_flow_x_1_10} & \\
  & \Hp \sum_{\ell \in V} x_{ij}^\ell = x_{ij}                          &   & (i,j) \in A                 & \label{OMT_flow_x_1_11}  & \\
  & \Hp \sum_{(i,j) \in A} x_{ij}^\ell = 1                              &   & \ell \in V                  & \label{OMT_flow_x_1_12}  & \\
  & \Hp \sum_{(i,j) \in A} c_{ij}x_{ij}^\ell \leq \sum_{(i,j) \in A} c_{ij}x_{ij}^{\ell+1}                & & \ell \in V: \ell < |V| & \label{OMT_flow_x_1_13} & \\   
  & \Hp x_{ij} \in \{0,1\}                                              &   & (i,j) \in A \\
  & \Hp x_{ij}^\ell \in \{0,1\}                                         &   & (i,j) \in A, \ell \in V \\
  & \Hp r_i \in \{0,1\}                                                 &   & i \in V \\
  & \Hp z_{ij} \in \{0,1\}                                              &   & (i,j) \in E \\
  & \Hp f_{ij} \geq 0                                                   &   & (i,j) \in A: i \neq j.
\end{align}
\end{subequations}
\endgroup

\begin{itemize}[noitemsep]
  \item[] (\ref{OMT_flow_x_1_0}): Weighted minimization of the total compensated allocation cost of the system plus the design cost of the tree of facilities (objective).
  \item[] (\ref{OMT_flow_x_1_1}): Exactly $p$ facilities (allocation).
  \item[] (\ref{OMT_flow_x_1_2}): Each client is allocated to exactly one facility (allocation).
  \item[] (\ref{OMT_flow_x_1_3}): No client is allocated to a non-facility (allocation).
  \item[] (\ref{OMT_flow_x_1_4}): Edges of the tree of facilities can only be selected if both nodes are facilities (allocation).
  \item[] (\ref{OMT_flow_x_1_5}): Exactly $p-1$ edges connecting facilities (connectivity).
  \item[] (\ref{OMT_flow_x_1_6}): Only one root node selected (connectivity).
  \item[] (\ref{OMT_flow_x_1_7}): Only facilities nodes can be activated as a root (connectivity).
  \item[] (\ref{OMT_flow_x_1_8}): Flow is properly distributed along the tree of facilities (connectivity).
  \item[] (\ref{OMT_flow_x_1_9})-(\ref{OMT_flow_x_1_10}): All variables sending flow must be activated (connectivity).
  \item[] (\ref{OMT_flow_x_1_11}): Relationship between sorting and allocation variables (sorting).
  \item[] (\ref{OMT_flow_x_1_12}): Every sort position must be consider (sorting).
  \item[] (\ref{OMT_flow_x_1_13}): Correct sorting of the costs sequence (sorting).
\end{itemize}

\begin{itemize}
  \item \textbf{Formulation without using additional variables}
\end{itemize}

\vspace{-0.5cm}

\begingroup
\allowdisplaybreaks
\begin{subequations}
\begin{align}
  & \hhp \hspace{1cm} F1_x^{flow2}: \min \quad \frac{1}{\sum_{\ell \in V} \lambda_\ell} \sum_{\ell \in V} \sum_{(i,j) \in A} \lambda_\ell c_{ij} x_{ij}^\ell + \frac{1}{p-1} \sum_{(i,j) \in E} c_{ij} z_{ij} & & & \label{OMT_flow_x_2_0} & \\
  & \hp \sum_{i \in V} x_{ii} = p                                       &   &                        & \label{OMT_flow_x_2_1}  & \\
  & \Hp \sum_{j \in V} x_{ij} = 1                                       &   & i \in V                & \label{OMT_flow_x_2_2}  & \\
  & \Hp x_{ij} \leq x_{jj}                                              &   & (i,j) \in A: i \neq j  & \label{OMT_flow_x_2_3}  & \\
  & \Hp 2z_{ij} \leq x_{ii} + x_{jj}                                    &   & (i,j) \in E            & \label{OMT_flow_x_2_4}  & \\
  & \Hp \sum_{(i,j) \in E} z_{ij} = p-1                                 &   &                        & \label{OMT_flow_x_2_5}  & \\
  & \Hp \sum_{(i,j) \in \delta^+(i)}f_{ij} - \sum_{(j,i) \in \delta^-(i)}f_{ji} = p x_{ri} - x_{ii}  &   & i \in V & \label{OMT_flow_x_2_6} & \\
  & \Hp f_{ij} \leq (p-1)z_{ij}                                         &   & (i,j) \in E            & \label{OMT_flow_x_2_7}  & \\
  & \Hp f_{ji} \leq (p-1)z_{ij}                                         &   & (i,j) \in E            & \label{OMT_flow_x_2_8}  & \\
  & \Hp \sum_{\ell \in V} x_{ij}^\ell = x_{ij}                          &   & (i,j) \in A            & \label{OMT_flow_x_2_9}  & \\
  & \Hp \sum_{(i,j) \in A} x_{ij}^\ell = 1                              &   & \ell \in V             & \label{OMT_flow_x_2_10} & \\
  & \Hp \sum_{(i,j) \in A} c_{ij}x_{ij}^\ell \leq \sum_{(i,j) \in A} c_{ij}x_{ij}^{\ell+1}           &   & \ell \in V: \ell < |V| & \label{OMT_flow_x_2_11} & \\  
  & \Hp x_{ij} \in \{0,1\}                                              &   & (i,j) \in A \\
  & \Hp x_{ij}^\ell \in \{0,1\}                                         &   & (i,j) \in A, \ell \in V \\
  & \Hp z_{ij} \in \{0,1\}                                              &   & (i,j) \in E \\
  & \Hp f_{ij} \geq 0                                                   &   & (i,j) \in A: i \neq j.
\end{align}
\end{subequations}
\endgroup

\begin{itemize}[noitemsep]
  \item[] (\ref{OMT_flow_x_2_0}): Weighted minimization of the total compensated allocation cost plus the design cost of the tree of facilities (objective).
  \item[] (\ref{OMT_flow_x_2_1}): Exactly $p$ facilities (allocation).
  \item[] (\ref{OMT_flow_x_2_2}): Each client is allocated to exactly one facility (allocation).
  \item[] (\ref{OMT_flow_x_2_3}): No client is allocated to a non-facility (allocation).
  \item[] (\ref{OMT_flow_x_2_4}): Edges of the tree of facilities can only be activated if both nodes are facilities (allocation).
  \item[] (\ref{OMT_flow_x_2_5}): Exactly $p-1$ edges connecting facilities (connectivity).
  \item[] (\ref{OMT_flow_x_2_6}): Flow is properly distributed along the tree of facilities (connectivity).
  \item[] (\ref{OMT_flow_x_2_7})-(\ref{OMT_flow_x_2_8}): All variables sending flow must be activated (connectivity).
  \item[] (\ref{OMT_flow_x_2_9}): Relationship between sorting and allocation variables (sorting).
  \item[] (\ref{OMT_flow_x_2_10}): Every sort position must be consider (sorting).
  \item[] (\ref{OMT_flow_x_2_11}): Correct sorting of the costs sequence (sorting).
\end{itemize}

The flow based formulation modeling a tree in $V$ can be expressed as follows:

\vspace{-0.5cm}

\begingroup
\allowdisplaybreaks
\begin{subequations}
\begin{align}
  & \hhp \hspace{1cm} F2_x^{flow}: \min \quad \frac{1}{\displaystyle{\sum_{\ell \in V} \lambda_\ell}} \sum_{\ell \in V} \sum_{(i,j) \in A} \lambda_\ell c_{ij} x_{ij}^\ell + \frac{1}{p-1} \sum_{(i,j) \in E} c_{ij} (z_{ij}-x_{ij}-x_{ji})  & & & \label{OMT_flow_x_3_0} & \\
  & \hp \sum_{i \in V} x_{ii} = p                       &   &                               & \label{OMT_flow_x_3_1}  & \\
  & \Hp \sum_{j \in V} x_{ij} = 1                       &   & i \in V                       & \label{OMT_flow_x_3_2}  & \\
  & \Hp 2x_{ij} \leq 1 - x_{ii} + x_{jj}                &   & (i,j) \in A: i\neq j          & \label{OMT_flow_x_3_3}  & \\
  & \Hp x_{ij} + x_{ji} \leq z_{ij}                     &   & (i,j) \in E                   & \label{OMT_flow_x_3_4}  & \\
  & \Hp 2z_{ij} \leq x_{ii} + x_{jj} + x_{ij} + x_{ji}  &   & (i,j) \in E                   & \label{OMT_flow_x_3_5}  & \\
  & \Hp \sum_{(i,j) \in E} z_{ij} = N-1                 &   &                               & \label{OMT_flow_x_3_6}  & \\
  & \Hp \sum_{(r,j) \in \delta^+(r)}f_{rj} - \sum_{(j,r) \in \delta^-(r)}f_{jr} = N-1 &     & & \label{OMT_flow_x_3_7} & \\
  & \Hp \sum_{(i,j) \in \delta^+(i)}f_{ij} - \sum_{(j,i) \in \delta^-(i)}f_{ji} = -1  &     & i \in V \setminus{r} & \label{OMT_flow_x_3_8} & \\
  & \Hp f_{ij} \leq (p-1)z_{ij}                         &   & (i,j) \in E                   & \label{OMT_flow_x_3_9}  & \\
  & \Hp f_{ji} \leq (p-1)z_{ij}                         &   & (i,j) \in E                   & \label{OMT_flow_x_3_10}  & \\
  & \Hp \sum_{\ell \in V} x_{ij}^\ell = x_{ij}          &   & (i,j) \in A                   & \label{OMT_flow_x_3_11}  & \\
  & \Hp \sum_{(i,j) \in A} x_{ij}^\ell = 1              &   & \ell \in V                    & \label{OMT_flow_x_3_12} & \\
  & \Hp \sum_{(i,j) \in A} c_{ij}x_{ij}^\ell \leq \sum_{(i,j) \in A} c_{ij}x_{ij}^{\ell+1}  & & \ell \in V: \ell < |V|  & \label{OMT_flow_x_3_13} & \\  
  & \Hp x_{ij} \in \{0,1\}                              &   & (i,j) \in A \\
  & \Hp x_{ij}^\ell \in \{0,1\}                         &   & (i,j) \in A,\ell \in V \\
  & \Hp z_{ij} \in \{0,1\}                              &   & (i,j) \in E \\
  & \Hp f_{ij} \geq 0                                   &   & (i,j) \in A: i \neq j.
\end{align}
\end{subequations}
\endgroup

\begin{itemize}[noitemsep]
  \item[] (\ref{OMT_flow_x_3_0}): Weighted minimization of the total compensated allocation cost plus the design cost of the tree of facilities (objective).
  \item[] (\ref{OMT_flow_x_3_1}): Exactly $p$ facilities (allocation).
  \item[] (\ref{OMT_flow_x_3_2}): Each client is allocated to exactly one facility (allocation).
  \item[] (\ref{OMT_flow_x_3_3}): Client-facility allocations are only possible if $i$ is a client and $j$ is a facility (allocation).
  \item[] (\ref{OMT_flow_x_3_4}): Only one allocation, whether client $i$ to facility $j$ or viceversa, can be considered if there exists an edge connecting $i,j \in V$ (allocation).
  \item[] (\ref{OMT_flow_x_3_5}): Only possibilities in which an edge can be selected is if both $i$ or $j$ are facilities or if either $i$, or $j$, is a client and $j$, or $i$ respectively, its facility (allocation).
  \item[] (\ref{OMT_flow_x_3_6}): Exactly $p-1$ edges connecting facilities (connectivity).
  \item[] (\ref{OMT_flow_x_3_7})-(\ref{OMT_flow_x_3_8}): Flow is properly distributed along the tree of facilities (connectivity).
  \item[] (\ref{OMT_flow_x_3_9})-(\ref{OMT_flow_x_3_10}): All variables sending flow must be activated (connectivity).
  \item[] (\ref{OMT_flow_x_3_11}): Relationship between sorting and allocation variables (sorting).
  \item[] (\ref{OMT_flow_x_3_12}): Every sort position must be consider (sorting).
  \item[] (\ref{OMT_flow_x_3_13}): Correct sorting of the costs sequence (sorting).
\end{itemize}

Formulations $F1_{u}^{flow1}$, $F1_{u}^{flow2}$ and $F2_{u}^{flow}$ are left to the reader.

\subsection{\OMT KM formulation}
\label{subsec:OMTKMformulation}

\begingroup
\allowdisplaybreaks
\begin{subequations}
\begin{align}
  & \hhp \hspace{1cm} F1_x^{km}: \min \quad \frac{1}{\sum_{\ell \in V} \lambda_\ell} \sum_{\ell \in V} \sum_{(i,j) \in A} \lambda_\ell c_{ij} x_{ij}^\ell + \frac{1}{p-1} \sum_{(i,j) \in E} c_{ij} z_{ij} & & & \label{OMT_km_x_1_0} & \\
  & \hp \sum_{i \in V} x_{ii} = p                       &   &                               & \label{OMT_km_x_1_1}  & \\
  & \Hp \sum_{j \in V} x_{ij} = 1                       &   & i \in V                       & \label{OMT_km_x_1_2}  & \\
  & \Hp x_{ij} \leq x_{jj}                              &   & (i,j) \in A: i \neq j         & \label{OMT_km_x_1_3}  & \\
  & \Hp 2 z_{ij} \leq x_{ii} + x_{jj}                   &   & (i,j) \in E                   & \label{OMT_km_x_1_4}  & \\
  & \Hp \sum_{(i,j) \in E} z_{ij} = p-1                 &   &                               & \label{OMT_km_x_1_5}  & \\
  & \Hp q_{kij} + q_{kji} = z_{ij}                      &   & k \in V, (i,j) \in E          & \label{OMT_km_x_1_6}  & \\
  & \Hp \sum_{(k,j) \in \delta^{+}(k)}q_{kkj} \leq 0    &   & k \in V                       & \label{OMT_km_x_1_7}  & \\
  & \Hp \sum_{(i,j) \in \delta^{+}(i)}q_{kij} \leq 1    &   & k,i \in V: i \neq k           & \label{OMT_km_x_1_8}  & \\
  & \Hp \sum_{\ell \in V} x_{ij}^\ell = x_{ij}          &   & (i,j) \in A                   & \label{OMT_km_x_1_9}  & \\
  & \Hp \sum_{(i,j) \in A} x_{ij}^\ell = 1              &   & \ell \in V                    & \label{OMT_km_x_1_10}  & \\
  & \Hp \sum_{(i,j) \in A} c_{ij}x_{ij}^\ell \leq \sum_{(i,j) \in A} c_{ij}x_{ij}^{\ell+1}  & & \ell \in V: \ell < |V|  & \label{OMT_km_x_1_11} & \\
  & \Hp x_{ij} \in \{0,1\}                              &   & (i,j) \in A \\
  & \Hp x_{ij}^\ell \in \{0,1\}                         &   & (i,j) \in A, \ell \in V \\
  & \Hp z_{ij} \in \{0,1\}                              &   & (i,j) \in E \\
  & \Hp q_{kij} \geq 0                                  &   & (i,j) \in A, k \in V.
\end{align}
\end{subequations}
\endgroup

\begin{itemize}[noitemsep]
  \item[] (\ref{OMT_km_x_1_0}): Weighted minimization of the compensated allocation cost of the system plus the design cost of the tree of facilities (objective).
  \item[] (\ref{OMT_km_x_1_1}): Exactly $p$ facilities (allocation).
  \item[] (\ref{OMT_km_x_1_2}): Each client is allocated to exactly one facility (allocation).
  \item[] (\ref{OMT_km_x_1_3}): No client is allocated to a non-facility (allocation).
  \item[] (\ref{OMT_km_x_1_4}): Edges of the tree of facilities can only be selected if both nodes are facilities (allocation).
  \item[] (\ref{OMT_km_x_1_5}): Exactly $p-1$ edges connecting facilities (connectivity).
  \item[] (\ref{OMT_km_x_1_6}): If $(i,j) \in E$ is a tree edge, then there is only one variable related to the arcs underlying that can be selected (connectivity).
  \item[] (\ref{OMT_km_x_1_7}): Forbids any arc leaving the root node $k$ (connectivity).
  \item[] (\ref{OMT_km_x_1_8}): Impose that no more that one arc leaves any node different from the root $k$ (connectivity).
  \item[] (\ref{OMT_km_x_1_9}): Relationship between sorting and allocation variables (sorting).
  \item[] (\ref{OMT_km_x_1_10}): Every sort position must be consider (sorting).
  \item[] (\ref{OMT_km_x_1_11}): Correct sorting of the costs sequence (sorting).
\end{itemize}

Formulations $F1_{u}^{km}$, $F2_{x}^{km}$ and $F2_{u}^{km}$ are left to the reader.

\newpage

\section{Appendix D: Preliminary \OMT results}
\label{AppendixD}

\subsection{Median criterion}

\vspace{0.5cm}

\begingroup
\tiny
\begin{longtable}{rrrrrrrrrrrrrrr}
\caption{Instances results table for model $F1_{x^{\ell}}^{mtz}$} \\ 
$|V|$ & p & ins & cpu & objU & objL & objR & obj$\overline{\text{U}}$ & obj$\overline{\text{L}}$ & gUL & g$\overline{\text{U}}$L & gU$\overline{\text{L}}$ & g$\overline{\text{U}}$R & opt & nod \\ 
  \hline
    20 & 5 & 1 & 6.6     & 15831.2 & 15831.2 & 8678.3   & 15831.2 & 15831.2 & 0.0  & 0.0 & 0.0  & 45.2  & 1 & 1711 \\ 
    20 & 5 & 2 & 8.3     & 16273.0 & 16273.0 & 11464.7  & 16273.0 & 16273.0 & 0.0  & 0.0 & 0.0  & 29.5  & 1 & 934 \\ 
    20 & 5 & 3 & 28.0    & 15039.1 & 15039.1 & 8966.5   & 15039.1 & 15039.1 & 0.0  & 0.0 & 0.0  & 40.4  & 1 & 1469 \\ 
    20 & 5 & 4 & 12.1    & 9076.2  & 9076.2  & 5449.0   & 9076.2  & 9076.2  & 0.0  & 0.0 & 0.0  & 40.0  & 1 & 641 \\ 
    20 & 5 & 5 & 3.9     & 16943.6 & 16943.6 & 11867.1  & 16943.6 & 16943.6 & 0.0  & 0.0 & 0.0  & 30.0  & 1 & 1 \\ 
    20 & 6 & 1 & 4.2     & 11799.8 & 11799.8 & 7620.6   & 11799.8 & 11799.8 & 0.0  & 0.0 & 0.0  & 35.4  & 1 & 1 \\
\hline
\end{longtable}
\endgroup

\begingroup
\tiny
\begin{longtable}{rrrrrrrrr}
\caption{Summary results table for model $F1_{x^{\ell}}^{mtz}$} \\ 
$|V|$ & p & $|$\#$|$ & cpu & g$\overline{\text{U}}$R & g$\overline{\text{U}}$L & gU$\overline{\text{L}}$ & gUL & nod \\ 
  \hline
  20 &  5 & 5 &   11.8 & 37.0 & 0.0 & 0.0 & 0.0 & 951 \\ 
  20 &  6 & 5 &   12.2 & 35.0 & 0.0 & 0.0 & 0.0 & 762 \\ 
  20 & 10 & 5 &    6.4 & 28.9 & 0.0 & 0.0 & 0.0 & 516 \\ 
  30 &  7 & 5 &  127.5 & 39.6 & 0.0 & 0.0 & 0.0 & 1005 \\ 
  30 & 10 & 5 &   59.0 & 33.1 & 0.0 & 0.0 & 0.0 & 1110 \\ 
  30 & 15 & 5 &   47.3 & 32.9 & 0.0 & 0.0 & 0.0 & 2090 \\ 
  40 & 10 & 5 &  380.3 & 33.3 & 0.0 & 0.0 & 0.0 & 1669 \\ 
  40 & 13 & 5 &  448.3 & 35.4 & 0.0 & 0.0 & 0.0 & 3080 \\ 
  40 & 20 & 5 &  141.2 & 28.6 & 0.0 & 0.0 & 0.0 & 1293 \\ 
  50 & 12 & 2 & 2944.8 & 35.8 & 1.4 & 14.0 & 15.1 & 2707 \\ 
  50 & 16 & 5 & 1342.5 & 35.7 & 0.0 & 0.0 & 0.0 & 1908 \\ 
  50 & 25 & 5 &  473.8 & 28.6 & 0.0 & 0.0 & 0.0 & 3616 \\ 
\hline
\end{longtable}
\endgroup

\newpage

\begingroup
\tiny
\begin{longtable}{rrrrrrrrrrrrrrr}
\caption{Instances results table for model $F1_{x^{\ell}}^{flow1}$} \\ 
$|V|$ & p & ins & cpu & objU & objL & objR & obj$\overline{\text{U}}$ & obj$\overline{\text{L}}$ & gUL & g$\overline{\text{U}}$L & gU$\overline{\text{L}}$ & g$\overline{\text{U}}$R & opt & nod \\  
  \hline
  20 & 5 & 1 & 3.2     & 15831.2 & 15831.2 & 8678.2    & 15831.2 & 15831.2 & 0.0 & 0.0 & 0.0   & 45.2 & 1 & 47 \\ 
  20 & 5 & 2 & 6.8     & 16273.0 & 16273.0 & 11464.7   & 16273.0 & 16273.0 & 0.0 & 0.0 & 0.0   & 29.5 & 1 & 419 \\ 
  20 & 5 & 3 & 11.1    & 15039.1 & 15039.1 & 8966.5    & 15039.1 & 15039.1 & 0.0 & 0.0 & 0.0   & 40.4 & 1 & 952 \\ 
  20 & 5 & 4 & 6.1     & 9076.2  & 9076.2  & 5449.0    & 9076.2  & 9076.2  & 0.0 & 0.0 & 0.0   & 40.0 & 1 & 1 \\ 
  20 & 5 & 5 & 4.3     & 16943.6 & 16943.6 & 11867.1   & 16943.6 & 16943.6 & 0.0 & 0.0 & 0.0   & 30.0 & 1 & 1 \\ 
  20 & 6 & 1 & 19.3    & 11799.8 & 11799.8 & 7620.6    & 11799.8 & 11799.8 & 0.0 & 0.0 & 0.0   & 35.4 & 1 & 1062 \\
\hline
\end{longtable}
\endgroup

\begingroup
\tiny
\begin{longtable}{rrrrrrrrr}
\caption{Summary results table for model $F1_{x^{\ell}}^{flow1}$} \\ 
$|V|$ & p & $|$\#$|$ & cpu & g$\overline{\text{U}}$R & g$\overline{\text{U}}$L & gU$\overline{\text{L}}$ & gUL & nod \\ 
  \hline
  20 &  5 & 5 & 6.3    & 37.0  & 0.0 & 0.0 & 0.0 & 284 \\ 
  20 &  6 & 5 & 10.2   & 35.0  & 0.0 & 0.0 & 0.0 & 741 \\ 
  20 & 10 & 5 & 5.7    & 28.9  & 0.0 & 0.0 & 0.0 & 404 \\ 
  30 &  7 & 5 & 106.2  & 39.6  & 0.0 & 0.0 & 0.0 & 949 \\ 
  30 & 10 & 5 & 62.4   & 33.1  & 0.0 & 0.0 & 0.0 & 1531 \\ 
  30 & 15 & 5 & 44.2   & 33.0  & 0.0 & 0.0 & 0.0 & 710 \\ 
  40 & 10 & 5 & 502.2  & 33.3  & 0.0 & 0.0 & 0.0 & 2353 \\ 
  40 & 13 & 5 & 701.1  & 35.4  & 0.0 & 0.0 & 0.0 & 6339 \\ 
  40 & 20 & 5 & 88.9   & 28.6  & 0.0 & 0.0 & 0.0 & 1611 \\ 
  50 & 12 & 4 & 2913.2 & 35.8  & 0.5 & 4.5 & 4.9 & 2874 \\ 
  50 & 16 & 4 & 1563.5 & 35.7  & 0.1 & 0.9 & 1.0 & 2769 \\ 
  50 & 25 & 4 & 927.2  & 28.6  & 0.5 & 0.0 & 0.5 & 12136 \\ 
\hline
\end{longtable}
\endgroup

\newpage

\begingroup
\tiny
\begin{longtable}{rrrrrrrrrrrrrrr}
\caption{Instances results table for model $F1_{x^{\ell}}^{flow2}$} \\ 
$|V|$ & p & ins & cpu & objU & objL & objR & obj$\overline{\text{U}}$ & obj$\overline{\text{L}}$ & gUL & g$\overline{\text{U}}$L & gU$\overline{\text{L}}$ & g$\overline{\text{U}}$R & opt & nod \\  
  \hline
  20 &  5 & 1 &    5.4 & 15831.2 & 15831.2 & 9767.5   & 15831.2 & 15831.2 &  0.0 & 0.0 &  0.0 & 38.3 & 1 & 776 \\ 
  20 &  5 & 2 &    7.2 & 16273.0 & 16273.0 & 11950.3  & 16273.0 & 16273.0 &  0.0 & 0.0 &  0.0 & 26.6 & 1 & 729 \\ 
  20 &  5 & 3 &    6.2 & 15039.1 & 15039.1 & 9701.2   & 15039.1 & 15039.1 &  0.0 & 0.0 &  0.0 & 35.5 & 1 & 1 \\ 
  20 &  5 & 4 &   12.6 &  9076.2 &  9076.2 & 5819.3   &  9076.2 &  9076.2 &  0.0 & 0.0 &  0.0 & 35.9 & 1 & 1098 \\ 
  20 &  5 & 5 &    9.3 & 16943.6 & 16943.6 & 12351.4  & 16943.6 & 16943.6 &  0.0 & 0.0 &  0.0 & 27.1 & 1 & 82 \\ 
  20 &  6 & 1 &   11.2 & 11799.8 & 11799.8 & 8334.0   & 11799.8 & 11799.8 &  0.0 & 0.0 &  0.0 & 29.4 & 1 & 135 \\
\hline
\end{longtable}
\endgroup

\begingroup
\tiny
\begin{longtable}{rrrrrrrrr}
\caption{Summary results table for model $F1_{x^{\ell}}^{flow2}$} \\ 
$|V|$ & p & $|$\#$|$ & cpu & g$\overline{\text{U}}$R & g$\overline{\text{U}}$L & gU$\overline{\text{L}}$ & gUL & nod \\ 
  \hline
  20 &  5 & 5 &    8.1 & 32.7 & 0.0 & 0.0 & 0.0 & 537 \\ 
  20 &  6 & 5 &    7.1 & 29.0 & 0.0 & 0.0 & 0.0 & 166 \\ 
  20 & 10 & 5 &    2.9 & 25.8 & 0.0 & 0.0 & 0.0 & 376 \\ 
  30 &  7 & 5 &  102.5 & 33.9 & 0.0 & 0.0 & 0.0 & 898 \\ 
  30 & 10 & 5 &   43.7 & 29.3 & 0.0 & 0.0 & 0.0 & 238 \\ 
  30 & 15 & 5 &   19.2 & 30.6 & 0.0 & 0.0 & 0.0 & 454 \\ 
  40 & 10 & 5 &  426.6 & 30.7 & 0.0 & 0.0 & 0.0 & 1422 \\ 
  40 & 13 & 5 &  395.2 & 31.5 & 0.0 & 0.0 & 0.0 & 750 \\ 
  40 & 20 & 5 &   67.1 & 25.4 & 0.0 & 0.0 & 0.0 & 14 \\ 
  50 & 12 & 2 & 3318.3 & 33.6 & 0.3 & 7.6 & 7.8 & 1996 \\ 
  50 & 16 & 5 & 1417.1 & 33.0 & 0.0 & 0.0 & 0.0 & 1741 \\ 
  50 & 25 & 5 &  349.9 & 26.8 & 0.0 & 0.0 & 0.0 & 1908 \\ 
\hline
\end{longtable}
\endgroup

\newpage

\begingroup
\tiny
\begin{longtable}{rrrrrrrrrrrrrrr}
\caption{Instances results table for model $F1_{x^{\ell}}^{km}$} \\ 
$|V|$ & p & ins & cpu & objU & objL & objR & obj$\overline{\text{U}}$ & obj$\overline{\text{L}}$ & gUL & g$\overline{\text{U}}$L & gU$\overline{\text{L}}$ & g$\overline{\text{U}}$R & opt & nod \\  
  \hline
  20 & 5 & 1  & 1.7    & 15831.2 & 15831.2 & 8678.3  & 15831.2 & 15831.2 & 0.0 & 0.0 & 0.0 & 45.2 & 1 & 1 \\ 
  20 & 5 & 2  & 12.7   & 16273.0 & 16273.0 & 11464.7 & 16273.0 & 16273.0 & 0.0 & 0.0 & 0.0 & 29.5 & 1 & 1045 \\ 
  20 & 5 & 3  & 7.3    & 15039.1 & 15039.1 & 8966.5  & 15039.1 & 15039.1 & 0.0 & 0.0 & 0.0 & 40.4 & 1 & 165 \\ 
  20 & 5 & 4  & 9.6    & 9076.2  & 9076.2  & 5448.9  & 9076.2  & 9076.2  & 0.0 & 0.0 & 0.0 & 40.0 & 1 & 1 \\ 
  20 & 5 & 5  & 1.4    & 16943.6 & 16943.6 & 11867.1 & 16943.6 & 16943.6 & 0.0 & 0.0 & 0.0 & 30.0 & 1 & 1 \\ 
  20 & 6 & 1  & 1.9    & 11799.8 & 11799.8 & 7620.6  & 11799.8 & 11799.8 & 0.0 & 0.0 & 0.0 & 35.4 & 1 & 1 \\
\hline
\end{longtable}
\endgroup

\begingroup
\tiny
\begin{longtable}{rrrrrrrrr}
\caption{Summary results table for model $F1_{x^{\ell}}^{km}$} \\ 
$|V|$ & p & $|$\#$|$ & cpu & g$\overline{\text{U}}$R & g$\overline{\text{U}}$L & gU$\overline{\text{L}}$ & gUL & nod \\ 
  \hline
  20 & 5 & 5 & 6.5     & 37.0 & 0.0 & 0.0 & 0.0 & 243 \\ 
  20 & 6 & 5 & 2.6     & 35.0 & 0.0 & 0.0 & 0.0 & 1 \\ 
  20 & 10 & 5 & 2.1    & 28.5 & 0.0 & 0.0 & 0.0 & 1 \\ 
  30 & 7 & 5 & 94.3    & 39.6 & 0.0 & 0.0 & 0.0 & 687 \\ 
  30 & 10 & 5 & 46.4   & 33.1 & 0.0 & 0.0 & 0.0 & 22 \\ 
  30 & 15 & 5 & 31.9   & 32.6 & 0.0 & 0.0 & 0.0 & 477 \\ 
  40 & 10 & 5 & 370.6  & 33.3 & 0.0 & 0.0 & 0.0 & 756 \\ 
  40 & 13 & 5 & 165.4  & 35.4 & 0.0 & 0.0 & 0.0 & 422 \\ 
  40 & 20 & 5 & 84.3   & 28.6 & 0.0 & 0.0 & 0.0 & 581 \\ 
  50 & 12 & 5 & 1414.4 & 35.8 & 0.0 & 0.0 & 0.0 & 1035 \\ 
  50 & 16 & 5 & 1460.9 & 35.7 & 0.0 & 0.0 & 0.0 & 377 \\ 
  50 & 25 & 5 & 528.0  & 28.5 & 0.0 & 0.0 & 0.0 & 320 \\ 
\hline
\end{longtable}
\endgroup

\newpage

\begingroup
\tiny
\begin{longtable}{rrrrrrrrrrrrrrr}
\caption{Instances results table for model $F1_{x^{\ell}}^{sub1}$} \\ 
$|V|$ & p & ins & cpu & objU & objL & objR & obj$\overline{\text{U}}$ & obj$\overline{\text{L}}$ & gUL & g$\overline{\text{U}}$L & gU$\overline{\text{L}}$ & g$\overline{\text{U}}$R & opt & nod \\  
  \hline
  20 &  5 & 1 & 36.1   & 15831.2 & 15831.2 & 8678.2  & 15831.2 & 15831.2 &  0.0 &  0.0 &  0.0 & 45.2 & 1 & 6659 \\ 
  20 &  5 & 2 & 29.7   & 16273.0 & 16273.0 & 11464.7 & 16273.0 & 16273.0 &  0.0 &  0.0 &  0.0 & 29.5 & 1 & 1266 \\ 
  20 &  5 & 3 & 30.6   & 15039.1 & 15039.1 & 8966.5  & 15039.1 & 15039.1 &  0.0 &  0.0 &  0.0 & 40.4 & 1 & 1380 \\ 
  20 &  5 & 4 & 20.0   & 9076.2  & 9076.2  & 5449.0  & 9076.2  & 9076.2  &  0.0 &  0.0 &  0.0 & 40.0 & 1 & 1529 \\ 
  20 &  5 & 5 & 31.7   & 16943.6 & 16943.6 & 11867.1 & 16943.6 & 16943.6 &  0.0 &  0.0 &  0.0 & 30.0 & 1 & 1359 \\ 
  20 &  6 & 1 & 33.7   & 11799.8 & 11799.8 & 7620.6  & 11799.8 & 11799.8 &  0.0 &  0.0 &  0.0 & 35.4 & 1 & 1469 \\
\hline
\end{longtable}
\endgroup

\begingroup
\tiny
\begin{longtable}{rrrrrrrrr}
\caption{Summary results table for model $F1_{x^{\ell}}^{sub1}$} \\ 
$|V|$ & p & $|$\#$|$ & cpu & g$\overline{\text{U}}$R & g$\overline{\text{U}}$L & gU$\overline{\text{L}}$ & gUL & nod \\ 
  \hline
  20 & 5 & 5 & 29.6    & 37.0 & 0.0 & 0.0 & 0.0 & 2439 \\ 
  20 & 6 & 5 & 29.6    & 35.0 & 0.0 & 0.0 & 0.0 & 1977 \\ 
  20 & 10 & 5 & 39.2   & 28.9 & 0.0 & 0.0 & 0.0 & 9244 \\ 
  30 & 7 & 5 & 900.3   & 39.6 & 0.0 & 0.0 & 0.0 & 36150 \\ 
  30 & 10 & 5 & 1053.8 & 33.1 & 0.0 & 0.0 & 0.0 & 44437 \\ 
  30 & 15 & 4 & 1236.2 & 33.0 & 2.2 & 4.3 & 6.0 & 89108 \\ 
  40 & 10 & 2 & 2920.6 & 33.3 & 4.5 & 27.6 & 30.2 & 57710 \\ 
  40 & 13 & 0 & 3600.1 & 35.4 & 8.0 & 51.0 & 54.3 & 71740 \\ 
  40 & 20 & 1 & 3197.8 & 28.6 & 5.1 & 48.2 & 49.8 & 140556 \\ 
  50 & 12 & 0 & 3600.3 & 35.8 & 4.8 & 87.7 & 88.3 & 10781 \\ 
  50 & 16 & 0 & 3600.3 & 35.7 & 3.8 & 92.2 & 92.5 & 19514 \\ 
  50 & 25 & 0 & 3600.4 & 28.6 & 4.3 & 83.6 & 84.5 & 38469 \\ 
\hline
\end{longtable}
\endgroup

\newpage

\begingroup
\tiny
\begin{longtable}{rrrrrrrrrrrrrrr}
\caption{Instances results table for model $F1_{x^{\ell}}^{sub2}$} \\ 
$|V|$ & p & ins & cpu & objU & objL & objR & obj$\overline{\text{U}}$ & obj$\overline{\text{L}}$ & gUL & g$\overline{\text{U}}$L & gU$\overline{\text{L}}$ & g$\overline{\text{U}}$R & opt & nod \\ 
  \hline
  20 & 5 & 1 & 62.8    & 15831.2 & 15831.2 & 8678.2   & 15831.2 & 15831.2 &  0.0 &  0.0 &  0.0 & 45.2 & 1 & 10033 \\ 
  20 & 5 & 2 & 29.3    & 16273.0 & 16273.0 & 11464.7  & 16273.0 & 16273.0 &  0.0 &  0.0 &  0.0 & 29.5 & 1 & 1266 \\ 
  20 & 5 & 3 & 30.9    & 15039.1 & 15039.1 & 8966.5   & 15039.1 & 15039.1 &  0.0 &  0.0 &  0.0 & 40.4 & 1 & 1485 \\ 
  20 & 5 & 4 & 19.7    & 9076.2  & 9076.2  & 5449.0   & 9076.2  & 9076.2  &  0.0 &  0.0 &  0.0 & 40.0 & 1 & 1529 \\ 
  20 & 5 & 5 & 31.8    & 16943.6 & 16943.6 & 11867.1  & 16943.6 & 16943.6 &  0.0 &  0.0 &  0.0 & 30.0 & 1 & 1359 \\ 
  20 & 6 & 1 & 33.7    & 11799.8 & 11799.8 & 7620.6   & 11799.8 & 11799.8 &  0.0 &  0.0 &  0.0 & 35.4 & 1 & 1469 \\
\hline
\end{longtable}
\endgroup

\begingroup
\tiny
\begin{longtable}{rrrrrrrrr}
\caption{Summary results table for model $F1_{x^{\ell}}^{sub2}$} \\ 
$|V|$ & p & $|$\#$|$ & cpu & g$\overline{\text{U}}$R & g$\overline{\text{U}}$L & gU$\overline{\text{L}}$ & gUL & nod \\ 
  \hline
  20 & 5 & 5 & 34.9    & 37.0 & 0.0 & 0.0 & 0.0 & 3134 \\ 
  20 & 6 & 5 & 31.5    & 35.0 & 0.0 & 0.0 & 0.0 & 2549 \\ 
  20 & 10 & 5 & 67.1   & 28.9 & 0.0 & 0.0 & 0.0 & 13133 \\ 
  30 & 7 & 5 & 1047.7  & 39.6 & 0.0 & 0.0 & 0.0 & 42684 \\ 
  30 & 10 & 5 & 915.5  & 33.1 & 0.0 & 0.0 & 0.0 & 41757 \\ 
  30 & 15 & 5 & 622.6  & 33.0 & 0.0 & 0.0 & 0.0 & 34092 \\ 
  40 & 10 & 1 & 3418.0 & 33.3 & 4.8 & 50.0 & 52.5 & 63279 \\ 
  40 & 13 & 0 & 3600.2 & 35.4 & 8.0 & 66.3 & 68.8 & 67439 \\ 
  40 & 20 & 1 & 3418.8 & 28.6 & 2.6 & 26.0 & 27.6 & 111976 \\ 
  50 & 12 & 0 & 3600.3 & 35.8 & 4.8 & 87.7 & 88.3 & 10857 \\ 
  50 & 16 & 0 & 3600.3 & 35.7 & 3.8 & 92.2 & 92.5 & 21540 \\ 
  50 & 25 & 0 & 3600.3 & 28.6 & 4.4 & 92.7 & 93.1 & 37979 \\ 
\hline
\end{longtable}
\endgroup

\newpage

\begingroup
\tiny
\begin{longtable}{rrrrrrrrrrrrrrr}
\caption{Instances results table for model $F2_{x^{\ell}}^{mtz}$} \\ 
$|V|$ & p & ins & cpu & objU & objL & objR & obj$\overline{\text{U}}$ & obj$\overline{\text{L}}$ & gUL & g$\overline{\text{U}}$L & gU$\overline{\text{L}}$ & g$\overline{\text{U}}$R & opt & nod \\ 
  \hline
  20 & 5 & 1 & 6.4 & 15831.2 & 15831.2   & 11628.1 & 15831.2  & 15831.2 & 0.0 & 0.0 & 0.0    & 26.5   & 1 & 550 \\ 
  20 & 5 & 2 & 7.3 & 16273.0 & 16273.0   & 13320.5 & 16273.0  & 16273.0 & 0.0 & 0.0 & 0.0    & 18.1   & 1 & 156 \\ 
  20 & 5 & 3 & 14.8 & 15039.1 & 15039.1  & 11682.6 & 15039.1  & 15039.1 & 0.0 & 0.0 & 0.0    & 22.3   & 1 & 3356 \\ 
  20 & 5 & 4 & 7.9 & 9076.2 & 9076.2     & 6951.0  & 9076.2   & 9076.2 & 0.0 & 0.0 & 0.0     & 23.4   & 1 & 1 \\ 
  20 & 5 & 5 & 1.5 & 16943.6 & 16943.6   & 14144.9 & 16943.6  & 16943.6 & 0.0 & 0.0 & 0.0    & 16.5   & 1 & 1 \\ 
  20 & 6 & 1 & 3.1 & 11799.8 & 11799.8   & 10139.9 & 11799.8  & 11799.8 & 0.0 & 0.0 & 0.0    & 14.1   & 1 & 1 \\
\hline
\end{longtable}
\endgroup

\begingroup
\tiny
\begin{longtable}{rrrrrrrrr}
\caption{Summary results table for model $F2_{x^{\ell}}^{mtz}$} \\ 
$|V|$ & p & $|$\#$|$ & cpu & g$\overline{\text{U}}$R & g$\overline{\text{U}}$L & gU$\overline{\text{L}}$ & gUL & nod \\ 
  \hline
  20 & 5 & 5 & 7.6     &  21.4   & 0.0 & 0.0 & 0.0 & 813 \\ 
  20 & 6 & 5 & 6.4     &  12.6   & 0.0 & 0.0 & 0.0 & 226 \\ 
  20 & 10 & 5 & 3.5    &  8.7    & 0.0 & 0.0 & 0.0 & 136 \\ 
  30 & 7 & 5 & 67.6    &  20.6   & 0.0 & 0.0 & 0.0 & 544 \\ 
  30 & 10 & 5 & 45.7   &  10.7   & 0.0 & 0.0 & 0.0 & 847 \\ 
  30 & 15 & 5 & 62.8   &  7.2    & 0.0 & 0.0 & 0.0 & 2831 \\ 
  40 & 10 & 5 & 636.1  &  17.6   & 0.0 & 0.0 & 0.0 & 947 \\ 
  40 & 13 & 4 & 1412.4 &  15.4   & 1.1 & 0.8 & 1.8 & 4244 \\ 
  40 & 20 & 5 & 43.5   &  6.8    & 0.0 & 0.0 & 0.0 & 525 \\ 
  50 & 12 & 3 & 2501.8 &  17.9   & 0.8 & 13.7 & 14.2 & 1367 \\ 
  50 & 16 & 4 & 1728.2 &  12.7   & 0.3 & 2.4 & 2.6 & 566 \\ 
  50 & 25 & 5 & 176.7  &  6.3    & 0.0 & 0.0 & 0.0 & 554 \\ 
\hline
\end{longtable}
\endgroup

\newpage

\begingroup
\tiny
\begin{longtable}{rrrrrrrrrrrrrrr}
\caption{Instances results table for model $F2_{x^{\ell}}^{flow}$} \\ 
$|V|$ & p & ins & cpu & objU & objL & objR & obj$\overline{\text{U}}$ & obj$\overline{\text{L}}$ & gUL & g$\overline{\text{U}}$L & gU$\overline{\text{L}}$ & g$\overline{\text{U}}$R & opt & nod \\ 
  \hline
  20 &  5 & 1 & 48.0   & 15831.2 & 15831.2  & 11628.2  & 15831.2 & 15831.2 & 0.0  & 0.0  & 0.0  & 26.5 & 1 & 7231 \\ 
  20 &  5 & 2 & 9.6    & 16273.0 & 16273.0  & 13320.5  & 16273.0 & 16273.0 & 0.0  & 0.0  & 0.0  & 18.1 & 1 & 907 \\ 
  20 &  5 & 3 & 9.2    & 15039.1 & 15039.1  & 11682.6  & 15039.1 & 15039.1 & 0.0  & 0.0  & 0.0  & 22.3 & 1 & 736 \\ 
  20 &  5 & 4 & 7.4    & 9076.2  & 9076.2   & 6951.0   & 9076.2  & 9076.2  & 0.0  & 0.0  & 0.0  & 23.4 & 1 & 541 \\ 
  20 &  5 & 5 & 1.3    & 16943.6 & 16943.6  & 14127.3  & 16943.6 & 16943.6 & 0.0  & 0.0  & 0.0  & 16.6 & 1 & 1 \\ 
  20 &  6 & 1 & 2.3    & 11799.8 & 11799.8  & 10139.9  & 11799.8 & 11799.8 & 0.0  & 0.0  & 0.0  & 14.1 & 1 & 1 \\
\hline
\end{longtable}
\endgroup

\begingroup
\tiny
\begin{longtable}{rrrrrrrrr}
\caption{Summary results table for model $F2_{x^{\ell}}^{flow}$} \\ 
$|V|$ & p & $|$\#$|$ & cpu & g$\overline{\text{U}}$R & g$\overline{\text{U}}$L & gU$\overline{\text{L}}$ & gUL & nod \\ 
  \hline
  20 & 5 & 5 & 15.1    & 21.4 & 0.0 & 0.0 & 0.0 & 1883 \\ 
  20 & 6 & 5 & 5.2     & 12.6 & 0.0 & 0.0 & 0.0 & 77 \\ 
  20 & 10 & 5 & 38.2   &  9.7 & 0.0 & 0.0 & 0.0 & 6454 \\ 
  30 & 7 & 5 & 217.7   & 20.6 & 0.0 & 0.0 & 0.0 & 2485 \\ 
  30 & 10 & 5 & 111.5  & 11.0 & 0.0 & 0.0 & 0.0 & 5838 \\ 
  30 & 15 & 5 & 31.9   &  9.7 & 0.0 & 0.0 & 0.0 & 1865 \\ 
  40 & 10 & 4 & 1513.0 & 18.8 & 2.6 & 0.9 & 3.4 & 5035 \\ 
  40 & 13 & 2 & 2391.4 & 16.8 & 4.9 & 2.2 & 6.9 & 13620 \\ 
  40 & 20 & 5 & 114.5  &  8.0 & 0.0 & 0.0 & 0.0 & 2043 \\ 
  50 & 12 & 0 & 3600.3 & 18.4 & 3.9 & 21.1 & 24.2 & 1404 \\ 
  50 & 16 & 3 & 2519.7 & 13.7 & 1.7 & 3.0 & 4.6 & 3770 \\ 
  50 & 25 & 4 & 1280.2 &  7.0 & 0.8 & 0.0 & 0.8 & 16303 \\ 
\hline
\end{longtable}
\endgroup

\newpage

\begingroup
\tiny
\begin{longtable}{rrrrrrrrrrrrrrr}
\caption{Instances results table for model $F2_{x^{\ell}}^{km}$} \\ 
$|V|$ & p & ins & cpu & objU & objL & objR & obj$\overline{\text{U}}$ & obj$\overline{\text{L}}$ & gUL & g$\overline{\text{U}}$L & gU$\overline{\text{L}}$ & g$\overline{\text{U}}$R & opt & nod \\ 
  \hline 
  20 &  5 & 1 & 12.1   & 15831.2 & 15831.2 & 11820.0 & 15831.2 & 15831.2 &  0.0 & 0.0 &  0.0 & 25.3 & 1 & 655 \\ 
  20 &  5 & 2 & 26.6   & 16273.0 & 16273.0 & 13586.0 & 16273.0 & 16273.0 &  0.0 & 0.0 &  0.0 & 16.5 & 1 & 53 \\ 
  20 &  5 & 3 & 16.9   & 15039.1 & 15039.1 & 11919.1 & 15039.1 & 15039.1 &  0.0 & 0.0 &  0.0 & 20.8 & 1 & 131 \\ 
  20 &  5 & 4 & 35.6   & 9076.2  & 9076.2  & 7166.6  & 9076.2  & 9076.2  &  0.0 & 0.0 &  0.0 & 21.0 & 1 & 248 \\ 
  20 &  5 & 5 & 4.1    & 16943.6 & 16943.6 & 14722.7 & 16943.6 & 16943.6 &  0.0 & 0.0 &  0.0 & 13.1 & 1 & 1 \\ 
  20 &  6 & 1 & 14.9   & 11799.8 & 11799.8 & 10272.4 & 11799.8 & 11799.8 &  0.0 & 0.0 &  0.0 & 12.9 & 1 & 96 \\
\hline
\end{longtable}
\endgroup

\begingroup
\tiny
\begin{longtable}{rrrrrrrrr}
\caption{Summary results table for model $F2_{x^{\ell}}^{km}$} \\ 
$|V|$ & p & $|$\#$|$ & cpu & g$\overline{\text{U}}$R & g$\overline{\text{U}}$L & gU$\overline{\text{L}}$ & gUL & nod \\ 
  \hline
  20 & 5 & 5 & 19.1    & 19.3 & 0.0 & 0.0 & 0.0 & 218 \\ 
  20 & 6 & 5 & 18.9    & 10.7 & 0.0 & 0.0 & 0.0 & 477 \\ 
  20 & 10 & 5 & 7.0    & 3.5  & 0.0 & 0.0 & 0.0 & 61 \\ 
  30 & 7 & 5 & 246.3   & 18.6 & 0.0 & 0.0 & 0.0 & 471 \\ 
  30 & 10 & 5 & 101.0  & 7.5  & 0.0 & 0.0 & 0.0 & 506 \\ 
  30 & 15 & 5 & 56.7   & 3.8  & 0.0 & 0.0 & 0.0 & 235 \\ 
  40 & 10 & 4 & 1355.2 & 13.7 & 0.3 & 3.1 & 3.3 & 259 \\ 
  40 & 13 & 4 & 1334.9 & 12.5 & 0.2 & 2.6 & 2.8 & 237 \\ 
  40 & 20 & 5 & 271.4  & 2.5  & 0.0 & 0.0 & 0.0 & 152 \\ 
  50 & 12 & 2 & 2629.2 & 15.6 & 0.4 & 25.5 & 25.7 & 22 \\ 
  50 & 16 & 4 & 1236.4 & 10.5 & 0.0 & 2.0 & 2.0 & 66 \\ 
  50 & 25 & 5 & 583.2  & 3.4  & 0.0 & 0.0 & 0.0 & 253 \\ 
\hline
\end{longtable}
\endgroup

\newpage

\begingroup
\tiny
\begin{longtable}{rrrrrrrrrrrrrrr}
\caption{Instances results table for model $F2_{x^{\ell}}^{sub}$} \\ 
$|V|$ & p & ins & cpu & objU & objL & objR & obj$\overline{\text{U}}$ & obj$\overline{\text{L}}$ & gUL & g$\overline{\text{U}}$L & gU$\overline{\text{L}}$ & g$\overline{\text{U}}$R & opt & nod \\  
  \hline
  20 &  5 & 1 & 174.1  & 15831.2 & 15831.2 & 11628.1 & 15831.2 & 15831.2 &  0.0 &  0.0 &  0.0 & 26.6 & 1 & 27592 \\ 
  20 &  5 & 2 & 46.8   & 16273.0 & 16273.0 & 13320.5 & 16273.0 & 16273.0 &  0.0 &  0.0 &  0.0 & 18.1 & 1 & 3351 \\ 
  20 &  5 & 3 & 49.6   & 15039.1 & 15039.1 & 11682.6 & 15039.1 & 15039.1 &  0.0 &  0.0 &  0.0 & 22.3 & 1 & 4509 \\ 
  20 &  5 & 4 & 33.0   & 9076.2  & 9076.2  & 6951.0  & 9076.2  & 9076.2  &  0.0 &  0.0 &  0.0 & 23.4 & 1 & 1452 \\ 
  20 &  5 & 5 & 33.5   & 16943.6 & 16943.6 & 14127.3 & 16943.6 & 16943.6 &  0.0 &  0.0 &  0.0 & 16.6 & 1 & 1571 \\ 
  20 &  6 & 1 & 31.7   & 11799.8 & 11799.8 & 10139.9 & 11799.8 & 11799.8 &  0.0 &  0.0 &  0.0 & 14.1 & 1 & 1141 \\
\hline
\end{longtable}
\endgroup

\begingroup
\tiny
\begin{longtable}{rrrrrrrrr}
\caption{Summary results table for model $F2_{x^{\ell}}^{sub}$} \\ 
$|V|$ & p & $|$\#$|$ & cpu & g$\overline{\text{U}}$R & g$\overline{\text{U}}$L & gU$\overline{\text{L}}$ & gUL & nod \\ 
  \hline
  20 & 5 & 5 & 67.4    & 21.4 & 0.0 & 0.0 & 0.0 & 7695 \\ 
  20 & 6 & 5 & 27.3    & 12.6 & 0.0 & 0.0 & 0.0 & 1468 \\ 
  20 & 10 & 5 & 76.4   &  9.9 & 0.0 & 0.0 & 0.0 & 19984 \\ 
  30 & 7 & 5 & 1090.4  & 20.7 & 0.0 & 0.0 & 0.0 & 40945 \\ 
  30 & 10 & 4 & 1297.2 & 11.1 & 1.1 & 3.4 & 4.3 & 64765 \\ 
  30 & 15 & 5 & 1427.2 &  9.7 & 0.0 & 0.0 & 0.0 & 93664 \\ 
  40 & 10 & 0 & 3600.3 & 18.8 & 4.7 & 81.5 & 82.1 & 30934 \\ 
  40 & 13 & 0 & 3600.1 & 16.9 & 8.4 & 72.6 & 74.8 & 40147 \\ 
  40 & 20 & 1 & 3321.2 &  8.5 & 4.3 & 44.4 & 45.5 & 107928 \\ 
  50 & 12 & 0 & 3600.2 & 18.4 & 4.9 & 87.7 & 88.3 & 4867 \\ 
  50 & 16 & 0 & 3600.3 & 13.8 & 3.9 & 92.2 & 92.5 & 11042 \\ 
  50 & 25 & 0 & 3600.3 & 7.1 & 4.4 & 92.7 & 93.1 & 20888 \\ 
\hline
\end{longtable}
\endgroup

\newpage

\begingroup
\tiny
\begin{longtable}{rrrrrrrrrrrrrrr}
\caption{Instances results table for model OMT Benders modern Kruskal} \\ 
$|V|$ & p & ins & cpu & objU & objL & objR & obj$\overline{\text{U}}$ & obj$\overline{\text{L}}$ & gUL & g$\overline{\text{U}}$L & gU$\overline{\text{L}}$ & g$\overline{\text{U}}$R & opt & nod \\  
  \hline
  20 &  5 & 1 & 3600.0 & 17846.0 & 9219.5  & 5459.2 & 15831.2 & 15831.2 & 48.3 & 41.8 & 11.3 & 65.5 & 0 & 1269394 \\ 
  20 &  5 & 2 & 3600.0 & 16502.0 & 9869.6  & 6947.0 & 16273.0 & 16273.0 & 40.2 & 39.4 & 1.4  & 57.3 & 0 & 1098198 \\ 
  20 &  5 & 3 & 3600.0 & 15144.0 & 9493.5  & 6024.8 & 15039.1 & 15039.1 & 37.3 & 36.9 & 0.7  & 59.9 & 0 & 914845 \\ 
  20 &  5 & 4 & 3600.1 &  9327.2 & 6684.3  & 3857.6 & 9076.2  & 9076.2  & 28.3 & 26.4 & 2.7  & 57.5 & 0 & 810724 \\ 
  20 &  5 & 5 & 3600.0 & 16943.6 & 11289.4 & 7852.6 & 16943.6 & 16943.6 & 33.4 & 33.4 & 0.0  & 53.6 & 0 & 765388 \\ 
  20 &  6 & 1 & 3600.1 & 14580.5 & 6206.5  & 4648.1 & 11799.8 & 11799.8 & 57.4 & 47.4 & 19.1 & 60.6 & 0 & 690567 \\
\hline
\end{longtable}
\endgroup

\begingroup
\tiny
\begin{longtable}{rrrrrrrrr}
\caption{Summary results table for model OMT Benders modern Kruskal} \\
$|V|$ & p & $|$\#$|$ & cpu & g$\overline{\text{U}}$R & g$\overline{\text{U}}$L & gU$\overline{\text{L}}$ & gUL & nod \\ 
  \hline
  20 &  5 & 0 & 3600.0 & 58.8 & 35.6 & 3.2 & 37.5 & 971710 \\ 
  20 &  6 & 0 & 3600.1 & 58.3 & 43.5 & 11.9 & 49.7 & 886817 \\ 
  20 & 10 & 0 & 3600.2 & 79.4 & 70.8 & 16.8 & 75.7 & 671882 \\ 
  30 &  7 & 0 & 3600.2 & 58.3 & 54.6 & 41.7 & 73.5 & 279547 \\ 
  30 & 10 & 0 & 3600.3 & 64.5 & 63.0 & 40.7 & 77.9 & 312307 \\ 
  30 & 15 & 0 & 3600.4 & 83.2 & 81.5 & 36.7 & 88.3 & 306955 \\ 
  40 & 10 & 0 & 3600.2 & 52.7 & 51.4 & 58.9 & 80.0 & 118585 \\ 
  40 & 13 & 0 & 3600.0 & 64.5 & 64.3 & 50.7 & 82.4 & 210418 \\ 
  40 & 20 & 0 & 3600.1 & 80.2 & 80.0 & 47.1 & 89.4 & 244894 \\ 
  50 & 12 & 0 & 3600.3 & 57.8 & 56.3 & 56.6 & 81.1 & 10193 \\ 
  50 & 16 & 0 & 3600.3 & 62.6 & 62.4 & 64.0 & 86.5 & 21578 \\ 
  50 & 25 & 0 & 3600.2 & 80.9 & 80.9 & 50.7 & 90.6 & 90455 \\ 
\hline
\end{longtable}
\endgroup

\newpage

\begingroup
\tiny
\begin{longtable}{rrrrrrrrrrrrrrr}
\caption{Instances results table for model OMT Benders modern $km$} \\ 
$|V|$ & p & ins & cpu & objU & objL & objR & obj$\overline{\text{U}}$ & obj$\overline{\text{L}}$ & gUL & g$\overline{\text{U}}$L & gU$\overline{\text{L}}$ & g$\overline{\text{U}}$R & opt & nod \\  
  \hline
  20 &  5 & 1 &  528.5 & 15831.2 & 15831.2  & 5459.2 & 15831.2 & 15831.2 &  0.0 &  0.0 &  0.0 & 65.5 & 1 & 197137 \\ 
  20 &  5 & 2 &  478.9 & 16273.0 & 16273.0  & 6947.0 & 16273.0 & 16273.0 &  0.0 &  0.0 &  0.0 & 57.3 & 1 & 198687 \\ 
  20 &  5 & 3 &  752.1 & 15039.1 & 15039.1  & 6024.8 & 15039.1 & 15039.1 &  0.0 &  0.0 &  0.0 & 59.9 & 1 & 223528 \\ 
  20 &  5 & 4 &  770.9 &  9076.2 &  9076.2  & 3857.6 &  9076.2 &  9076.2 &  0.0 &  0.0 &  0.0 & 57.5 & 1 & 166283 \\ 
  20 &  5 & 5 &  651.0 & 16943.6 & 16943.6  & 7852.6 & 16943.6 & 16943.6 &  0.0 &  0.0 &  0.0 & 53.6 & 1 & 138776 \\ 
  20 &  6 & 1 &  826.0 & 11799.8 & 11799.8  & 4648.1 & 11799.8 & 11799.8 &  0.0 &  0.0 &  0.0 & 60.6 & 1 & 214848 \\
\hline
\end{longtable}
\endgroup

\begingroup
\tiny
\begin{longtable}{rrrrrrrrr}
\caption{Summary results table for model OMT Benders modern $km$} \\ 
$|V|$ & p & $|$\#$|$ & cpu & g$\overline{\text{U}}$R & g$\overline{\text{U}}$L & gU$\overline{\text{L}}$ & gUL & nod \\ 
  \hline
  20 & 5 & 5 & 636.3   & 58.8 & 0.0 & 0.0 & 0.0 & 184882 \\ 
  20 & 6 & 5 & 666.1   & 58.3 & 0.0 & 0.0 & 0.0 & 167086 \\ 
  20 & 10 & 5 & 940.1  & 79.4 & 0.0 & 0.0 & 0.0 & 200740 \\ 
  30 & 7 & 0 & 3601.6  & 58.2 & 53.4 & 46.1 & 74.6 & 277577 \\ 
  30 & 10 & 0 & 3605.7 & 64.5 & 60.7 & 41.9 & 77.1 & 178372 \\ 
  30 & 15 & 0 & 3617.4 & 83.2 & 80.1 & 39.2 & 87.8 & 81018 \\ 
  40 & 10 & 0 & 3602.2 & 52.7 & 51.2 & 59.2 & 79.8 & 97581 \\ 
  40 & 13 & 0 & 3609.0 & 64.5 & 64.0 & 50.1 & 82.0 & 91645 \\ 
  40 & 20 & 0 & 3623.8 & 80.2 & 79.6 & 51.6 & 90.1 & 50921 \\ 
  50 & 12 & 0 & 3600.2 & 57.7 & 56.2 & 56.6 & 81.1 & 5467 \\ 
  50 & 16 & 0 & 3600.3 & 62.6 & 62.2 & 65.5 & 87.0 & 18136 \\ 
  50 & 25 & 0 & 3953.6 & 80.9 & 80.7 & 54.3 & 91.2 & 22866 \\ 
\hline
\end{longtable}
\endgroup

\newpage

\begingroup
\tiny
\begin{longtable}{rrrrrrrrrrrrrrr}
\caption{Instances results table for model OMT Benders classic $km$} \\ 
$|V|$ & p & ins & cpu & objU & objL & objR & obj$\overline{\text{U}}$ & obj$\overline{\text{L}}$ & gUL & g$\overline{\text{U}}$L & gU$\overline{\text{L}}$ & g$\overline{\text{U}}$R & opt & nod \\  
  \hline
  20 & 5 & 1 & 3600.8 & 15831.2 & 13564.2 & 5918.1 & 15831.2 & 15831.2 & 14.3 & 14.3 & 0.0 & 62.6 & 0 &  \\ 
  20 & 5 & 2 & 3600.5 & 16414.7 & 13791.2 & 10878.9 & 16273.0 & 16273.0 & 16.0 & 15.2 & 0.9 & 33.1 & 0 &  \\ 
  20 & 5 & 3 & 3600.5 & 15526.1 & 10876.5 & 5974.5 & 15039.1 & 15039.1 & 30.0 & 27.7 & 3.1 & 60.3 & 0 &  \\ 
  20 & 5 & 4 & 3600.5 & 9076.2 & 7059.9 & 3916.0 & 9076.2 & 9076.2 & 22.2 & 22.2 & 0.0 & 56.9 & 0 &  \\ 
  20 & 5 & 5 & 3600.5 & 17423.8 & 14708.0 & 8284.4 & 16943.6 & 16943.6 & 15.6 & 13.2 & 2.8 & 51.1 & 0 &  \\ 
  20 & 6 & 1 & 3600.5 & 15756.0 & 7396.0 & 4648.2 & 11799.8 & 11799.8 & 53.1 & 37.3 & 25.1 & 60.6 & 0 &  \\
\hline
\end{longtable}
\endgroup

\begingroup
\tiny
\begin{longtable}{rrrrrrrrr}
\caption{Summary results table for model OMT Benders classic $km$} \\ 
$|V|$ & p & $|$\#$|$ & cpu & g$\overline{\text{U}}$R & g$\overline{\text{U}}$L & gU$\overline{\text{L}}$ & gUL & nod \\ 
  \hline
  20 & 5 & 0 & 3600.6 & 52.8 & 18.5 & 1.4 & 19.6 &  \\ 
  20 & 6 & 1 & 3335.3 & 50.4 & 20.1 & 9.0 & 26.0 &  \\ 
  20 & 10 & 0 & 3600.7 & 50.5 & 23.0 & 6.3 & 26.4 &  \\ 
  30 & 7 & 0 & 3604.8 & 58.1 & 50.0 & 37.4 & 68.8 &  \\ 
  30 & 10 & 0 & 3607.2 & 64.0 & 58.2 & 36.3 & 73.4 &  \\ 
  30 & 15 & 0 & 3607.1 & 82.5 & 77.5 & 41.1 & 86.7 &  \\ 
  40 & 10 & 0 & 3629.1 & 52.7 & 49.0 & 50.6 & 74.6 &  \\ 
  40 & 13 & 0 & 3644.4 & 64.5 & 62.7 & 46.0 & 80.0 &  \\ 
  40 & 20 & 0 & 3626.5 & 79.4 & 77.6 & 43.9 & 87.4 &  \\ 
  50 & 12 & 0 & 3757.5 & 57.7 & 55.3 & 53.4 & 79.2 &  \\ 
  50 & 16 & 0 & 3729.2 & 62.5 & 61.2 & 59.2 & 84.1 &  \\ 
  50 & 25 & 0 & 3738.2 & 80.7 & 80.2 & 48.0 & 89.7 &  \\ 
\hline
\end{longtable}
\endgroup

\newpage

\begingroup
\tiny
\begin{longtable}{rrrrrrrrrrrrrrr}
\caption{Instances results table for model $F1_{u}^{mtz}$} \\ 
$|V|$ & p & ins & cpu & objU & objL & objR & obj$\overline{\text{U}}$ & obj$\overline{\text{L}}$ & gUL & g$\overline{\text{U}}$L & gU$\overline{\text{L}}$ & g$\overline{\text{U}}$R & opt & nod \\ 
  \hline
  20 & 5  & 1 & 1.6   & 15831.2 & 15831.2 & 8678.3  & 15831.2 & 15831.2 & 0.0 & 0.0 & 0.0  & 45.2 & 1 & 1 \\ 
  20 & 5  & 2 & 2.2   & 16273.0 & 16273.0 & 11464.7 & 16273.0 & 16273.0 & 0.0 & 0.0 & 0.0  & 29.5 & 1 & 1 \\ 
  20 & 5  & 3 & 1.5   & 15039.1 & 15039.1 & 8966.5  & 15039.1 & 15039.1 & 0.0 & 0.0 & 0.0  & 40.4 & 1 & 1 \\ 
  20 & 5  & 4 & 1.3   & 9076.2  & 9076.2  & 5449.0  & 9076.2  & 9076.2  & 0.0 & 0.0 & 0.0  & 40.0 & 1 & 1 \\ 
  20 & 5  & 5 & 1.1   & 16943.6 & 16943.6 & 11867.1 & 16943.6 & 16943.6 & 0.0 & 0.0 & 0.0  & 30.0 & 1 & 1 \\ 
  20 & 6  & 1 & 0.8   & 11799.8 & 11799.8 & 7620.6  & 11799.8 & 11799.8 & 0.0 & 0.0 & 0.0  & 35.4 & 1 & 1 \\
\hline
\end{longtable}
\endgroup

\begingroup
\tiny
\begin{longtable}{rrrrrrrrr}
\caption{Summary results table for model $F1_{u}^{mtz}$} \\ 
$|V|$ & p & $|$\#$|$ & cpu & g$\overline{\text{U}}$R & g$\overline{\text{U}}$L & gU$\overline{\text{L}}$ & gUL & nod \\ 
  \hline
  20 &  5 & 5 &   1.5 & 37.0 & 0.0 & 0.0 & 0.0 & 1 \\ 
  20 &  6 & 5 &   1.3 & 35.0 & 0.0 & 0.0 & 0.0 & 1 \\ 
  20 & 10 & 5 &   1.2 & 28.9 & 0.0 & 0.0 & 0.0 & 34 \\ 
  30 &  7 & 5 &  16.2 & 39.6 & 0.0 & 0.0 & 0.0 & 477 \\ 
  30 & 10 & 5 &   8.4 & 33.1 & 0.0 & 0.0 & 0.0 & 237 \\ 
  30 & 15 & 5 &  16.4 & 32.9 & 0.0 & 0.0 & 0.0 & 2213 \\ 
  40 & 10 & 5 &  62.7 & 33.3 & 0.0 & 0.0 & 0.0 & 715 \\ 
  40 & 13 & 5 &  66.4 & 35.4 & 0.0 & 0.0 & 0.0 & 2400 \\ 
  40 & 20 & 5 &  40.9 & 28.6 & 0.0 & 0.0 & 0.0 & 2148 \\ 
  50 & 12 & 5 & 276.8 & 35.8 & 0.0 & 0.0 & 0.0 & 1165 \\ 
  50 & 16 & 5 & 145.1 & 35.7 & 0.0 & 0.0 & 0.0 & 1615 \\ 
  50 & 25 & 5 &  83.9 & 28.6 & 0.0 & 0.0 & 0.0 & 2028 \\ 
\hline
\end{longtable}
\endgroup

\newpage

\begingroup
\tiny
\begin{longtable}{rrrrrrrrrrrrrrr}
\caption{Instances results table for model $F1_{u}^{flow1}$} \\ 
$|V|$ & p & ins & cpu & objU & objL & objR & obj$\overline{\text{U}}$ & obj$\overline{\text{L}}$ & gUL & g$\overline{\text{U}}$L & gU$\overline{\text{L}}$ & g$\overline{\text{U}}$R & opt & nod \\  
  \hline
  20 & 5 & 1 & 1.5     & 15831.2 & 15831.2 & 8678.2    & 15831.2 & 15831.2 & 0.0 & 0.0 & 0.0  & 45.2 & 1 & 128 \\ 
  20 & 5 & 2 & 0.6     & 16273.0 & 16273.0 & 11464.7   & 16273.0 & 16273.0 & 0.0 & 0.0 & 0.0  & 29.5 & 1 & 1 \\ 
  20 & 5 & 3 & 1.7     & 15039.1 & 15039.1 & 8966.5    & 15039.1 & 15039.1 & 0.0 & 0.0 & 0.0  & 40.4 & 1 & 1 \\ 
  20 & 5 & 4 & 0.5     & 9076.2  & 9076.2  & 5449.0    & 9076.2  & 9076.2 & 0.0 & 0.0 & 0.0   & 40.0 & 1 & 1 \\ 
  20 & 5 & 5 & 0.3     & 16943.6 & 16943.6 & 11867.1   & 16943.6 & 16943.6 & 0.0 & 0.0 & 0.0  & 30.0 & 1 & 1 \\ 
  20 & 6 & 1 & 0.5     & 11799.8 & 11799.8 & 7620.6    & 11799.8 & 11799.8 & 0.0 & 0.0 & 0.0  & 35.4 & 1 & 1 \\
\hline
\end{longtable}
\endgroup

\begingroup
\tiny
\begin{longtable}{rrrrrrrrr}
\caption{Summary results table for model $F1_{u}^{flow1}$} \\ 
$|V|$ & p & $|$\#$|$ & cpu & g$\overline{\text{U}}$R & g$\overline{\text{U}}$L & gU$\overline{\text{L}}$ & gUL & nod \\
  \hline
  20 &  5 & 5 & 0.9   & 37.0 & 0.0 & 0.0 & 0.0 & 26 \\ 
  20 &  6 & 5 & 0.6   & 35.0 & 0.0 & 0.0 & 0.0 & 1 \\ 
  20 & 10 & 5 & 1.6   & 28.9 & 0.0 & 0.0 & 0.0 & 219 \\ 
  30 &  7 & 5 & 8.9   & 39.6 & 0.0 & 0.0 & 0.0 & 1 \\ 
  30 & 10 & 5 & 8.4   & 33.1 & 0.0 & 0.0 & 0.0 & 712 \\ 
  30 & 15 & 5 & 5.7   & 33.0 & 0.0 & 0.0 & 0.0 & 109 \\ 
  40 & 10 & 5 & 46.3  & 33.3 & 0.0 & 0.0 & 0.0 & 1215 \\ 
  40 & 13 & 5 & 127.7 & 35.4 & 0.0 & 0.0 & 0.0 & 7650 \\ 
  40 & 20 & 5 & 16.5  & 28.6 & 0.0 & 0.0 & 0.0 & 1 \\ 
  50 & 12 & 5 & 243.3 & 35.8 & 0.0 & 0.0 & 0.0 & 1530 \\ 
  50 & 16 & 5 & 119.1 & 35.7 & 0.0 & 0.0 & 0.0 & 956 \\ 
  50 & 25 & 5 & 275.0 & 28.6 & 0.0 & 0.0 & 0.0 & 15674 \\ 
\hline
\end{longtable}
\endgroup

\newpage

\begingroup
\tiny
\begin{longtable}{rrrrrrrrrrrrrrr}
\caption{Instances results table for model $F1_{u}^{flow2}$} \\ 
$|V|$ & p & ins & cpu & objU & objL & objR & obj$\overline{\text{U}}$ & obj$\overline{\text{L}}$ & gUL & g$\overline{\text{U}}$L & gU$\overline{\text{L}}$ & g$\overline{\text{U}}$R & opt & nod \\  
  \hline
  20 &  5 & 1 &   1.1 & 15831.2 & 15831.2 & 9767.5  & 15831.2 & 15831.2 & 0.0 & 0.0 & 0.0 & 38.3 & 1 & 1 \\ 
  20 &  5 & 2 &   1.0 & 16273.0 & 16273.0 & 11950.3 & 16273.0 & 16273.0 & 0.0 & 0.0 & 0.0 & 26.6 & 1 & 1 \\ 
  20 &  5 & 3 &   1.0 & 15039.1 & 15039.1 & 9701.2  & 15039.1 & 15039.1 & 0.0 & 0.0 & 0.0 & 35.5 & 1 & 1 \\ 
  20 &  5 & 4 &   0.7 &  9076.2 &  9076.2 & 5819.3  &  9076.2 &  9076.2 & 0.0 & 0.0 & 0.0 & 35.9 & 1 & 1 \\ 
  20 &  5 & 5 &   0.3 & 16943.6 & 16943.6 & 12351.4 & 16943.6 & 16943.6 & 0.0 & 0.0 & 0.0 & 27.1 & 1 & 1 \\ 
  20 &  6 & 1 &   0.5 & 11799.8 & 11799.8 & 8334.0  & 11799.8 & 11799.8 & 0.0 & 0.0 & 0.0 & 29.4 & 1 & 1 \\
\hline
\end{longtable}
\endgroup

\begingroup
\tiny
\begin{longtable}{rrrrrrrrr}
\caption{Summary results table for model $F1_{u}^{flow2}$} \\ 
$|V|$ & p & $|$\#$|$ & cpu & g$\overline{\text{U}}$R & g$\overline{\text{U}}$L & gU$\overline{\text{L}}$ & gUL & nod \\ 
  \hline
  20 &  5 & 5 &   0.8 & 32.7 & 0.0 & 0.0 & 0.0 & 1 \\ 
  20 &  6 & 5 &   0.5 & 29.0 & 0.0 & 0.0 & 0.0 & 1 \\ 
  20 & 10 & 5 &   1.0 & 25.8 & 0.0 & 0.0 & 0.0 & 28 \\ 
  30 &  7 & 5 &   6.4 & 33.9 & 0.0 & 0.0 & 0.0 & 1 \\ 
  30 & 10 & 5 &   4.0 & 29.3 & 0.0 & 0.0 & 0.0 & 23 \\ 
  30 & 15 & 5 &   3.7 & 30.6 & 0.0 & 0.0 & 0.0 & 1 \\ 
  40 & 10 & 5 &  29.3 & 30.7 & 0.0 & 0.0 & 0.0 & 809 \\ 
  40 & 13 & 5 &  23.2 & 31.5 & 0.0 & 0.0 & 0.0 & 370 \\ 
  40 & 20 & 5 &  10.0 & 25.4 & 0.0 & 0.0 & 0.0 & 1 \\ 
  50 & 12 & 5 & 119.9 & 33.6 & 0.0 & 0.0 & 0.0 & 644 \\ 
  50 & 16 & 5 &  63.8 & 33.0 & 0.0 & 0.0 & 0.0 & 157 \\ 
  50 & 25 & 5 &  28.7 & 26.8 & 0.0 & 0.0 & 0.0 & 686 \\ 
\hline
\end{longtable}
\endgroup

\newpage

\begingroup
\tiny
\begin{longtable}{rrrrrrrrrrrrrrr}
\caption{Instances results table for model $F1_{u}^{km}$} \\ 
$|V|$ & p & ins & cpu & objU & objL & objR & obj$\overline{\text{U}}$ & obj$\overline{\text{L}}$ & gUL & g$\overline{\text{U}}$L & gU$\overline{\text{L}}$ & g$\overline{\text{U}}$R & opt & nod \\ 
  \hline
  20 &  5 & 1 &  0.9  & 15831.2 & 15831.2  & 8678.3  & 15831.2 & 15831.2 & 0.0 & 0.0 & 0.0 & 45.2 & 1 & 1 \\ 
  20 &  5 & 2 &  1.9  & 16273.0 & 16273.0  & 11464.7 & 16273.0 & 16273.0 & 0.0 & 0.0 & 0.0 & 29.5 & 1 & 1 \\ 
  20 &  5 & 3 &  1.0  & 15039.1 & 15039.1  & 8966.5  & 15039.1 & 15039.1 & 0.0 & 0.0 & 0.0 & 40.4 & 1 & 1 \\ 
  20 &  5 & 4 &  0.6  & 9076.2  & 9076.2   & 5448.9  & 9076.2  & 9076.2  & 0.0 & 0.0 & 0.0 & 40.0 & 1 & 1 \\ 
  20 &  5 & 5 &  0.3  & 16943.6 & 16943.6  & 11867.1 & 16943.6 & 16943.6 & 0.0 & 0.0 & 0.0 & 30.0 & 1 & 0 \\ 
  20 &  6 & 1 &  0.3  & 11799.8 & 11799.8  & 7620.6  & 11799.8 & 11799.8 & 0.0 & 0.0 & 0.0 & 35.4 & 1 & 0 \\
\hline
\end{longtable}
\endgroup

\begingroup
\tiny
\begin{longtable}{rrrrrrrrr}
\caption{Summary results table for model $F1_{u}^{km}$} \\ 
$|V|$ & p & $|$\#$|$ & cpu & g$\overline{\text{U}}$R & g$\overline{\text{U}}$L & gU$\overline{\text{L}}$ & gUL & nod \\ 
  \hline
  20 & 5 & 5 & 0.9    & 37.0 & 0.0 & 0.0 & 0.0 & 1 \\ 
  20 & 6 & 5 & 0.7    & 35.0 & 0.0 & 0.0 & 0.0 & 0 \\ 
  20 & 10 & 5 & 0.4   & 28.5 & 0.0 & 0.0 & 0.0 & 0 \\ 
  30 & 7 & 5 & 11.9   & 39.6 & 0.0 & 0.0 & 0.0 & 1 \\ 
  30 & 10 & 5 & 8.2   & 33.1 & 0.0 & 0.0 & 0.0 & 1 \\ 
  30 & 15 & 5 & 8.4   & 32.6 & 0.0 & 0.0 & 0.0 & 1 \\ 
  40 & 10 & 5 & 41.5  & 33.3 & 0.0 & 0.0 & 0.0 & 462 \\ 
  40 & 13 & 5 & 30.1  & 35.4 & 0.0 & 0.0 & 0.0 & 1 \\ 
  40 & 20 & 5 & 22.3  & 28.6 & 0.0 & 0.0 & 0.0 & 1 \\ 
  50 & 12 & 5 & 153.7 & 35.8 & 0.0 & 0.0 & 0.0 & 116 \\ 
  50 & 16 & 5 & 51.5  & 35.7 & 0.0 & 0.0 & 0.0 & 1 \\ 
  50 & 25 & 5 & 44.6  & 28.5 & 0.0 & 0.0 & 0.0 & 1 \\ 
\hline
\end{longtable}
\endgroup

\newpage

\begingroup
\tiny
\begin{longtable}{rrrrrrrrrrrrrrr}
\caption{Instances results table for model $F1_{u}^{sub1}$} \\ 
$|V|$ & p & ins & cpu & objU & objL & objR & obj$\overline{\text{U}}$ & obj$\overline{\text{L}}$ & gUL & g$\overline{\text{U}}$L & gU$\overline{\text{L}}$ & g$\overline{\text{U}}$R & opt & nod \\  
  \hline
  20 & 5 & 1 & 2.1     & 15831.2 & 15831.2 & 8678.2  & 15831.2 & 15831.2 &  0.0 &  0.0 &  0.0 & 45.2 & 1 & 570 \\ 
  20 & 5 & 2 & 1.5     & 16273.0 & 16273.0 & 11464.7 & 16273.0 & 16273.0 &  0.0 &  0.0 &  0.0 & 29.5 & 1 & 214 \\ 
  20 & 5 & 3 & 2.2     & 15039.1 & 15039.1 & 8966.5  & 15039.1 & 15039.1 &  0.0 &  0.0 &  0.0 & 40.4 & 1 & 499 \\ 
  20 & 5 & 4 & 1.9     & 9076.2  & 9076.2  & 5449.0  & 9076.2  & 9076.2  &  0.0 &  0.0 &  0.0 & 40.0 & 1 & 36 \\ 
  20 & 5 & 5 & 0.9     & 16943.6 & 16943.6 & 11867.1 & 16943.6 & 16943.6 &  0.0 &  0.0 &  0.0 & 30.0 & 1 & 1 \\ 
  20 & 6 & 1 & 1.3     & 11799.8 & 11799.8 & 7620.6  & 11799.8 & 11799.8 &  0.0 &  0.0 &  0.0 & 35.4 & 1 & 236 \\
\hline
\end{longtable}
\endgroup

\begingroup
\tiny
\begin{longtable}{rrrrrrrrr}
\caption{Summary results table for model $F1_{u}^{sub1}$} \\ 
$|V|$ & p & $|$\#$|$ & cpu & g$\overline{\text{U}}$R & g$\overline{\text{U}}$L & gU$\overline{\text{L}}$ & gUL & nod \\ 
  \hline
  20 & 5 & 5 & 1.7     & 37.0 & 0.0 & 0.0 & 0.0 & 264 \\ 
  20 & 6 & 5 & 1.4     & 35.0 & 0.0 & 0.0 & 0.0 & 169 \\ 
  20 & 10 & 5 & 3.2    & 28.9 & 0.0 & 0.0 & 0.0 & 1266 \\ 
  30 & 7 & 5 & 17.7    & 39.6 & 0.0 & 0.0 & 0.0 & 973 \\ 
  30 & 10 & 5 & 10.4   & 33.1 & 0.0 & 0.0 & 0.0 & 786 \\ 
  30 & 15 & 5 & 247.4  & 33.0 & 0.0 & 0.0 & 0.0 & 76092 \\ 
  40 & 10 & 5 & 76.6   & 33.3 & 0.0 & 0.0 & 0.0 & 5141 \\ 
  40 & 13 & 5 & 586.0  & 35.4 & 0.0 & 0.0 & 0.0 & 94447 \\ 
  40 & 20 & 3 & 1597.6 & 28.6 & 1.7 & 3.0 & 4.5 & 163936 \\ 
  50 & 12 & 4 & 1341.9 & 35.8 & 0.3 & 0.0 & 0.3 & 37016 \\ 
  50 & 16 & 4 & 890.9  & 35.7 & 1.4 & 5.2 & 6.3 & 29338 \\ 
  50 & 25 & 3 & 2174.6 & 28.6 & 1.2 & 0.9 & 2.1 & 131109 \\ 
\hline
\end{longtable}
\endgroup

\newpage

\begingroup
\tiny
\begin{longtable}{rrrrrrrrrrrrrrr}
\caption{Instances results table for model $F1_{u}^{sub2}$} \\ 
$|V|$ & p & ins & cpu & objU & objL & objR & obj$\overline{\text{U}}$ & obj$\overline{\text{L}}$ & gUL & g$\overline{\text{U}}$L & gU$\overline{\text{L}}$ & g$\overline{\text{U}}$R & opt & nod \\  
  \hline
  20 & 5 & 1 & 7.6     & 15831.2 & 15831.2 & 8678.2  & 15831.2 & 15831.2 &  0.0  & 0.0  & 0.0  & 45.2 & 1 & 2152 \\ 
  20 & 5 & 2 & 1.8     & 16273.0 & 16273.0 & 11464.7 & 16273.0 & 16273.0 &  0.0  & 0.0  & 0.0  & 29.5 & 1 & 208 \\ 
  20 & 5 & 3 & 2.2     & 15039.1 & 15039.1 & 8966.5  & 15039.1 & 15039.1 &  0.0  & 0.0  & 0.0  & 40.4 & 1 & 520 \\ 
  20 & 5 & 4 & 1.8     & 9076.2  & 9076.2  & 5449.0  & 9076.2  & 9076.2  &  0.0  & 0.0  & 0.0  & 40.0 & 1 & 324 \\ 
  20 & 5 & 5 & 0.9     & 16943.6 & 16943.6 & 11867.1 & 16943.6 & 16943.6 &  0.0  & 0.0  & 0.0  & 30.0 & 1 & 1 \\ 
  20 & 6 & 1 & 1.3     & 11799.8 & 11799.8 & 7620.6  & 11799.8 & 11799.8 &  0.0  & 0.0  & 0.0  & 35.4 & 1 & 90 \\
\hline
\end{longtable}
\endgroup

\begingroup
\tiny
\begin{longtable}{rrrrrrrrr}
\caption{Summary results table for model $F1_{u}^{sub2}$} \\ 
$|V|$ & p & $|$\#$|$ & cpu & g$\overline{\text{U}}$R & g$\overline{\text{U}}$L & gU$\overline{\text{L}}$ & gUL & nod \\ 
  \hline
  20 & 5 & 5 & 2.9     & 37.0 & 0.0 & 0.0 & 0.0 & 641 \\ 
  20 & 6 & 5 & 1.5     & 35.0 & 0.0 & 0.0 & 0.0 & 191 \\ 
  20 & 10 & 5 & 7.9    & 28.9 & 0.0 & 0.0 & 0.0 & 4403 \\ 
  30 & 7 & 5 & 31.1    & 39.6 & 0.0 & 0.0 & 0.0 & 3763 \\ 
  30 & 10 & 5 & 32.2   & 33.1 & 0.0 & 0.0 & 0.0 & 5429 \\ 
  30 & 15 & 5 & 27.6   & 33.0 & 0.0 & 0.0 & 0.0 & 5379 \\ 
  40 & 10 & 4 & 780.5  & 33.3 & 1.7 & 0.0 & 1.7 & 75466 \\ 
  40 & 13 & 3 & 1543.3 & 35.4 & 2.8 & 0.0 & 2.8 & 152208 \\ 
  40 & 20 & 5 & 88.2   & 28.6 & 0.0 & 0.0 & 0.0 & 1520 \\ 
  50 & 12 & 3 & 2373.4 & 35.8 & 0.8 & 1.1 & 1.9 & 91614 \\ 
  50 & 16 & 4 & 1010.4 & 35.7 & 1.3 & 1.4 & 2.6 & 49093 \\ 
  50 & 25 & 4 & 1278.1 & 28.6 & 0.7 & 0.1 & 0.7 & 88801 \\ 
\hline
\end{longtable}
\endgroup

\newpage

\begingroup
\tiny
\begin{longtable}{rrrrrrrrrrrrrrr} 
\caption{Instances results table for model $F2_{u}^{mtz}$} \\ 
$|V|$ & p & ins & cpu & objU & objL & objR & obj$\overline{\text{U}}$ & obj$\overline{\text{L}}$ & gUL & g$\overline{\text{U}}$L & gU$\overline{\text{L}}$ & g$\overline{\text{U}}$R & opt & nod \\  
  \hline
  20 & 5 & 1 & 2.3 & 15831.2 & 15831.2   & 11628.1   & 15831.2 & 15831.2 & 0.0 & 0.0 & 0.0   & 26.5   & 1 & 37 \\ 
  20 & 5 & 2 & 2.0 & 16273.0 & 16273.0   & 13320.5   & 16273.0 & 16273.0 & 0.0 & 0.0 & 0.0   & 18.1   & 1 & 1 \\ 
  20 & 5 & 3 & 3.8 & 15039.1 & 15039.1   & 11682.6   & 15039.1 & 15039.1 & 0.0 & 0.0 & 0.0   & 22.3   & 1 & 465 \\ 
  20 & 5 & 4 & 1.6 & 9076.2 & 9076.2     & 6951.0    & 9076.2 & 9076.2 & 0.0 & 0.0 & 0.0     & 23.4   & 1 & 1 \\ 
  20 & 5 & 5 & 0.5 & 16943.6 & 16943.6   & 14144.9   & 16943.6 & 16943.6 & 0.0 & 0.0 & 0.0   & 16.5   & 1 & 1 \\ 
  20 & 6 & 1 & 0.6 & 11799.8 & 11799.8   & 10139.9   & 11799.8 & 11799.8 & 0.0 & 0.0 & 0.0   & 14.1   & 1 & 1 \\
\hline
\end{longtable}
\endgroup

\begingroup
\tiny
\begin{longtable}{rrrrrrrrr}
\caption{Summary results table for model $F2_{u}^{mtz}$} \\ 
$|V|$ & p & $|$\#$|$ & cpu & g$\overline{\text{U}}$R & g$\overline{\text{U}}$L & gU$\overline{\text{L}}$ & gUL & nod \\  
  \hline
  20 & 5 & 5 & 2.0     &  21.4   & 0.0 & 0.0 & 0.0 & 101 \\ 
  20 & 6 & 5 & 1.4     &  12.6   & 0.0 & 0.0 & 0.0 & 1 \\ 
  20 & 10 & 5 & 0.9    &  8.7    & 0.0 & 0.0 & 0.0 & 12 \\ 
  30 & 7 & 5 & 18.3    &  20.6   & 0.0 & 0.0 & 0.0 & 75 \\ 
  30 & 10 & 5 & 7.9    &  10.7   & 0.0 & 0.0 & 0.0 & 44 \\ 
  30 & 15 & 5 & 4.5    &  7.2    & 0.0 & 0.0 & 0.0 & 47 \\ 
  40 & 10 & 5 & 57.6   &  17.6   & 0.0 & 0.0 & 0.0 & 375 \\ 
  40 & 13 & 5 & 67.3   &  15.4   & 0.0 & 0.0 & 0.0 & 1494 \\ 
  40 & 20 & 5 & 10.7   &  6.8    & 0.0 & 0.0 & 0.0 & 15 \\ 
  50 & 12 & 5 & 176.6  &  17.9   & 0.0 & 0.0 & 0.0 & 237 \\ 
  50 & 16 & 5 & 97.5   &  12.7   & 0.0 & 0.0 & 0.0 & 1087 \\ 
  50 & 25 & 5 & 51.8   &  6.3    & 0.0 & 0.0 & 0.0 & 8883 \\ 
\hline
\end{longtable}
\endgroup

\newpage

\begingroup
\tiny
\begin{longtable}{rrrrrrrrrrrrrrr}
\caption{Instances results table for model $F2_{u}^{flow}$} \\ 
$|V|$ & p & ins & cpu & objU & objL & objR & obj$\overline{\text{U}}$ & obj$\overline{\text{L}}$ & gUL & g$\overline{\text{U}}$L & gU$\overline{\text{L}}$ & g$\overline{\text{U}}$R & opt & nod \\  
  \hline
  20 &  5 & 1 & 4.3    & 15831.2 & 15831.2  & 11628.2  & 15831.2 & 15831.2 & 0.0  & 0.0  & 0.0  & 26.5 & 1 & 1329 \\ 
  20 &  5 & 2 & 3.1    & 16273.0 & 16273.0  & 13320.5  & 16273.0 & 16273.0 & 0.0  & 0.0  & 0.0  & 18.1 & 1 & 271 \\ 
  20 &  5 & 3 & 2.7    & 15039.1 & 15039.1  & 11682.6  & 15039.1 & 15039.1 & 0.0  & 0.0  & 0.0  & 22.3 & 1 & 1 \\ 
  20 &  5 & 4 & 1.6    & 9076.2  & 9076.2   & 6951.0   & 9076.2  & 9076.2  & 0.0  & 0.0  & 0.0  & 23.4 & 1 & 1 \\ 
  20 &  5 & 5 & 1.3    & 16943.6 & 16943.6  & 14127.3  & 16943.6 & 16943.6 & 0.0  & 0.0  & 0.0  & 16.6 & 1 & 1 \\ 
  20 &  6 & 1 & 0.7    & 11799.8 & 11799.8  & 10139.9  & 11799.8 & 11799.8 & 0.0  & 0.0  & 0.0  & 14.1 & 1 & 1 \\
\hline
\end{longtable}
\endgroup

\begingroup
\tiny
\begin{longtable}{rrrrrrrrr}
\caption{Summary results table for model $F2_{u}^{flow}$} \\ 
$|V|$ & p & $|$\#$|$ & cpu & g$\overline{\text{U}}$R & g$\overline{\text{U}}$L & gU$\overline{\text{L}}$ & gUL & nod \\
  \hline
  20 & 5 & 5 & 2.6     & 21.4 & 0.0 & 0.0 & 0.0 & 321 \\ 
  20 & 6 & 5 & 1.1     & 12.6 & 0.0 & 0.0 & 0.0 & 1 \\ 
  20 & 10 & 5 & 9.0    &  9.7 & 0.0 & 0.0 & 0.0 & 4303 \\ 
  30 & 7 & 5 & 18.0    & 20.6 & 0.0 & 0.0 & 0.0 & 722 \\ 
  30 & 10 & 5 & 12.7   & 11.0 & 0.0 & 0.0 & 0.0 & 1928 \\ 
  30 & 15 & 5 & 13.4   &  9.7 & 0.0 & 0.0 & 0.0 & 1048 \\ 
  40 & 10 & 4 & 772.2  & 18.8 & 0.9 & 0.0 & 0.9 & 30383 \\ 
  40 & 13 & 3 & 1476.2 & 16.8 & 2.9 & 0.0 & 2.9 & 89230 \\ 
  40 & 20 & 5 & 16.6   &  8.0 & 0.0 & 0.0 & 0.0 & 376 \\ 
  50 & 12 & 5 & 500.4  & 18.4 & 0.0 & 0.0 & 0.0 & 13912 \\ 
  50 & 16 & 5 & 590.4  & 13.7 & 0.0 & 0.0 & 0.0 & 37986 \\ 
  50 & 25 & 4 & 753.9  &  7.0 & 0.3 & 0.0 & 0.3 & 147065 \\ 
\hline
\end{longtable}
\endgroup

\newpage

\begingroup
\tiny
\begin{longtable}{rrrrrrrrrrrrrrr}
\caption{Instances results table for model $F2_{u}^{km}$} \\ 
$|V|$ & p & ins & cpu & objU & objL & objR & obj$\overline{\text{U}}$ & obj$\overline{\text{L}}$ & gUL & g$\overline{\text{U}}$L & gU$\overline{\text{L}}$ & g$\overline{\text{U}}$R & opt & nod \\  
  \hline
  20 &  5 & 1 & 5.2    & 15831.2 & 15831.2 & 11820.0 & 15831.2 & 15831.2 & 0.0 & 0.0  & 0.0  & 25.3 & 1 & 1 \\ 
  20 &  5 & 2 & 5.6    & 16273.0 & 16273.0 & 13586.0 & 16273.0 & 16273.0 & 0.0 & 0.0  & 0.0  & 16.5 & 1 & 1 \\ 
  20 &  5 & 3 & 6.9    & 15039.1 & 15039.1 & 11919.1 & 15039.1 & 15039.1 & 0.0 & 0.0  & 0.0  & 20.8 & 1 & 9 \\ 
  20 &  5 & 4 & 4.7    & 9076.2  & 9076.2  & 7166.6  & 9076.2  & 9076.2  & 0.0 & 0.0  & 0.0  & 21.0 & 1 & 1 \\ 
  20 &  5 & 5 & 1.1    & 16943.6 & 16943.6 & 14722.7 & 16943.6 & 16943.6 & 0.0 & 0.0  & 0.0  & 13.1 & 1 & 1 \\ 
  20 &  6 & 1 & 1.1    & 11799.8 & 11799.8 & 10272.4 & 11799.8 & 11799.8 & 0.0 & 0.0  & 0.0  & 12.9 & 1 & 1 \\
\hline
\end{longtable}
\endgroup

\begingroup
\tiny
\begin{longtable}{rrrrrrrrr}
\caption{Summary results table for model $F2_{u}^{km}$} \\ 
$|V|$ & p & $|$\#$|$ & cpu & g$\overline{\text{U}}$R & g$\overline{\text{U}}$L & gU$\overline{\text{L}}$ & gUL & nod \\
  \hline
  20 & 5 & 5  & 4.7   & 19.3 & 0.0 & 0.0 & 0.0 & 3 \\ 
  20 & 6 & 5  & 2.4   & 10.7 & 0.0 & 0.0 & 0.0 & 67 \\ 
  20 & 10 & 5 & 1.2   & 3.5  & 0.0 & 0.0 & 0.0 & 1 \\ 
  30 & 7 & 5  & 37.7  & 18.6 & 0.0 & 0.0 & 0.0 & 77 \\ 
  30 & 10 & 5 & 14.6  & 7.5  & 0.0 & 0.0 & 0.0 & 28 \\ 
  30 & 15 & 5 & 11.8  & 3.8  & 0.0 & 0.0 & 0.0 & 1 \\ 
  40 & 10 & 5 & 258.6 & 13.7 & 0.0 & 0.0 & 0.0 & 141 \\ 
  40 & 13 & 5 & 88.9  & 12.5 & 0.0 & 0.0 & 0.0 & 14 \\ 
  40 & 20 & 5 & 22.4  & 2.5  & 0.0 & 0.0 & 0.0 & 1 \\ 
  50 & 12 & 5 & 702.2 & 15.6 & 0.0 & 0.0 & 0.0 & 62 \\ 
  50 & 16 & 5 & 167.5 & 10.5 & 0.0 & 0.0 & 0.0 & 1 \\ 
  50 & 25 & 5 & 63.7  & 3.4  & 0.0 & 0.0 & 0.0 & 1 \\ 
\hline
\end{longtable}
\endgroup

\newpage

\begingroup
\tiny
\begin{longtable}{rrrrrrrrrrrrrrr}
\caption{Instances results table for model $F2_{u}^{sub}$} \\ 
$|V|$ & p & ins & cpu & objU & objL & objR & obj$\overline{\text{U}}$ & obj$\overline{\text{L}}$ & gUL & g$\overline{\text{U}}$L & gU$\overline{\text{L}}$ & g$\overline{\text{U}}$R & opt & nod \\   
  \hline
  20 &  5 & 1 & 11.5   & 15831.2 & 15831.2 & 11628.1 & 15831.2 & 15831.2 &  0.0 &  0.0 &  0.0 & 26.6 & 1 & 3378 \\ 
  20 &  5 & 2 & 2.8    & 16273.0 & 16273.0 & 13320.5 & 16273.0 & 16273.0 &  0.0 &  0.0 &  0.0 & 18.1 & 1 & 1214 \\ 
  20 &  5 & 3 & 2.5    & 15039.1 & 15039.1 & 11682.6 & 15039.1 & 15039.1 &  0.0 &  0.0 &  0.0 & 22.3 & 1 & 868 \\ 
  20 &  5 & 4 & 2.0    & 9076.2  & 9076.2  & 6951.0  & 9076.2  & 9076.2  &  0.0 &  0.0 &  0.0 & 23.4 & 1 & 118 \\ 
  20 &  5 & 5 & 0.7    & 16943.6 & 16943.6 & 14127.3 & 16943.6 & 16943.6 &  0.0 &  0.0 &  0.0 & 16.6 & 1 & 1 \\ 
  20 &  6 & 1 & 0.4    & 11799.8 & 11799.8 & 10139.9 & 11799.8 & 11799.8 &  0.0 &  0.0 &  0.0 & 14.1 & 1 & 1 \\
\hline
\end{longtable}
\endgroup

\begingroup
\tiny
\begin{longtable}{rrrrrrrrr}
\caption{Summary results table for model $F2_{u}^{sub}$} \\ 
$|V|$ & p & $|$\#$|$ & cpu & g$\overline{\text{U}}$R & g$\overline{\text{U}}$L & gU$\overline{\text{L}}$ & gUL & nod \\
  \hline
  20 & 5 & 5 & 3.9     & 21.4 & 0.0 & 0.0 & 0.0 & 1116 \\ 
  20 & 6 & 5 & 1.2     & 12.6 & 0.0 & 0.0 & 0.0 & 66 \\ 
  20 & 10 & 5 & 6.0    &  9.9 & 0.0 & 0.0 & 0.0 & 5004 \\ 
  30 & 7 & 5 & 38.7    & 20.7 & 0.0 & 0.0 & 0.0 & 5613 \\ 
  30 & 10 & 5 & 23.5   & 11.1 & 0.0 & 0.0 & 0.0 & 5012 \\ 
  30 & 15 & 5 & 27.8   &  9.7 & 0.0 & 0.0 & 0.0 & 3712 \\ 
  40 & 10 & 4 & 760.4  & 18.8 & 1.9 & 0.1 & 2.0 & 34009 \\ 
  40 & 13 & 3 & 1493.0 & 16.9 & 2.4 & 0.0 & 2.4 & 118491 \\ 
  40 & 20 & 5 & 47.5   &  8.5 & 0.0 & 0.0 & 0.0 & 1146 \\ 
  50 & 12 & 4 & 2218.1 & 18.4 & 0.4 & 0.0 & 0.4 & 78321 \\ 
  50 & 16 & 4 & 893.1  & 13.8 & 1.4 & 0.9 & 2.3 & 33043 \\ 
  50 & 25 & 4 & 887.9  & 7.1 & 0.7 & 0.0 & 0.7 & 70353 \\ 
\hline
\end{longtable}
\endgroup

\newpage

\begingroup
\tiny
\input{results/instances_table_model_24_1.tex}
\endgroup

\begingroup
\tiny
\input{results/summary_table_model_24_1.tex}
\endgroup

\newpage

\begingroup
\tiny
\input{results/instances_table_model_25_1.tex}
\endgroup

\begingroup
\tiny
\input{results/summary_table_model_25_1.tex}
\endgroup

\newpage

\subsection{$k$-center criterion}

\begingroup
\tiny
\input{results/instances_table_model_1_2.tex}
\endgroup

\begingroup
\tiny
\input{results/summary_table_model_1_2.tex}
\endgroup

\newpage

\begingroup
\tiny
\input{results/instances_table_model_2_2.tex}
\endgroup

\begingroup
\tiny
\input{results/summary_table_model_2_2.tex}
\endgroup

\newpage

\begingroup
\tiny
\input{results/instances_table_model_3_2.tex}
\endgroup

\begingroup
\tiny
\input{results/summary_table_model_3_2.tex}
\endgroup

\newpage

\begingroup
\tiny
\input{results/instances_table_model_4_2.tex}
\endgroup

\begingroup
\tiny
\input{results/summary_table_model_4_2.tex}
\endgroup

\newpage

\begingroup
\tiny
\input{results/instances_table_model_5_2.tex}
\endgroup

\begingroup
\tiny
\input{results/summary_table_model_5_2.tex}
\endgroup

\newpage

\begingroup
\tiny
\input{results/instances_table_model_6_2.tex}
\endgroup

\begingroup
\tiny
\input{results/summary_table_model_6_2.tex}
\endgroup

\newpage

\begingroup
\tiny
\input{results/instances_table_model_7_2.tex}
\endgroup

\begingroup
\tiny
\input{results/summary_table_model_7_2.tex}
\endgroup

\newpage

\begingroup
\tiny
\input{results/instances_table_model_8_2.tex}
\endgroup

\begingroup
\tiny
\input{results/summary_table_model_8_2.tex}
\endgroup

\newpage

\begingroup
\tiny
\input{results/instances_table_model_9_2.tex}
\endgroup

\begingroup
\tiny
\input{results/summary_table_model_9_2.tex}
\endgroup

\newpage

\begingroup
\tiny
\input{results/instances_table_model_10_2.tex}
\endgroup

\begingroup
\tiny
\input{results/summary_table_model_10_2.tex}
\endgroup

\newpage

\begingroup
\tiny
\input{results/instances_table_model_11_2.tex}
\endgroup

\begingroup
\tiny
\input{results/summary_table_model_11_2.tex}
\endgroup

\newpage

\begingroup
\tiny
\input{results/instances_table_model_12_2.tex}
\endgroup

\begingroup
\tiny
\input{results/summary_table_model_12_2.tex}
\endgroup

\newpage

\begingroup
\tiny
\input{results/instances_table_model_13_2.tex}
\endgroup

\begingroup
\tiny
\input{results/summary_table_model_13_2.tex}
\endgroup

\newpage

\begingroup
\tiny
\input{results/instances_table_model_14_2.tex}
\endgroup

\begingroup
\tiny
\input{results/summary_table_model_14_2.tex}
\endgroup

\newpage

\begingroup
\tiny
\input{results/instances_table_model_15_2.tex}
\endgroup

\begingroup
\tiny
\input{results/summary_table_model_15_2.tex}
\endgroup

\newpage

\begingroup
\tiny
\input{results/instances_table_model_16_2.tex}
\endgroup

\begingroup
\tiny
\input{results/summary_table_model_16_2.tex}
\endgroup

\newpage

\begingroup
\tiny
\input{results/instances_table_model_17_2.tex}
\endgroup

\begingroup
\tiny
\input{results/summary_table_model_17_2.tex}
\endgroup

\newpage

\begingroup
\tiny
\input{results/instances_table_model_18_2.tex}
\endgroup

\begingroup
\tiny
\input{results/summary_table_model_18_2.tex}
\endgroup

\newpage

\begingroup
\tiny
\input{results/instances_table_model_19_2.tex}
\endgroup

\begingroup
\tiny
\input{results/summary_table_model_19_2.tex}
\endgroup

\newpage

\begingroup
\tiny
\input{results/instances_table_model_20_2.tex}
\endgroup

\begingroup
\tiny
\input{results/summary_table_model_20_2.tex}
\endgroup

\newpage

\begingroup
\tiny
\input{results/instances_table_model_21_2.tex}
\endgroup

\begingroup
\tiny
\input{results/summary_table_model_21_2.tex}
\endgroup

\newpage

\begingroup
\tiny
\input{results/instances_table_model_22_2.tex}
\endgroup

\begingroup
\tiny
\input{results/summary_table_model_22_2.tex}
\endgroup

\newpage

\begingroup
\tiny
\input{results/instances_table_model_23_2.tex}
\endgroup

\begingroup
\tiny
\input{results/summary_table_model_23_2.tex}
\endgroup

\newpage

\begingroup
\tiny
\input{results/instances_table_model_24_2.tex}
\endgroup

\begingroup
\tiny
\input{results/summary_table_model_24_2.tex}
\endgroup

\newpage

\begingroup
\tiny
\input{results/instances_table_model_25_2.tex}
\endgroup

\begingroup
\tiny
\input{results/summary_table_model_25_2.tex}
\endgroup

\newpage

\subsection{$k$-trimmed mean criterion}

\begingroup
\tiny
\input{results/instances_table_model_1_3.tex}
\endgroup

\begingroup
\tiny
\input{results/summary_table_model_1_3.tex}
\endgroup

\newpage

\begingroup
\tiny
\input{results/instances_table_model_2_3.tex}
\endgroup

\begingroup
\tiny
\input{results/summary_table_model_2_3.tex}
\endgroup

\newpage

\begingroup
\tiny
\input{results/instances_table_model_3_3.tex}
\endgroup

\begingroup
\tiny
\input{results/summary_table_model_3_3.tex}
\endgroup

\newpage

\begingroup
\tiny
\input{results/instances_table_model_4_3.tex}
\endgroup

\begingroup
\tiny
\input{results/summary_table_model_4_3.tex}
\endgroup

\newpage

\begingroup
\tiny
\input{results/instances_table_model_5_3.tex}
\endgroup

\begingroup
\tiny
\input{results/summary_table_model_5_3.tex}
\endgroup

\newpage

\begingroup
\tiny
\input{results/instances_table_model_6_3.tex}
\endgroup

\begingroup
\tiny
\input{results/summary_table_model_6_3.tex}
\endgroup

\newpage

\begingroup
\tiny
\input{results/instances_table_model_7_3.tex}
\endgroup

\begingroup
\tiny
\input{results/summary_table_model_7_3.tex}
\endgroup

\newpage

\begingroup
\tiny
\input{results/instances_table_model_8_3.tex}
\endgroup

\begingroup
\tiny
\input{results/summary_table_model_8_3.tex}
\endgroup

\newpage

\begingroup
\tiny
\input{results/instances_table_model_9_3.tex}
\endgroup

\begingroup
\tiny
\input{results/summary_table_model_9_3.tex}
\endgroup

\newpage

\begingroup
\tiny
\input{results/instances_table_model_10_3.tex}
\endgroup

\begingroup
\tiny
\input{results/summary_table_model_10_3.tex}
\endgroup

\newpage

\begingroup
\tiny
\input{results/instances_table_model_11_3.tex}
\endgroup

\begingroup
\tiny
\input{results/summary_table_model_11_3.tex}
\endgroup

\newpage

\begingroup
\tiny
\input{results/instances_table_model_12_3.tex}
\endgroup

\begingroup
\tiny
\input{results/summary_table_model_12_3.tex}
\endgroup

\newpage

\begingroup
\tiny
\input{results/instances_table_model_13_3.tex}
\endgroup

\begingroup
\tiny
\input{results/summary_table_model_13_3.tex}
\endgroup

\newpage

\begingroup
\tiny
\input{results/instances_table_model_14_3.tex}
\endgroup

\begingroup
\tiny
\input{results/summary_table_model_14_3.tex}
\endgroup

\newpage

\begingroup
\tiny
\input{results/instances_table_model_15_3.tex}
\endgroup

\begingroup
\tiny
\input{results/summary_table_model_15_3.tex}
\endgroup

\newpage

\begingroup
\tiny
\input{results/instances_table_model_16_3.tex}
\endgroup

\begingroup
\tiny
\input{results/summary_table_model_16_3.tex}
\endgroup

\newpage

\begingroup
\tiny
\input{results/instances_table_model_17_3.tex}
\endgroup

\begingroup
\tiny
\input{results/summary_table_model_17_3.tex}
\endgroup

\newpage

\begingroup
\tiny
\input{results/instances_table_model_18_3.tex}
\endgroup

\begingroup
\tiny
\input{results/summary_table_model_18_3.tex}
\endgroup

\newpage

\begingroup
\tiny
\input{results/instances_table_model_19_3.tex}
\endgroup

\begingroup
\tiny
\input{results/summary_table_model_19_3.tex}
\endgroup

\newpage

\begingroup
\tiny
\input{results/instances_table_model_20_3.tex}
\endgroup

\begingroup
\tiny
\input{results/summary_table_model_20_3.tex}
\endgroup

\newpage

\begingroup
\tiny
\input{results/instances_table_model_21_3.tex}
\endgroup

\begingroup
\tiny
\input{results/summary_table_model_21_3.tex}
\endgroup

\newpage

\begingroup
\tiny
\input{results/instances_table_model_22_3.tex}
\endgroup

\begingroup
\tiny
\input{results/summary_table_model_22_3.tex}
\endgroup

\newpage

\begingroup
\tiny
\input{results/instances_table_model_23_3.tex}
\endgroup

\begingroup
\tiny
\input{results/summary_table_model_23_3.tex}
\endgroup

\newpage

\begingroup
\tiny
\input{results/instances_table_model_24_3.tex}
\endgroup

\begingroup
\tiny
\input{results/summary_table_model_24_3.tex}
\endgroup

\newpage

\begingroup
\tiny
\input{results/instances_table_model_25_3.tex}
\endgroup

\begingroup
\tiny
\input{results/summary_table_model_25_3.tex}
\endgroup

\newpage

\bibliography{OMT_extended}

\end{document}